\documentclass[preprintnumbers,nofootinbib,showpacs,floatfix,letterpaper]{revtex4}

\usepackage{graphicx}
\usepackage{latexsym}
\usepackage[english]{babel}
\usepackage{amsmath}
\usepackage{amsthm}
\usepackage{epsfig}
\usepackage{color}
\usepackage{amssymb}
\usepackage{epsfig}
\usepackage{verbatim}
\usepackage{psfrag}
\usepackage{bm}
\usepackage{bbm}
\usepackage{amssymb}

\begin{document}

\title{\Large Decoherence in an Interacting Quantum Field Theory: \\ Thermal Case}

\preprint{ITP-UU-11/06, SPIN-11/04}

\preprint{HD-THEP-11-4}


\author{Jurjen F. Koksma}
\affiliation{Institute for Theoretical Physics (ITP) \& Spinoza
Institute, Utrecht University, Postbus 80195, 3508 TD Utrecht, The
Netherlands and \\
Dept. of Physics and MIT Kavli Institute,
Massachusetts Institute of Technology, 77 Massachusetts Ave.,
Cambridge, MA 02139, USA
\\ \texttt{\textup{J.F.Koksma@uu.nl}}}

\author{Tomislav Prokopec}
\affiliation{Institute for Theoretical Physics (ITP) \& Spinoza
Institute, Utrecht University, Postbus 80195, 3508 TD Utrecht, The
Netherlands \\ \texttt{\textup{T.Prokopec@uu.nl}}}

\author{Michael G. Schmidt}
\affiliation{Institut f\"ur Theoretische Physik, Heidelberg
University, Philosophenweg 16, D-69120 Heidelberg, Germany
\\ \texttt{\textup{M.G.Schmidt@thphys.uni-heidelberg.de}}}

\begin{abstract}
We study the decoherence of a renormalised quantum field
theoretical system. We consider our novel correlator approach to
decoherence where entropy is generated by neglecting
observationally inaccessible correlators. Using out-of-equilibrium
field theory techniques at finite temperatures, we show that the
Gaussian von Neumann entropy for a pure quantum state asymptotes
to the interacting thermal entropy. The decoherence rate can be
well described by the single particle decay rate in our model.
Connecting to electroweak baryogenesis scenarios, we moreover
study the effects on the entropy of a changing mass of the system
field. Finally, we compare our correlator approach to existing
approaches to decoherence in the simple quantum mechanical
analogue of our field theoretical model. The entropy following
from the perturbative master equation suffers from physically
unacceptable secular growth.
\end{abstract}

\maketitle

\section{Introduction}
\label{Introduction}

Recently, we have advocated a new decoherence program
\cite{Koksma:2009wa, Koksma:2010zi, Koksma:2010dt, Koksma:2011fx}
particularly designed for applications in quantum field theory.
Similar ideas have been proposed by Giraud and Serreau
\cite{Giraud:2009tn} independently. Older work can already be
interpreted in a similar spirit \cite{Prokopec:1992ia,
Brandenberger:1992jh, Kiefer:2006je, Campo:2008ju, Campo:2008ij}.
Like in the conventional approach to decoherence we assume the
existence of a distinct system, environment and observer (see e.g.
\cite{Zeh:1970, Zurek:1981xq, Joos:1984uk, Joos:book,
Zurek:2003zz}). Rather than tracing over the unaccessible
environmental degrees of freedom of the density matrix to obtain
the reduced density matrix
$\hat{\rho}_{\mathrm{red}}=\mathrm{Tr}_{E}[\hat{\rho}]$, we use
the well known idea that loss of information about a system leads
to an entropy increase as perceived by the observer. If an
observer performs a measurement on a quantum system, he or she
measures $n$-point correlators or correlation functions. Note that
these $n$-point correlators can also be mixed and contain
information about the correlation between the system and
environment. A ``perfect observer'' would in principle be able to
detect the infinite hierarchy of correlation functions up to
arbitrary order. In reality, our observer is of course limited by
the sensitivity of its measurement device. Also, higher order
correlation functions become more and more difficult to measure
due to their non-local character. Therefore, neglecting the
information stored in these unaccessible correlators will give
rise to an increase in entropy.

In other words, our system and environment evolve unitarily,
however to our observer it seems that the system evolves into a
mixed state with positive entropy as information about the system
is dispersed in inaccessible correlation functions. The total von
Neumann entropy $S_{\mathrm{vN}}$ can be subdivided as:
\begin{equation}\label{vNeumanEntropySplit}
S_{\mathrm{vN}} = S^{\mathrm{g}}(t) + S^{\mathrm{ng}}(t) =
S^{\mathrm{g}}_S + S^{\mathrm{g}}_E + S^{\mathrm{ng}}\,.
\end{equation}
In unitary theories $S_{\mathrm{vN}}$ is conserved. In the
equation above $S^{\mathrm{g}}$ is the total Gaussian von Neumann
entropy, that contains information about both the system
$S^{\mathrm{g}}_S$, environment $S^{\mathrm{g}}_E$ and their
correlations at the Gaussian level $S^{\mathrm{g}}_{SE}$ (which
vanish in this paper), and $S^{\mathrm{ng}}$ is the total
non-Gaussian von Neumann entropy which consists again of
contributions from the system, environment and their correlations.
Although $S_{\mathrm{vN}}$ is conserved in unitary theories,
$S^{\mathrm{g}}_S(t)$ can increase at the expense of other
decreasing contributions to the total von Neumann entropy, such as
$S^{\mathrm{ng}}(t)$.

In the conventional approach one attempts to solve for the reduced
density matrix by making use of a non-unitary perturbative
``master equation'' \cite{Paz:2000le}. It suffers from several
drawbacks. In the conventional approach to decoherence it is
extremely challenging to solve for the dynamics of the reduced
density matrix in a realistic interacting, out-of-equilibrium,
finite temperature quantum field theoretical setting that moreover
captures perturbative corrections arising from renormalisation. In
fact, we are not aware of any solution to the perturbative master
equation that meets these basic requirements\footnote{For example
in \cite{Burgess:2006jn} the decoherence of inflationary
primordial fluctuations is studied using the master equation
however renormalisation is not addressed. In \cite{Hu:1993vs,
Hu:1993qa, Calzetta:book} however, perturbative corrections to a
density matrix are calculated in various quantum mechanical
cases.}. Secondly, it is important to note that our approach does
not rely on non-unitary physics. Although the von Neumann equation
for the full density matrix is of course unitary, the perturbative
master equation is not. From a theoretical point of view it is
disturbing that the reduced density matrix should follow from a
non-unitary equation despite of the fact that the underlying
theory is unitary and hence the implications should be carefully
checked.

\subsection{Outline}
\label{Outline}

In this work, we study entropy generation in an interacting,
out-of-equilibrium, finite temperature field theory. We consider
the following action \cite{Koksma:2009wa}:
\begin{equation}\label{action:tree1}
S[\phi,\chi] = \int \mathrm{d}^{\scriptscriptstyle{D}}\!x {\cal
L}[\phi,\chi] = \int \mathrm{d}^{\scriptscriptstyle{D}}\! x {\cal
L}_{0}[\phi] + {\cal L}_{0}[\chi]+ {\cal
L}_{\mathrm{int}}[\phi,\chi] \,,
\end{equation}
where:
\begin{subequations}
\label{action:tree2}
\begin{eqnarray}
{\cal L}_{0}[\phi] &=& -\frac{1}{2} \partial_\mu\phi(x)
\partial_\nu \phi(x) \eta^{\mu\nu} - \frac{1}{2} m^{2}_{\phi}(t)
\phi^{2}(x)
\label{action:tree2a}\\
{\cal L}_{0}[\chi] &=& -\frac{1}{2} \partial_\mu\chi(x)
\partial_\nu \chi(x)\eta^{\mu\nu} - \frac{1}{2} m^{2}_{\chi} \chi^{2}(x)
\label{action:tree2b}\\
{\cal L}_{\mathrm{int}}[\phi,\chi] &=& -
\frac{\lambda}{3!}\chi^{3}(x) -\frac{1}{2}h \chi^{2}(x) \phi(x)
\,, \label{action:tree2c}
\end{eqnarray}
\end{subequations}
where $\eta_{\mu\nu}=\mathrm{diag}(-1,1,1,\cdots)$ is the
$D$-dimensional Minkowski metric. Here, $\phi(x)$ plays the role
of the system, interacting with an environment $\chi(x)$, where we
assume that $\lambda \gg h$ such that the environment is in
thermal equilibrium at temperature $T$. In \cite{Koksma:2009wa} we
studied an environment at temperature $T=0$, i.e.: an environment
in its vacuum state. In the present work, we study finite
temperature effects. We assume that $\langle\hat{\phi}\rangle = 0
= \langle\hat{\chi}\rangle$, which can be realised by suitably
renormalising the tadpoles.

Let us at this point explicitly state the two main assumptions of
our work. Firstly, we assume that the observer can only detect
Gaussian correlators or two-point functions and consequently
neglects the information stored in all higher order non-Gaussian
correlators (of both $\phi$ and of the correlation between $\phi$
and $\chi$). This assumption can of course be generalised to
incorporate knowledge of e.g. three- or four-point functions in
the definition of the entropy \cite{Koksma:2010zi}. Secondly, we
neglect the backreaction from the system field on the environment
field, i.e.: we assume that we can neglect the self-mass
corrections due to the $\phi$-field on the environment $\chi$.
This assumption is perturbatively well justified
\cite{Koksma:2009wa} and thus implies that the environment remains
in thermal equilibrium at temperature $T$. Consequently, the
counterterms introduced to renormalise the tadpoles do not depend
on time too such that we can remove these terms in a consistent
manner. In fact, the presence of the $\lambda \chi^{3}$
interaction will introduce perturbative thermal corrections to the
tree-level thermal state, which we neglect for simplicity in this
work.

The calculation we are about to embark on can be outlined as
follows. The first assumption above implies that we only use the
three Gaussian correlators to calculate the (Gaussian) von Neumann
entropy: $\langle
\hat{\phi}(\vec{x},t)\hat{\phi}(\vec{y},t)\rangle$, $\langle
\hat{\pi}(\vec{x},t) \hat{\pi}(\vec{y},t)\rangle$ and
$\frac{1}{2}\langle \{\hat{\phi}(\vec{x},t),
\hat{\pi}(\vec{y},t)\}\rangle$. Rather than attempting to solve
for the dynamics of these three correlators separately, we solve
for the statistical propagator from which these three Gaussian
correlators can be straightforwardly extracted. Starting from the
action in equation (\ref{action:tree1}), we thus calculate the 2PI
(two particle irreducible) effective action that captures the
perturbative loop corrections to the various propagators of our
system field. Most of our attention is thus devoted to calculate
the self-masses, renormalise the vacuum contributions to the
self-masses and deal with the memory integrals as a result of the
interaction between the two fields. Once we have the statistical
propagator at our disposal, our life becomes easier. Various
coincidence expressions of the statistical propagator and
derivatives thereof fix the Gaussian entropy of our system field
uniquely \cite{Koksma:2009wa, Koksma:2010zi,Campo:2008ij}.

In section \ref{Decoherence in an Interacting Quantum Field
Theory: Vacuum Case} we recall how to evaluate the Gaussian von
Neumann entropy by making use of the statistical propagator. We
moreover present the main results from \cite{Koksma:2009wa}. In
section \ref{Finite Temperature Contributions to the Self-Masses}
we evaluate the finite temperature contributions to the
self-masses. In section \ref{Entropy Generation in Quantum
Mechanics} we study the time evolution of the Gaussian von Neumann
entropy in quantum mechanical model analogous to equation
(\ref{action:tree1}). This allows to quantitatively compare our
results for the entropy evolution to existing approaches. In
section \ref{Results: Entropy Generation in Quantum Field Theory}
we study the time evolution of the Gaussian von Neumann entropy in
the field theoretic case and we discuss our main results.

\subsection{Applications}
\label{Applications}

The work presented in this paper is important for electroweak
baryogenesis scenarios. The attentive reader will have appreciated
that we allow for a changing mass of the system field in the
Lagrangian (\ref{action:tree2a}): $m_{\phi} = m_{\phi}(t)$.
Theories invoking new physics at the electroweak scale that try to
explain the observed baryon-antibaryon asymmetry in the universe
are usually collectively referred to as electroweak baryogenesis.
During a first order phase transition at the electroweak scale,
bubbles of the true vacuum emerge and expand in the sea of the
false vacuum. Particles thus experience a rapid change in their
mass as a bubble's wall passes by. Sakharov's conditions are
fulfilled during this violent process such that a baryon asymmetry
is supposed to be generated. The problem is to calculate axial
vector currents generated by a CP violating advancing phase
interface. These currents then feed in hot sphalerons, thus
biasing baryon production. The axial currents are difficult to
calculate, as it requires a controlled calculation of
non-equilibrium dynamics in a finite temperature plasma, taking a
non-adiabatically changing mass parameter into account. In this
work we do not consider fermions but scalar fields, yet the set-up
of our theory features many of the properties relevant for
electroweak baryogenesis: our interacting scalar field model
closely resembles the Yukawa part of the standard model
Lagrangian, where one scalar field plays the role of the Higgs
field and the other generalises to a heavy fermion (e.g. a top
quark or a chargino of a supersymmetric theory). The entropy is,
just as the axial vector current, sensitive to quantum coherence.
The relevance of scattering processes for electroweak baryogenesis
has been treated in several papers in the 1990s
\cite{Farrar:1993hn, Farrar:1993sp, Gavela:1993ts, Huet:1994jb,
Gavela:1994ds, Gavela:1994dt}, however no satisfactory solution to
the problem has been found so far. Quantum coherence also plays a
role in models where CP violating particle mixing is invoked to
source baryogenesis \cite{Balazs:2004bu, Konstandin:2005cd,
Huber:2006wf, Fromme:2006cm, Chung:2009qs}. More recently,
Herranen, Kainulainen and Rahkila \cite{Herranen:2008hu,
Herranen:2008hi, Herranen:2008di} observed that the constraint
equations for fermions and scalars admit a third shell at $k_0=0$.
The authors show that this third shell can be used to correctly
reproduce the Klein paradox both for fermions and bosons in a step
potential, and hope that their intrinsically off-shell formulation
can be used to include interactions in a field theoretical setting
for which off-shell physics is essential. The relevance of
coherence in baryogenesis for a phase transition at the end of
inflation has been addressed in \cite{Garbrecht:2003mn,
Garbrecht:2004gv, Garbrecht:2005rr}.

A second application is of course the study of out-of-equilibrium
quantum fields from a theoretical perspective. In recent years,
out-of-equilibrium dynamics of quantum fields has enjoyed a
considerable attention as the calculations involved become more
and more tractable (for an excellent review we refer to
\cite{Berges:2004yj}). Many calculations have been performed in
non-equilibrium $\lambda\phi^4(x)$, see e.g. \cite{Aarts:2002dj,
Juchem:2004cs, Arrizabalaga:2005tf}, however also see
\cite{Berges:2002wr, Anisimov:2008dz, Jackson:2010cw}. The
renormalisation of the Kadanoff-Baym equations has also received
considerable attention \cite{vanHees:2002bv, Borsanyi:2009zza,
Blaizot:2003br}. Calzetta and Hu \cite{Calzetta:2003dk} prove an
$H$-theorem for a quantum mechanical $O(N)$-model, also see
\cite{Nishiyama:2010wc}, and refer to ``correlation entropy'' what
we would call ``Gaussian von Neumann entropy''. A very interesting
study has been performed by Garny and M\"uller \cite{Garny:2009ni}
where renormalised Kadanoff-Baym equations in $\lambda\phi^4(x)$
are numerically integrated by imposing non-Gaussian initial
conditions at some initial time $t_0$. We differ in our approach
as we include memory effects before $t_0$ such that our evolution,
like Garny and M\"uller's, is divergence free at $t_0$.

Finally, we can expect that a suitable generalisation of our setup
in an expanding Universe can also be applied to the decoherence of
cosmological perturbations \cite{Brandenberger:1990bx,
Polarski:1995jg, Calzetta:1995ys, Lesgourgues:1996jc,
Kiefer:1998qe, Campo:2005sy, Lombardo:2005iz, Burgess:2006jn,
Martineau:2006ki, Lyth:2006qz, Prokopec:2006fc, Sharman:2007gi,
Kiefer:2007zza, Campo:2008ju, Sudarsky:2009za}. Undoubtedly the
most interesting aspect of inflation is that it provides us with a
causal mechanism to create the initial inhomogeneities of the
Universe by means of a quantum process that later grow out to
become the structure we observe today in the form of galaxies and
clusters of galaxies. Decoherence should bridge the gap between
the intrinsically quantum nature of the initial inhomogeneities
during inflation and the classical stochastic behaviour as assumed
in cosmological perturbation theory.

\section{Kadanoff-Baym Equation for the Statistical Propagator}
\label{Decoherence in an Interacting Quantum Field Theory: Vacuum
Case}

This section not only aims at summarising the main results of
\cite{Koksma:2009wa} we also rely upon in the present paper, but
we also extend the analysis of \cite{Koksma:2009wa} to incorporate
finite temperature effects.

There is a connection between the statistical propagator and the
Gaussian von Neumann entropy of a system. The Gaussian von Neumann
entropy per mode $S_{k}$ of a certain translationary invariant
quantum system is uniquely fixed by the phase space area
$\Delta_{k}$ the state occupies:
\begin{equation}\label{entropy}
S_{k}(t) = \frac{ \Delta_{k}(t)+1}{2}
\log\left(\frac{\Delta_{k}(t)+1}{2}\right) - \frac{
\Delta_{k}(t)-1}{2} \log\left(\frac{\Delta_{k}(t)-1}{2}\right) \,.
\end{equation}
The phase space area, in turn, is determined from the statistical
propagator $F_{\phi}(k,t,t')$:
\begin{equation}\label{deltaareainphasespace}
\Delta_{k}^{2}(t) = 4\left. \left[
F(k,t,t')\partial_{t}\partial_{t'}F(k,t,t') -
\left\{\partial_{t}F(k,t,t')\right\}^{2} \right] \right|_{t=t'}
\,.
\end{equation}
Throughout the paper we set $\hbar=1$ and $c=1$. The phase space
area indeed corresponds to the phase space area of (an appropriate
slicing of) the Wigner function, defined as a Wigner transform of
the density matrix \cite{Koksma:2010zi}. For a pure state we have
$\Delta_{k}=1$, $S_{k}=0$, whereas for a mixed state $\Delta_{k}
>1$, $S_{k} >0$. The expression for the Gaussian von Neumann entropy is only valid
for pure or mixed Gaussian states, and not for a class of pure
excited state such as eigenstates of the number operator as these
states are non-Gaussian. The statistical propagator describes how
states are populated and is in the Heisenberg picture defined by:
\begin{equation}\label{statisticalpropagator}
F_{\phi}(x;x') = \frac{1}{2} \mathrm{Tr} \left[
\hat{\rho}(t_{0})\{ \hat{\phi}(x'), \hat{\phi}(x) \} \right]=
\frac{1}{2} \mathrm{Tr} \left[ \hat{\rho}(t_{0}) (
\hat{\phi}(x')\hat{\phi}(x) + \hat{\phi}(x)\hat{\phi}(x') )
\right] \,,
\end{equation}
given some initial density matrix operator $\hat{\rho}(t_{0})$. In
spatially homogeneous backgrounds, we can Fourier transform e.g.
the statistical propagator as follows:
\begin{equation}\label{statpropagatorFourier}
F_{\phi}(k,t,t')=\int \mathrm{d}(\vec{x}-\vec{x}') F_{\phi}(x;x')
e^{-\imath \vec{k}\cdot(\vec{x}-\vec{x}')} \,,
\end{equation}
which in the spatially translationary invariant case we consider
in this paper only depends on $k=\|\vec{k}\|$. Finally, it is
interesting to note that the phase space area can be related to an
effective phase space particle number density per mode or the
statistical particle number density per mode as:
\begin{equation}\label{particlenumber}
n_{k}(t) = \frac{ \Delta_{k}(t)-1}{2}\,,
\end{equation}
in which case the entropy per mode just reduces to the familiar
entropy equation for a collection of $n$ free Bose-particles (this
is of course an effective description). The three Gaussian
correlators are straightforwardly related to the statistical
propagator:
\begin{subequations}
\label{3 equal time correlators}
\begin{flalign}
\langle \hat{\phi}(\vec{x},t) \hat{\phi}(\vec{y},t) \rangle &=
F_{\phi}(\vec{x},t;\vec{y},t')|_{t=t'}\label{3 equal time
correlatorsa}
\\
\langle \hat{\pi}(\vec{x},t) \hat{\pi}(\vec{y},t) \rangle &=
\partial_{t} \partial_{t'} F_{\phi}(\vec{x},t;\vec{y},t')|_{t=t'} \label{3 equal time
correlatorsb}
\\
\langle \{ \hat{\phi}(\vec{x},t), \hat{\pi}(\vec{y},t) \}/2
\rangle &=
\partial_{t'} F_{\phi}(\vec{x},t;\vec{y},t')|_{t=t'} \label{3 equal time
correlatorsc}\,.
\end{flalign}
\end{subequations}

In order to deal with the difficulties arising in interacting
non-equilibrium quantum field theory, we work in the
Schwinger-Keldysh formalism \cite{Schwinger:1960qe,
Keldysh:1964ud, Koksma:2007uq}, in which we can define the
following propagators:
\begin{subequations}
\label{propagators}
\begin{eqnarray}
\imath\Delta^{++}_{\phi}(x;x^\prime) &=&
\mathrm{Tr}\left[\hat{\rho}(t_{0})
T[\hat\phi(x^\prime)\hat\phi(x)] \right]
\label{propagatorsa} \\
\imath\Delta^{--}_{\phi}(x;x^\prime) &=&
 \mathrm{Tr}\left[\hat{\rho}(t_{0}) \overline{T} [ \hat\phi(x^\prime)\hat\phi(x)]
\right]
\label{propagatorsb} \\
\imath\Delta^{-+}_{\phi}(x;x^\prime) &=&
 \mathrm{Tr}\left[\hat{\rho}(t_{0})  \hat\phi(x)\hat\phi(x^\prime)\right]
\label{propagatorsc}
\\
\imath\Delta^{+-}_{\phi}(x;x^\prime) &=&
 \mathrm{Tr}\left[\hat{\rho}(t_{0})\hat\phi(x^\prime)\hat\phi(x)\right]
 \,,\label{propagatorsd}
\end{eqnarray}
\end{subequations}
where $t_0$ denotes an initial time, $\overline{T}$ and $T$ denote
the anti-time ordering and time ordering operations, respectively.
We define the various propagators for the $\chi$-field
analogously. In equation (\ref{propagators}),
$\imath\Delta^{++}_{\phi} \equiv \imath\Delta^{F}_{\phi}$ denotes
the Feynman or time ordered propagator and
$\imath\Delta^{--}_{\phi}$ represents the anti-time ordered
propagator. The two Wightman functions are given by
$\imath\Delta^{-+}_{\phi}$ and $\imath\Delta^{+-}_{\phi}$. In this
work, we are primarily interested in the causal propagator
$\Delta_{\phi}^{c}$ and statistical propagator $F_{\phi}$, which
are defined as follows:
\begin{subequations}
\label{propagators2}
\begin{eqnarray}
\imath\Delta^{c}_{\phi} (x;x^\prime) &=& \mathrm{Tr}\left(
\hat{\rho}(t_{0})  [\hat\phi(x),\hat\phi(x^\prime)]\right)= \imath
\Delta^{-+}_{\phi}(x;x^\prime) - \imath
\Delta^{+-}_{\phi}(x;x^\prime) \label{propagatorcausal} \\
F_{\phi}(x;x') &=& \frac{1}{2} \mathrm{Tr}\left[ \hat{\rho}(t_{0})
\{\hat \phi(x'),\hat\phi(x)\} \right]=
\frac{1}{2}\Big(\imath\Delta^{-+}_{\phi}(x;x') +
\imath\Delta^{+-}_{\phi}(x;x')\Big) \label{propagatorsstatistical}
\,.
\end{eqnarray}
\end{subequations}
We can express all propagators $\imath \Delta^{ab}_{\phi}$ solely
in terms of the causal and statistical propagators:
\begin{subequations}
\label{reduction:F+Deltac}
\begin{eqnarray}
\imath \Delta^{+-}_{\phi}(x;x^\prime) &=& F_{\phi}(x;x^\prime)-
\frac{1}{2}\imath\Delta^{c}_{\phi}(x;x^\prime)
\label{reduction:F+Deltaca}\\
\imath \Delta^{-+}_{\phi}(x;x^\prime) &=& F_{\phi}(x;x^\prime) +
\frac{1}{2}\imath\Delta^{c}_{\phi}(x;x^\prime)
\label{reduction:F+Deltacb}\\
\imath \Delta^{++}_{\phi}(x;x^\prime) &=& F_{\phi}(x;x^\prime) +
\frac{1}{2}\mathrm{sgn}(t- t^\prime)
\imath\Delta^{c}_{\phi}(x;x^\prime)
\label{reduction:F+Deltacc}\\
\imath \Delta^{--}_{\phi}(x;x^\prime) &=& F_{\phi}(x;x^\prime)-
\frac{1}{2}\mathrm{sgn}(t-t^\prime)\imath\Delta^{c}_{\phi}(x;x^\prime)
\,. \label{reduction:F+Deltacd}
\end{eqnarray}
\end{subequations}

In order to study the effect of perturbative loop corrections on
classical expectation values, we consider the 2PI effective
action, using the Schwinger-Keldysh formalism outlined above.
Variation of the 2PI effective action with respect to the
propagators yields the so-called Kadanoff-Baym equations that
govern the dynamics of the propagators and contain the non-local
scalar self-energy corrections or self-mass corrections to the
propagators. The Kadanoff-Baym equations for the system field
read:
\begin{subequations}
\label{EOM3aExtended}
\begin{eqnarray}
(\partial_{x}^{2}-m^{2}_{\phi}) \imath
\Delta^{++}_{\phi}(x;x^\prime) - \int
\mathrm{d}^{\scriptscriptstyle{D}}\! y \left[\imath
M^{++}_{\phi}(x;y)\imath \Delta^{++}_{\phi}(y;x^\prime) - \imath
M^{+-}_{\phi}(x;y)\imath \Delta^{-+}_{\phi}(y;x^\prime)\right] &=&
\, \imath\delta^{\scriptscriptstyle{D}}\!(x-x^\prime)
\label{EOM3a++} \\
(\partial_{x}^{2} - m^{2}_{\phi})\imath
\Delta^{+-}_{\phi}(x;x^\prime)- \int
\mathrm{d}^{\scriptscriptstyle{D}}\!y \left[\imath
M^{++}_{\phi}(x;y)\imath \Delta^{+-}_{\phi}(y;x^\prime) - \imath
M^{+-}_{\phi}(x;y)\imath \Delta^{--}_{\phi}(y;x^\prime)\right] &=&
0
\label{EOM3a+-} \\
(\partial_{x}^{2} - m^{2}_{\phi})\imath
\Delta^{-+}_{\phi}(x;x^\prime) - \int
\mathrm{d}^{\scriptscriptstyle{D}}\!y \left[\imath
M^{-+}_{\phi}(x;y)\imath \Delta^{++}_{\phi}(y;x^\prime) - \imath
M^{--}_{\phi}(x;y)\imath \Delta^{-+}_{\phi}(y;x^\prime)\right] &=&
0
\label{EOM3a-+} \\
(\partial_{x}^{2}-m^{2}_{\phi})\imath
\Delta^{--}_{\phi}(x;x^\prime) - \int
\mathrm{d}^{\scriptscriptstyle{D}}\!y \left[\imath
M^{-+}_{\phi}(x;y)\imath \Delta^{+-}_{\phi}(y;x^\prime) - \imath
M^{--}_{\phi}(x;y)\imath \Delta^{--}_{\phi} (y;x^\prime)\right]
&=& -\imath \delta^{\scriptscriptstyle{D}}\!(x-x^\prime)  \,,
\label{EOM3a--}
\end{eqnarray}
\end{subequations}
where the self-masses at one loop have the form:
\begin{subequations}
\label{selfMass}
\begin{eqnarray}
\imath M^{ac}_{\phi}(x;x_1) &=&
 -\frac{\imath h^{2}}{2} \left(\imath
\Delta^{ac}_{\chi}(x;x_1)\right)^{2}
\label{selfMassa} \\
 \imath M^{ac}_{\chi}(x;x_1) &=&
 -\frac{\imath \lambda^{2}}{2} \left(\imath
\Delta^{ac}_{\chi}(x;x_1)\right)^{2} - \imath h^{2} \imath
\Delta^{ac}_{\chi}(x;x_1) \imath \Delta^{ac}_{\phi}(x;x_1)
\label{selfMassb} \,.
\end{eqnarray}
\end{subequations}
Note that we have another set of four equations of motion for the
$\chi$-field. We define a Fourier transform as:
\begin{subequations}
\label{Fouriertransformdef}
\begin{eqnarray}
\imath \Delta_{\phi}^{ab}(x;x')= \int
\frac{\mathrm{d}^{\scriptscriptstyle{D}-1}\vec{k}}{(2\pi)^{\scriptscriptstyle{D}-1}}
\imath \Delta_{\phi}^{ab}(\vec{k},t,t') e^{\imath
\vec{k}\cdot(\vec{x}-\vec{x}')}
\label{Fouriertransformdefa} \\
\imath \Delta_{\phi}^{ab}(\vec{k},t,t')= \int
\mathrm{d}^{\scriptscriptstyle{D}-1}(\vec{x}-\vec{x}') \imath
\Delta_{\phi}^{ab}(x ; x') e^{-\imath
\vec{k}\cdot(\vec{x}-\vec{x}')} \label{Fouriertransformdefb} \,,
\end{eqnarray}
\end{subequations}
such that equations (\ref{EOM3aExtended}) transform into:
\begin{subequations}
\label{EOM4aExtended}
\begin{eqnarray}
(\partial_{t}^{2}+k^{2}+m^{2}_{\phi}) \imath
\Delta^{++}_{\phi}(k,t,t') + \int_{-\infty}^{\infty}
\mathrm{d}t_{1} \left[\imath M^{++}_{\phi}(k,t,t_{1})\imath
\Delta^{++}_{\phi}(k,t_{1},t') - \imath
M^{+-}_{\phi}(k,t,t_{1})\imath
\Delta^{-+}_{\phi}(k,t_{1},t')\right] &=&
\label{EOM4a++} \\
&& -  \imath\delta(t-t')\nonumber \\
(\partial_{t}^{2}+k^{2}+m^{2}_{\phi})\imath
\Delta^{+-}_{\phi}(k,t,t')+ \int_{-\infty}^{\infty}
\mathrm{d}t_{1} \left[\imath M^{++}_{\phi}(k,t,t_{1})\imath
\Delta^{+-}_{\phi}(k,t_{1},t') - \imath
M^{+-}_{\phi}(k,t,t_{1})\imath
\Delta^{--}_{\phi}(k,t_{1},t')\right] &=& 0
\label{EOM4a+-} \\
(\partial_{t}^{2}+k^{2}+m^{2}_{\phi})\imath
\Delta^{-+}_{\phi}(k,t,t') + \int_{-\infty}^{\infty}
\mathrm{d}t_{1} \left[\imath M^{-+}_{\phi}(k,t,t_{1})\imath
\Delta^{++}_{\phi}(k,t_{1},t') - \imath
M^{--}_{\phi}(k,t,t')\imath \Delta^{-+}_{\phi}(k,t_{1},t')\right]
&=& 0
\label{EOM4a-+} \\
(\partial_{t}^{2}+k^{2}+m^{2}_{\phi})\imath
\Delta^{--}_{\phi}(k,t,t') + \int_{-\infty}^{\infty}
\mathrm{d}t_{1} \left[\imath M^{-+}_{\phi}(k,t,t_{1})\imath
\Delta^{+-}_{\phi}(k,t_{1},t') - \imath
M^{--}_{\phi}(k,t,t_{1})\imath \Delta^{--}_{\phi}
(k,t_{1},t')\right] &=& \label{EOM4a--} \\
&& \, \imath \delta(t-t')  \nonumber \,.
\end{eqnarray}
\end{subequations}
Note that we have extended the initial time $t_{0} \rightarrow
-\infty$ in the equation above. Again, we have an analogous set of
equations of motion for the $\chi$-field. In order to outline the
next simplifying assumption, we need to Fourier transform with
respect to the difference of the time variables:
\begin{subequations}
\label{Fouriertransformdef2}
\begin{eqnarray}
\imath \Delta_{\chi}^{ab}(x;x') &=& \int
\frac{\mathrm{d}^{\scriptscriptstyle{D}}k}{(2\pi)^{\scriptscriptstyle{D}}}
 \imath \Delta_{\chi}^{ab}(k^{\mu}){\rm e}^{\imath k \cdot (x-x')}
\label{Fouriertransformdef2a} \\
\imath \Delta_{\chi}^{ab}(k^{\mu}) &=& \int
\mathrm{d}^{\scriptscriptstyle{D}}(x-x')\imath
\Delta_{\chi}^{ab}(x;x')  {\rm e}^{-\imath k \cdot (x-x')} \,,
\label{Fouriertransformdef2b}
\end{eqnarray}
\end{subequations}
As already mentioned, we will not solve the dynamical equations
for both the system and environment propagators, but instead we
assume the following hierarchy of couplings:
\begin{equation}\label{h<<lambda}
h\ll\lambda
\end{equation}
We thus assume that $\lambda$ is large enough such that the
$\chi$-field is thermalised by its strong self-interaction which
allows us to approximate the solutions of the dynamical equations
for $\chi$ as thermal propagators \cite{LeBellac:1996}:
\begin{subequations}
\label{ThermalPropagator}
\begin{eqnarray}
\imath\Delta_{\chi}^{++}(k^{\mu}) &=&
\frac{-\imath}{k_{\mu}k^{\mu}+m_{\chi}^{2} -\imath\epsilon} + 2\pi
\delta(k_{\mu}k^{\mu}+m_{\chi}^{2})
n_{\chi}^{\mathrm{eq}}(|k_{0}|) \label{ThermalPropagator++}
\\
\imath\Delta_{\chi}^{--}(k^{\mu}) &=&
\frac{\imath}{k_{\mu}k^{\mu}+m_{\chi}^{2}+\imath\epsilon} +  2\pi
\delta(k_{\mu}k^{\mu}+m_{\chi}^{2}) n_{\chi}^{\mathrm{eq}} (|k^{0}|)\label{ThermalPropagator--}\\
\imath\Delta_{\chi}^{+-}(k^{\mu}) &=& 2\pi
\delta(k_{\mu}k^{\mu}+m_{\chi}^{2}) \left[ \theta(-k^{0}) +
n_{\chi}^{\mathrm{eq}}(|k^{0}|)\right] \label{ThermalPropagator+-}
\\
\imath\Delta_{\chi}^{-+}(k^{\mu}) &=& 2\pi
\delta(k_{\mu}k^{\mu}+m_{\chi}^{2}) \left[ \theta(k^{0}) +
n_{\chi}^{\mathrm{eq}}(|k^{0}|)\right] \,,
\label{ThermalPropagator-+}
\end{eqnarray}
\end{subequations}
where the Bose-Einstein distribution is given by:
\begin{equation}
n_\chi^{\rm eq}(k^0) = \frac{1}{{\rm e}^{\beta k^0} - 1} \,,\qquad
\beta = \frac{1}{k_BT} \,, \label{BoseEinstein}
\end{equation}
with $k_B$ denoting the Stefan-Boltzmann constant and $T$ the
temperature. Here we use the notation $k_{\mu}k^{\mu}=
-k_{0}^{2}+k^{2}$ to distinguish the four-vector length from the
spatial three-vector length $k=\|\vec{k}\|$. We thus neglect the
backreaction of the system field on the environment field, such
that the latter remains in thermal equilibrium at temperature $T$.
This assumption is perturbatively well justified
\cite{Koksma:2009wa}. Furthermore, we neglected for simplicity the
$\mathcal{O}(\lambda^{2})$ correction to the propagators above
that slightly changes the equilibrium state of the environment
field. Note finally that, in our approximation scheme, the
dynamics of the system-propagators is effectively influenced only
by the 1PI self-mass corrections.

In \cite{Koksma:2009wa}, we have considered an environment field
$\chi$ in its vacuum state at $T=0$ and in the present work we
investigate finite temperature effects. Divergences originate from
the vacuum contributions to the self-masses only. Since we already
discussed renormalisation extensively in \cite{Koksma:2009wa}, let
us just state that the renormalised self-masses are given by:
\begin{equation}
\imath M^{ab}_{\phi,\mathrm{ren}}(k,t,t') =
(\partial_{t}^{2}+k^{2})\imath Z^{ab}_{\phi}(k,t,t') + \imath
M^{ab}_{\phi,\mathrm{th}}(k,t,t') \label{structureselfmasses} \,,
\end{equation}
where the vacuum contributions $Z^{ab}_{\phi}(k,t,t')$ to the
self-masses are given by:
\begin{subequations}
\label{SelfMassFourierSpFinalResult}
\begin{eqnarray}
Z_{\phi}^{\pm \pm }(k,t,t') &=& \frac{h^{2}}{64 k \pi^{2}} \left[
e^{\mp \imath k |\Delta t|} \left( \gamma_{\mathrm{E}} +
\log\left[ \frac{k}{2 \mu^{2} |\Delta t|}\right] \mp \imath
\frac{\pi}{2}\right) + e^{\pm \imath k |\Delta t|} \Big(
\mathrm{ci}(2k|\Delta t|) \mp \imath \mathrm{si}(2k|\Delta
t|)\Big)\right] \label{SelfMassFourierSpFinalResulta}
\\
Z_{\phi}^{\mp\pm }(k,t,t') &=& \frac{h^{2}}{64 k \pi^{2}} \left[
e^{\mp \imath k \Delta t}\! \left( \gamma_{\mathrm{E}} +
\log\left[ \frac{k}{2 \mu^{2}|\Delta t|} \right]\! \mp \imath
\frac{\pi}{2} \mathrm{sgn}(\Delta t) \right) \!+ e^{\pm \imath k
\Delta t } \Big( \mathrm{ci}(2k|\Delta t|) \mp
\imath\mathrm{sgn}(\Delta t) \mathrm{si}(2k|\Delta t|)\!\Big)\!
\right] \! , \qquad\phantom{1}
\label{SelfMassFourierSpFinalResultb}
\end{eqnarray}
\end{subequations}
and where $\imath M^{ab}_{\phi,\mathrm{th}}(k,t,t')$ are the
thermal contributions to the self-masses that yet need to be
evaluated. In deriving (\ref{SelfMassFourierSpFinalResult}), we
made the simplifying assumption $m_{\chi}\rightarrow 0$. The
influence of the environment field on the system field is still
perturbatively under control \cite{Koksma:2009wa}. Furthermore,
$\mathrm{ci}(z)$ and $\mathrm{si}(z)$ are the cosine and sine
integral functions, defined by:
\begin{subequations}
\label{ciandsi}
\begin{eqnarray}
{\rm ci}(z) &\equiv&  - \int_z^\infty \mathrm{d}t
\frac{\cos(t)}{t} \label{ci} \\
{\rm si}(z) &\equiv& -\int_z^\infty \mathrm{d}t \frac{\sin(t)}{t}
\label{si} \,.
\end{eqnarray}
\end{subequations}
Note that the structure of the self-mass
(\ref{structureselfmasses}) is such that we can construct
relations analogous to equation (\ref{reduction:F+Deltac}):
\begin{subequations}
\label{reduction:selfMass:MF+Mc2}
\begin{eqnarray}
M^{+-}_{\phi}(k,t,t') &=& M^{F}_{\phi}(k,t,t') - \frac{1}{2}\imath
M^{c}_{\phi}(k,t,t')
\label{reduction:selfMass:MF+Mc2a}\\
M^{-+}_{\phi}(k,t,t') &=& M^{F}_{\phi}(k,t,t') + \frac{1}{2}\imath
M^{c}_{\phi}(k,t,t')
\label{reduction:selfMass:MF+Mc2b}\\
M^{++}_{\phi}(k,t,t') &=& M^{F}_{\phi}(k,t,t') + \frac{1}{2}{\rm
sgn}(t-t^\prime)\imath M^{c}_{\phi}(k,t,t')
\label{reduction:selfMass:MF+Mc2c}\\
M^{--}_{\phi}(k,t,t') &=& M^{F}_{\phi}(k,t,t')- \frac{1}{2}{\rm
sgn}(t-t^\prime)\imath M^{c}_{\phi}(k,t,t')
\label{reduction:selfMass:MF+Mc2d}\,.
\end{eqnarray}
\end{subequations}
This structure applies to both the vacuum and thermal
contributions separately. Thus, $Z^{F}_{\phi}(k,t,t')$ is the
vacuum contribution to the statistical self-mass and $\imath
Z^{c}_{\phi}(k,t,t')$ the vacuum contribution to the causal
self-mass. Similarly we can define the thermal contributions to
the statistical and causal self-masses
$M^{F}_{\phi,\mathrm{th}}(k,t,t')$ and $\imath
M^{c}_{\phi,\mathrm{th}}(k,t,t')$, respectively. Of course, we
still need to evaluate these expressions. The vacuum contributions
follow straightforwardly from equation
(\ref{SelfMassFourierSpFinalResult}):
\begin{subequations}
\label{selfmasses}
\begin{eqnarray}
Z^{F}_{\phi}(k,t,t') &=& \frac{1}{2} \left[Z^{-+}_{\phi}(k,t,t') + Z^{+-}_{\phi}(k,t,t')\right] \label{selfMass:MFphi}\\
&=& \frac{h^{2}}{64 k \pi^{2}} \left[ \cos(k\Delta
t)\left(\gamma_{\mathrm{E}} + \log\left[ \frac{k}{2 \mu^{2}|\Delta
t|} \right]+ \mathrm{ci}(2k|\Delta t|) \right)+\sin(k|\Delta
t|)\left(\mathrm{si}(2k|\Delta t|) - \frac{\pi}{2}\right)\right]
\nonumber \\
Z^{c}_{\phi}(k,t,t') &=& \imath \left[Z^{+-}_{\phi}(k,t,t')
-  Z^{-+}_{\phi}(k,t,t')\right]\label{selfMass:Mcphi}\\
&=& \frac{h^{2}}{64 k \pi^{2}} \left[ -2 \cos(k\Delta
t)\mathrm{sgn}(\Delta t) \left( \mathrm{si}(2k|\Delta t|) +
\frac{\pi}{2} \right)+2\sin(k\Delta t)\left( \mathrm{ci}(2k|\Delta
t|) - \gamma_{\mathrm{E}} - \log\left[ \frac{k}{2 \mu^{2}|\Delta
t|} \right]  \right)\right] \nonumber \,.
\end{eqnarray}
\end{subequations}

As before, we are primarily interested in the equations of motion
for the causal and statistical propagators, as it turns out they
yield a closed system of differential equations that can be
integrated by providing appropriate initial conditions. In order
to obtain the equation of motion for the causal propagator, we
subtract (\ref{EOM4a+-}) from (\ref{EOM4a-+}) and use equations
(\ref{structureselfmasses}) and (\ref{reduction:selfMass:MF+Mc2})
to find:
\begin{equation}
(\partial_{t}^{2}+k^{2}+m^{2}_{\phi}) \Delta^{c}_{\phi}(k,t,t') -
\left(\partial_{t}^{2}+k^{2}\right) \int_{t'}^{t} \mathrm{d}t_{1}
Z^{c}_{\phi}(k,t,t_{1}) \Delta^{c}_{\phi}(k,t_{1},t') -
\int_{t'}^{t} \mathrm{d}t_{1} M^{c}_{\phi,\mathrm{th}}(k,t,t_{1})
\Delta^{c}_{\phi}(k,t_{1},t') =0 \label{eomcausalprop2} \,.
\end{equation}
In order to get an equation for the statistical propagator, we add
equation (\ref{EOM4a+-}) to (\ref{EOM4a-+}), which we simplify to
get:
\begin{eqnarray}
(\partial_{t}^{2}+k^{2}+m^{2}_{\phi}) F_{\phi}(k,t,t') -
\left(\partial_{t}^{2}+k^{2}\right)\left[ \int_{-\infty}^{t}
\mathrm{d}t_{1} Z^{c}_{\phi}(k,t,t_{1}) F_{\phi}(k,t_{1},t') -
\int_{-\infty}^{t'} \mathrm{d}t_{1} Z^{F}_{\phi}(k,t,t_{1})
\Delta^{c}_{\phi}(k,t_{1},t') \right] && \label{eomstatprop2} \\
- \int_{-\infty}^{t} \mathrm{d}t_{1}
M^{c}_{\phi,\mathrm{th}}(k,t,t_{1}) F_{\phi}(k,t_{1},t') +
\int_{-\infty}^{t'} \mathrm{d}t_{1}
M^{F}_{\phi,\mathrm{th}}(k,t,t_{1}) \Delta^{c}_{\phi}(k,t_{1},t')
&=& 0 \nonumber \,.
\end{eqnarray}
Due to the non-locality inherent in any interacting quantum field
theory, the ``memory kernels'', the memory integrals in equation
(\ref{eomstatprop2}) above, range from negative past infinity to
either $t$ or $t'$. To make the numerical implementation feasible,
we insert a finite initial time $t_{0}$ by hand and approximate
the propagators in the memory kernels from the negative past to
$t_{0}$ with the free propagators inducing an error of the order
$\mathcal{O}(h^{4}/\omega_{\phi}^{4})$:
\begin{eqnarray}
(\partial_{t}^{2}+k^{2}+m^{2}_{\phi}) F_{\phi}(k,t,t') &-&
\left(\partial_{t}^{2}+k^{2}\right) \Bigg[ \int_{-\infty}^{t_{0}}
\mathrm{d}t_{1} Z^{c}_{\phi}(k,t,t_{1})
F_{\phi}^{\mathrm{free}}(k,t_{1},t') + \int_{t_{0}}^{t}
\mathrm{d}t_{1} Z^{c}_{\phi}(k,t,t_{1}) F_{\phi}(k,t_{1},t')
\label{eomnumericalc}\\
&& \qquad\qquad - \int_{-\infty}^{t_{0}} \mathrm{d}t_{1}
Z^{F}_{\phi}(k,t,t_{1})
\Delta^{c,\mathrm{free}}_{\phi}(k,t_{1},t') - \int_{t_{0}}^{t'}
\mathrm{d}t_{1} Z^{F}_{\phi}(k,t,t_{1})
\Delta^{c}_{\phi}(k,t_{1},t') \Bigg] \nonumber \\
&& - \int_{-\infty}^{t_{0}} \mathrm{d}t_{1}
M^{c}_{\phi,\mathrm{th}}(k,t,t_{1})
F_{\phi}^{\mathrm{free}}(k,t_{1},t') - \int_{t_{0}}^{t}
\mathrm{d}t_{1} M^{c}_{\phi,\mathrm{th}}(k,t,t_{1})
F_{\phi}(k,t_{1},t') \nonumber\\
&& + \int_{-\infty}^{t_{0}} \mathrm{d}t_{1}
M^{F}_{\phi,\mathrm{th}}(k,t,t_{1})
\Delta^{c,\mathrm{free}}_{\phi}(k,t_{1},t') + \int_{t_{0}}^{t'}
\mathrm{d}t_{1} M^{F}_{\phi,\mathrm{th}}(k,t,t_{1})
\Delta^{c}_{\phi}(k,t_{1},t') = 0 \nonumber \,.
\end{eqnarray}
Here, $F_{\phi}^{\mathrm{free}}(k,t,t')$ and
$\Delta^{c,\mathrm{free}}_{\phi}(k,t,t')$ are the free statistical
and causal propagators which, depending on the initial conditions
one imposes at $t_0$, should either be evaluated at $T=0$ or at
some finite temperature. The memory kernels need to be included to
remove the initial time singularity as discussed in
\cite{Koksma:2009wa, Garny:2009ni}. We postpone imposing initial
conditions to section \ref{Results: Entropy Generation in Quantum
Field Theory}, but let us at the moment just evaluate the memory
kernels in these two cases. The thermal propagators read:
\begin{subequations}
\label{thermalfreepropagators}
\begin{eqnarray}
F_{\phi,\mathrm{th}}^{\mathrm{free}}(k,t,t') &=& \frac{
\cos(\omega_{\phi}\left(t-t'\right))}{2\omega_{\phi}}\left(1+\frac{2}{\mathrm{e}^{\beta
\omega_{\phi}}-1}\right) =
\frac{\cos(\omega_{\phi}\left(t-t'\right))}{2\omega_{\phi}}
\coth\left(\frac{1}{2}\beta \omega_{\phi} \right)
\label{thermalfreepropagatorsstatistical} \\
\imath \Delta_{\phi}^{c,\mathrm{free}}(k,t,t') &=&  -
\frac{\imath}{\omega_{\phi}}
\sin\left(\omega_{\phi}\left(t-t'\right)\right)
\label{thermalfreepropagatorscausal}
\end{eqnarray}
\end{subequations}
Here, $\omega_{\phi}^{2}=k^{2}+m^{2}_{\phi,\mathrm{in}}$, where in
the case of a changing mass one should use the initial mass. Let
us now evaluate the ``infinite past memory kernels'' for the
vacuum contributions, i.e.: the memory kernels from negative past
infinity to $t_{0}$ using the two propagators above. The other
memory kernels in equation (\ref{eomnumericalc}) can only be
evaluated numerically, as soon as we have the actual expressions
of the thermal contributions to the self-masses. Let us thus
evaluate:
\begin{eqnarray}
&& \left(\partial_{t}^{2}+k^{2}\right) \left[
\int_{-\infty}^{t_{0}} \mathrm{d}t_{1} Z^{c}_{\phi}(k,t,t_{1})
F_{\phi,\mathrm{th}}^{\mathrm{free}}(k,t_{1},t')  - \int_{-\infty}^{t_{0}}
\mathrm{d}t_{1} Z^{F}_{\phi}(k,t,t_{1})
\Delta^{c,\mathrm{free}}_{\phi}(k,t_{1},t') \right]
\label{MemoryKernel}
\\
&& \qquad = \frac{h^{2}}{32 \pi^2} \int_{-\infty}^{t_{0}}
\mathrm{d}t_{1} \left[ \frac{\cos[k(t-t_{1})]}{t-t_{1}}\frac{\cos[
\omega_{\phi}(t_{1}-t')]}{\omega_{\phi}}\coth\left(\frac{1}{2}\beta
\omega_{\phi}\right) + \frac{\sin[k(t-t_{1})]}{t-t_{1}}\frac{\sin[
\omega_{\phi}(t_{1}-t')]}{\omega_{\phi}} \right] \nonumber \\
&& \qquad = - \frac{h^{2}}{64 \omega_{\phi} \pi^{2}} \Bigg[
\frac{\cos[\omega_{\phi}(t-t')]}{\sinh\left(\frac{1}{2}\beta
\omega_{\phi}\right)} \left\{ \mathrm{e}^{\frac{1}{2}\beta
\omega_{\phi}} \mathrm{ci}\left[(\omega_{\phi}+k)(t-t_{0})\right]
+ \mathrm{e}^{-\frac{1}{2}\beta \omega_{\phi}}
\mathrm{ci}\left[(\omega_{\phi}-k)(t-t_{0})\right] \right\}
\nonumber \\
&& \qquad\qquad\qquad\quad +
\frac{\sin[\omega_{\phi}(t-t')]}{\sinh\left(\frac{1}{2}\beta
\omega_{\phi}\right)} \left\{ \mathrm{e}^{\frac{1}{2}\beta
\omega_{\phi}} \mathrm{si}\left[(\omega_{\phi}+k)(t-t_{0})\right]
+ \mathrm{e}^{-\frac{1}{2}\beta \omega_{\phi}}
\mathrm{si}\left[(\omega_{\phi}-k)(t-t_{0})\right] \right\} \Bigg]
\nonumber\,.
\end{eqnarray}
Due to the fact that the free thermal statistical propagator
contains a temperature dependence, the corresponding ``infinite
past memory kernel'' is of course also affected. In case we would
need the $T=0$ vacuum propagators in the memory kernels only, one
can easily send $T\rightarrow 0$ in the expression above,
obtaining the same memory kernels as in \cite{Koksma:2009wa}:
\begin{eqnarray}
&& \left(\partial_{t}^{2}+k^{2}\right) \left[
\int_{-\infty}^{t_{0}} \mathrm{d}t_{1} Z^{c}_{\phi}(k,t,t_{1})
F_{\phi,\mathrm{vac}}^{\mathrm{free}}(k,t_{1},t')  - \int_{-\infty}^{t_{0}}
\mathrm{d}t_{1} Z^{F}_{\phi}(k,t,t_{1})
\Delta^{c,\mathrm{free}}_{\phi}(k,t_{1},t') \right]
\label{MemoryKernelvac}
\\
&& \qquad = - \frac{h^{2}}{32 \omega_{\phi} \pi^{2}} \left\{
\cos[\omega_{\phi}(t-t')]
\mathrm{ci}\left[(\omega_{\phi}+k)(t-t_{0})\right] +
\sin[\omega_{\phi}(t-t')]
\mathrm{si}\left[(\omega_{\phi}+k)(t-t_{0})\right] \right\}
\nonumber\,.
\end{eqnarray}

\section{Finite Temperature Contributions to the Self-Masses}
\label{Finite Temperature Contributions to the Self-Masses}

In this section, we evaluate all contributions to the self-masses
for a finite temperature.

\subsection{The Causal Self-Mass}
\label{The Causal Self-Mass}

Let us first evaluate the thermal contribution to the causal
self-mass. Formally, from equation (\ref{selfMass}), it reads:
\begin{eqnarray}
M^{c}_{\phi,\mathrm{th}}(k,\Delta t = t - t') &=& - \imath \left[
M_{\phi,\mathrm{th}}^{-+}(k,t,t') -
M_{\phi,\mathrm{th}}^{+-}(k,t,t')\right] \label{selfMass:causal1}
\\
&=& - h^{2} \int \frac{ \mathrm{d}^{\scriptscriptstyle{D}-1}
\vec{k}_{1}}{(2\pi)^{\scriptscriptstyle{D}-1}}
F_{\chi}^{\mathrm{th}}(k_{1},\Delta t)
\Delta_{\chi}^{c}(\|\vec{k}-\vec{k}_{1}\|,\Delta t) \nonumber\,,
\end{eqnarray}
where the superscript $F_{\chi}^{\mathrm{th}}$ denotes that we
should only keep the thermal contribution to the statistical
propagator as we have already evaluated the vacuum contribution.
The following change of variables is useful:
\begin{eqnarray}
\int \frac{\mathrm{d}^{\scriptscriptstyle{D}-1} \vec{k}_{1}}
{(2\pi)^{\scriptscriptstyle{D}-1}} &=&
\frac{1}{(2\pi)^{\scriptscriptstyle{D}-1}} \int_{0}^{\infty}
\mathrm{d} k_{1} k_{1}^{\scriptscriptstyle{D}-2} \int
\mathrm{d}\Omega_{D-2} =
\frac{\Omega_{D-3}}{(2\pi)^{\scriptscriptstyle{D}-1}}
 \int_{0}^{\infty} \mathrm{d} k_{1}
k_{1}^{\scriptscriptstyle{D}-2} \int_{-1}^1 \mathrm{d}\cos(\theta)
[\sin(\theta)]^{D-4} \label{angular integral:D-2} \\
&=&  \frac{\Omega_{D-3}}{(2\pi)^{\scriptscriptstyle{D}-1}}  \int
\mathrm{d} k_{1} k_{1}^{\scriptscriptstyle{D}-2}
\int_{\omega_-}^{\omega_+}\mathrm{d}\omega
\frac{2\omega}{(2kk_1)^{D-3}}
[(\omega_+^2-\omega^2)(\omega^2-\omega_-^2)]^\frac{D-4}{2}
\nonumber \,,
\end{eqnarray}
where we have chosen $\theta\equiv \angle(\vec k,\vec k_1)$. In
the final line we have changed variables to $\omega \equiv
\omega_{\chi}(\|\vec{k}-\vec{k}_1\|)=
(\|\vec{k}-\vec{k}_{1}\|^{2}+m_{\chi}^{2})^{1/2}$, which clearly
depends on $\theta$. Furthermore $\omega_\pm^2= (k\pm
k_1)^2+m_\chi^2$ and $\Omega_{D-3}$ denotes the area of the $D-3$
dimensional sphere $S^{D-3}$:
\begin{equation}\label{angularcontrib}
\Omega_{D-3} = \frac{2\pi^\frac{D-2}{2}}{\Gamma(\frac{D}{2}-1)}\,.
\end{equation}
Using equation (\ref{thermalfreepropagators}), we have:
\begin{eqnarray}
M^{c}_{\phi,\mathrm{th}}(k,\Delta t) &=& h^{2}
\frac{\Omega_{D-3}}{(2\pi)^{\scriptscriptstyle{D}-1}}
\int_{0}^{\infty} \mathrm{d} k_{1} k_{1}^{\scriptscriptstyle{D}-2}
\int_{\omega_-}^{\omega_+} \mathrm{d}\omega
\frac{2\omega}{(2kk_1)^{D-3}}
[(\omega_+^2-\omega^2)(\omega^2-\omega_-^2)]^\frac{D-4}{2}
\frac{1}{(k_{1}^{2}+m_{\chi}^{2})^{\frac{1}{2}}} \label{selfMass:causal2} \\
&& \qquad\qquad\qquad\qquad\qquad\qquad\qquad \times
\frac{1}{\omega}n_{\chi}^{\mathrm{eq}} \left
(\{k_{1}^{2}+m_{\chi}^{2}\}^{\frac{1}{2}} \right) \cos\left
(\{k_{1}^{2}+m_{\chi}^{2}\}^{\frac{1}{2}}\Delta t \right)
\sin(\omega\Delta t)\nonumber \,.
\end{eqnarray}
This contribution cannot contain any new divergences as the latter
all stem from the vacuum contribution, which allows us to let
$D\rightarrow 4$. Moreover, we are interested, as in section
\ref{Decoherence in an Interacting Quantum Field Theory: Vacuum
Case}, in the limit $m_{\chi} \rightarrow 0$. Equation
(\ref{selfMass:causal2}) thus simplifies to:
\begin{eqnarray}
M^{c}_{\phi,\mathrm{th}}(k,\Delta t) &=& \frac{h^{2}}{4\pi^{2}}
\frac{\sin(k\Delta t)}{k \Delta t} \int_{0}^{\infty}
\mathrm{d}k_{1} \frac{\sin(2k_{1}\Delta t)}{\mathrm{e}^{\beta
k_{1}}-1} \label{selfMass:causal3} \\
&=& \frac{h^{2}}{16\pi^{2}} \frac{\sin(k\Delta t)}{k (\Delta
t)^{2}} \left[\frac{2\pi \Delta t}{\beta}\coth\left(
\frac{2\pi\Delta t}{\beta}\right)-1\right]\nonumber \,.
\end{eqnarray}
At coincidence $\Delta t \rightarrow 0$, the thermal contribution
to the causal self-mass vanishes, as it should.

\subsection{The Statistical Self-Mass}
\label{The Statistical Self-Mass}

The thermal contribution to the statistical self-mass is somewhat
harder to obtain. It is given by:
\begin{eqnarray}
M^{F}_{\phi,\mathrm{th}}(k,\Delta t = t - t') &=& \frac{1}{2}
\left[ M_{\phi,\mathrm{th}}^{-+}(k,\Delta t) +
M_{\phi,\mathrm{th}}^{+-}(k,\Delta t)\right]
\label{selfMass:statistical1}
\\
&=& - \frac{h^{2}}{2} \int \frac{
\mathrm{d}^{\scriptscriptstyle{D}-1}
\vec{k}_{1}}{(2\pi)^{\scriptscriptstyle{D}-1}} \left[
F_{\chi}(k_{1},\Delta t) F_{\chi}(\|\vec{k}-\vec{k}_{1}\|,\Delta
t) - \frac{1}{4} \Delta_{\chi}^{c}(k_{1},\Delta t)
\Delta_{\chi}^{c}(\|\vec{k}-\vec{k}_{1}\|,\Delta t)
\right]\Bigg|_{\mathrm{th}} \nonumber \,,
\end{eqnarray}
where of course we are only interested in keeping the thermal
contributions. The second term in the integral consists of two
causal propagators that does not contribute at all at finite
temperature. It is convenient to split the thermal contributions
to the statistical self-mass as:
\begin{equation}\label{selfMass:statistical2}
M^{F}_{\phi,\mathrm{th}}(k,\Delta t) =
M^{F}_{\phi,\mathrm{vac-th}}(k,\Delta t) +
M^{F}_{\phi,\mathrm{th-th}}(k,\Delta t) \,,
\end{equation}
where, formally, we have:
\begin{subequations}
\label{selfMass:statistical3}
\begin{eqnarray}
M^{F}_{\phi,\mathrm{vac-th}}(k,\Delta t) &=& - \frac{h^{2}}{2}
\int \frac{ \mathrm{d}^{\scriptscriptstyle{D}-1}
\vec{k}_{1}}{(2\pi)^{\scriptscriptstyle{D}-1}}
\frac{\cos(\omega_{\chi}(k_{1})\Delta t)}{\omega_{\chi}(k_{1})}
\frac{\cos(\omega_{\chi}(\|\vec{k}-\vec{k}_{1}\|)\Delta
t)}{\omega_{\chi}(\|\vec{k}-\vec{k}_{1}\|)}
\frac{1}{\mathrm{e}^{\beta\omega_{\chi}(k_{1})}-1}
\label{selfMass:statistical3a}
\\
M^{F}_{\phi,\mathrm{th-th}}(k,\Delta t) &=& - \frac{h^{2}}{2} \int
\frac{ \mathrm{d}^{\scriptscriptstyle{D}-1}
\vec{k}_{1}}{(2\pi)^{\scriptscriptstyle{D}-1}}
\frac{\cos(\omega_{\chi}(k_{1})\Delta t)}{\omega_{\chi}(k_{1})}
\frac{1}{\mathrm{e}^{\beta\omega_{\chi}(k_{1})}-1}
\frac{\cos(\omega_{\chi}(\|\vec{k}-\vec{k}_{1}\|)\Delta
t)}{\omega_{\chi}(\|\vec{k}-\vec{k}_{1}\|)}
\frac{1}{\mathrm{e}^{\beta\omega_{\chi}(\|\vec{k}-\vec{k}_{1}\|)}-1}
\label{selfMass:statistical3b} \,.
\end{eqnarray}
\end{subequations}
Here, $M^{F}_{\phi,\mathrm{vac-th}}$ is the vacuum-thermal
contribution to the statistical self-mass and
$M^{F}_{\phi,\mathrm{th-th}}$ is the thermal-thermal contribution.
As before, we let $D\rightarrow 4$ and $m_{\chi} \rightarrow 0$.
Let us firstly evaluate the vacuum-thermal contribution. Equation
(\ref{selfMass:statistical3a}) thus simplifies to:
\begin{equation}\label{selfMass:statistical4}
M^{F}_{\phi,\mathrm{vac-th}}(k,\Delta t) = - \frac{h^{2}}{8\pi^{2}
k\Delta t} \int_{0}^{\infty} \mathrm{d} k_{1}
\frac{\cos(k_{1}\Delta t)}{\mathrm{e}^{\beta k_{1}}-1}\left[
\sin((k+k_{1})\Delta t) - \sin(|k-k_{1}|\Delta t) \right] \,.
\end{equation}
We have to take the absolute values in the equation above
correctly into account by making use of Heaviside step-functions
and we can moreover expand the exponential to find:
\begin{eqnarray}
M^{F}_{\phi,\mathrm{vac-th}}(k,\Delta t) &=& -\frac{h^{2}}{16\pi^2
k\Delta t} \int_0^\infty \mathrm{d} k_{1} \sum_{n=1}^\infty {\rm
e}^{-\beta n k_1} \Bigg\{ \sin\left((2k_1+k)\Delta t\right)+
2\theta(k_{1}-k) \sin\left(k\Delta t \right)
\label{selfMass:statistical4b}\\
&& \qquad\qquad\qquad\qquad\qquad\qquad + \left\{
\theta(k-k_1)-\theta(k_1-k) \right\} \sin\left((2k_1-k)\Delta
t\right) \Bigg\}\,. \nonumber
\end{eqnarray}
Integrating over $k_1$ and collecting the terms we get:
\begin{equation}
\! M^{F}_{\phi,\mathrm{vac-th}}(k,\Delta t) = -\frac{h^2}{8\pi^2
k\Delta t} \sum_{n=1}^\infty \!
 \Bigg\{ \! \cos(k\Delta t)\Big(1 \!-\!{\rm e}^{-\beta nk}\Big)\frac{2\Delta t}{(\beta n)^2\!+\!(2\Delta t)^2}
\! + \sin(k\Delta t){\mathrm e}^{-\beta nk}\!\left[\frac{1}{\beta
n} \!-\! \frac{\beta n}{(\beta n)^2 \!+\! (2\Delta t)^2} \right]\!
\Bigg\} \label{selfMass:statistical5} \! .
\end{equation}
The sum can be performed, resulting in:
\begin{eqnarray}
\! && \! M^{F}_{\phi,\mathrm{vac-th}}(k,\Delta t) \!=\!
\frac{h^2}{16\pi^2 k(\Delta t)^2} \Bigg [ \! \sin(k\Delta
t)\Bigg\{ \! \frac{2\Delta t}{\beta}\log \! \left(1 \!-\!{\rm
e}^{-\beta k}\right)\! +\! {\rm e}^{-\beta k} \!
\sum_{\pm}\frac{\pm \frac{\imath}{2}}{1\pm\frac{2\imath\Delta
t}{\beta}} {}_2F_1\left( \!2, \! 1 \! \pm\! \frac{2\imath\Delta
t}{\beta}; 2\! \pm\!\frac{2\imath\Delta t}{\beta};{\rm e}^{-\beta
k}\right) \!
\Bigg\} \nonumber \\
&& \qquad \qquad - \cos(k\Delta t)\Bigg\{ \frac{1}{2}\!
\left(\frac{2\pi\Delta t}{\beta}\coth \! \left(\frac{2\pi\Delta
t}{\beta}\right)-1\right) - {\rm e}^{-\beta k} \sum_{\pm}
\frac{\pm\frac{\imath\Delta t}{\beta}}{1\pm\frac{2\imath\Delta
t}{\beta}} {}_2F_1\left( \! 1,1\!\pm\!\frac{2\imath\Delta
t}{\beta}; 2\!\pm\! \frac{2\imath\Delta t}{\beta};{\rm e}^{-\beta
k}\right) \! \Bigg\} \! \Bigg] , \label{selfMass:statistical6}
\end{eqnarray}
where ${}_{2}F_{1}$ is the Gauss' hypergeometric function. For
convenience we quote the low temperature ($\beta k\gg 1$) and the
high temperature ($\beta k\ll 1$) limits of this expression. In
the low temperature limit~(\ref{selfMass:statistical6}) reduces
to:
\begin{eqnarray}
M^{F}_{\phi,\mathrm{vac-th}}(k,\Delta t) &\stackrel{\beta k \gg
1}{\longrightarrow} & -\frac{h^2}{16\pi^2 k(\Delta t)^2}
\Bigg\{\frac{\cos(k\Delta t)}{2}\left[\frac{2\pi\Delta
t}{\beta}\coth\left(\frac{2\pi\Delta t}{\beta}\right)-1\right]
\label{selfMass:statistical7} \\
&& \qquad\qquad\qquad\quad +\, {\rm e}^{-\beta
k}\left[\cos(k\Delta t)\frac{-(2\Delta t/\beta)^2}{1+(2\Delta
t/\beta)^2} + \sin(k\Delta t)\left(\frac{2\Delta t}{\beta}
-\frac{2 \Delta t/ \beta}{1+(2\Delta t/\beta)^2}\right) \right]
\Bigg\}\,, \nonumber
\end{eqnarray}
and its coincidence limit is finite:
\begin{equation}
\label{selfMass:statistical7coincidence}
\lim_{\Delta t \rightarrow 0} M^{F}_{\phi,\mathrm{vac-th}}(k,\Delta t) \stackrel{\beta k \gg
1}{\longrightarrow} - \frac{h^{2}(\pi^2-6 \mathrm{e}^{-k\beta})}{24\pi^2 k \beta^2}\,.
\end{equation}
In the high temperature limit (\ref{selfMass:statistical6})
reduces to:
\begin{eqnarray}
M^{F}_{\phi,\mathrm{vac-th}}(k,\Delta t) &\stackrel{\beta k \ll
1}{\longrightarrow}& \frac{h^2}{4\pi^2 \beta} \Bigg[ \cos(k\Delta
t)\left\{ \log(\beta k)+\gamma_{{E}}-1
+\frac{1}{2}\sum_{\pm}\psi\left(1\pm\frac{2\imath \Delta
t}{\beta}\right)\right\}
 \label{selfMass:statistical8} \\
&& \qquad\quad - \frac{\sin(k\Delta t)}{4 k\Delta t}
\left\{\sum_{\pm}\psi\left(1\pm\frac{2\imath \Delta
t}{\beta}\right) +2\gamma_E \right\} \Bigg] \,. \nonumber
\end{eqnarray}
There is a mild logarithmic divergence, $
M_{\phi,\mathrm{vac-th}}^F \propto \log(\beta k)$, in the limit
when $\beta k\rightarrow 0$. Also note that when we derived
equation (\ref{selfMass:statistical8}) above, we tacitly assumed
that also $\Delta t/\beta \ll 1$. We however only use equation
(\ref{selfMass:statistical8}) to calculate the coincidence limit
$\Delta t \rightarrow 0$ of the statistical self-mass in which
case this approximation is well justified:
\begin{equation}
\label{selfMass:statistical8coincidence}
\lim_{\Delta t \rightarrow 0} M^{F}_{\phi,\mathrm{vac-th}}(k,\Delta t) \stackrel{\beta k \ll
1}{\longrightarrow} \frac{h^{2}}{4\pi^2 \beta}(\log (\beta k) -1)\,.
\end{equation}
The final remaining contribution to the statistical self-mass is
$M^{F}_{\phi,\mathrm{th-th}}(k,\Delta t)$ in equation
(\ref{selfMass:statistical3b}) and is much harder to obtain. In
fact, it turns out we can only evaluate its high ($\beta k \ll 1$)
and low ($\beta k \gg 1$) temperature contributions in closed
form. For that reason we present the calculation in appendix
\ref{AppendixA}, and in the current section only state the main
results. The low temperature ($\beta k \gg 1$) limit of
$M^{F}_{\phi,\mathrm{th-th}}(k,\Delta t)$ is given by:
\begin{eqnarray}
M^{F}_{\phi,\mathrm{th-th}}(k,\Delta t)  &\stackrel{k\beta\gg
1}{\longrightarrow}& -\frac{h^2}{16\pi^2 k} \, {\rm e}^{-\beta k}
\left[\! \cos(k\Delta t) \left\{\frac{2 \pi \Delta t
\coth\left(\frac{2 \pi \Delta t}{\beta}\right)}{\beta
(\beta^2+(\Delta t)^2)} + \frac{\beta k}{\beta^2 + (\Delta t)^2} +
\frac{\beta^2(5\beta^2+11(\Delta t)^2)}{(\beta^2+(\Delta t)^{2})^2
(\beta^2+(2 \Delta t)^2)} \right\} \right. \nonumber \\
&&  \qquad \left. +\sin(k\Delta t)\left\{ \frac{2 \pi
\coth\left(\frac{2 \pi \Delta t}{\beta}\right)}{\beta^2+(\Delta
t)^2} - \frac{k\Delta t}{\beta^2 +(\Delta t)^2} - \frac{2\beta
\Delta t}{(\beta^2 +(\Delta t)^2)^2} -\frac{8 \Delta
t}{\beta(\beta^2+(2\Delta t)^2)} \right\} \right]
\label{MFphi:thth:4} \,.
\end{eqnarray}
Note that this expression is finite in the limit when $\Delta
t\rightarrow 0$:
\begin{equation}
\lim_{\Delta t \rightarrow 0} M^{F}_{\phi,\mathrm{th-th}}(k,\Delta
t) \stackrel{k\beta\gg 1}{\longrightarrow} - h^2
\frac{3+k\beta}{8\pi^2 k\beta^2} \mathrm{e}^{-k\beta}
 \label{MFphi:thth:4atcoincidence} \,.
\end{equation}
The high temperature ($\beta k \ll 1$) limit yields:
\begin{eqnarray}
\! && \! M^{F}_{\phi,\mathrm{th-th}}(k,\Delta t)
\stackrel{k\beta\ll 1}{\longrightarrow} \!
-\frac{h^{2}}{16\pi^{2}k\beta^{2}} \Bigg[ \!
\frac{\pi^{2}}{2}\!-\!4\left(\gamma_{\mathrm{E}} \!-\!
\mathrm{ci}(|k\Delta t |) \!+\!\log(|k\Delta t |) \right)
\!+\!\frac{(k\Delta t)^{2}}{2} \frac{\mathrm{d}}{\mathrm{d}\gamma}
\, {}_2F_3 \! \left(\! 1,1;2,2,1\!+\!\gamma;-\frac{(k\Delta
t)^2}{4}\right)\! \Bigg|_{\gamma=\frac12}
\nonumber\\
&& \qquad\qquad\qquad\qquad\qquad\quad + \frac{\beta\sin(k\Delta
t)}{\Delta t} \left(\mathrm{ci}(2|k\Delta t |) -
\gamma_{\mathrm{E}} -\log\left(\frac{2|\Delta t |}{k\beta^2}\right)-1
\right) - \frac{\beta \cos(k\Delta t) }{|\Delta t |} \left(
\mathrm{si}(2|k\Delta t |)+\frac{\pi}{2}\right)\! \label{MFphi:thth:hiT:FinalResult}\\
&& \qquad\qquad\qquad\qquad + k\beta \!\sum_{\pm} \!
\frac{\mathrm{e}^{-k\beta\pm \imath k \Delta t}}{2(1\mp\imath
\Delta t/\beta)} \Bigg[ {}_{2}F_{1} \!\left(\!
2,\!2\!\mp\!\frac{2\imath \Delta t}{\beta}\!;\!3\!\mp\!
\frac{2\imath \Delta t}{\beta}\!;\!
\mathrm{e}^{-\frac{k\beta}{2}}\right)\! + \frac{(k\beta)^2}{12}
{}_{2}F_{1}\! \left(\!4,\! 2\!\mp\! \frac{2\imath \Delta
t}{\beta}\!;\!3\! \mp \! \frac{2\imath \Delta t}{\beta}\!;\!
\mathrm{e}^{-\frac{k\beta}{2}}\right) \!\Bigg]\Bigg]\nonumber \,.
\end{eqnarray}
Clearly, the limit $\Delta t \rightarrow 0$ of the self-mass above
is finite too:
\begin{equation}
\lim_{\Delta t \rightarrow 0} M^{F}_{\phi,\mathrm{th-th}}(k,\Delta
t) \stackrel{k\beta\ll 1}{\longrightarrow} - \frac{h^{2}} {32
\pi^2 k \beta^2 } \left( 8 +\pi^2+4k\beta \log\left[\frac{1}{2}(k\beta)^{2}\right] \right) \,,
\label{MFphi:thth:hiT:FinalResultcoincidence}
\end{equation}
where we ignored the subleading term in equation
(\ref{MFphi:thth:hiT:FinalResult}) to derive the coincidence limit
above.

\section{Entropy Generation in Quantum Mechanics}
\label{Entropy Generation in Quantum Mechanics}

Now the stage is set to study entropy generation in our quantum
field theoretical model, let us digress somewhat and study entropy
generation in the analogous quantum mechanical model first. This
allows us to quantitatively compare the evolution of the entropy
resulting from the perturbative master equation and in our
correlator approach. A comparison in field theory is not possible
so far, due to the shortcomings of the conventional approach to
decoherence using the master equation as discussed in the
introduction. Let us consider the quantum mechanical system of
$N+1$ simple harmonic oscillators $x$ and $q_n$, $1 \leq n \leq
N$, coupled by an interaction term of the form $h_n x q_n^2$:
\begin{equation}\label{QMLagrangian}
L=L_{S}+L_{E}+L_{SE}=\frac{1}{2}\left(\dot{x}^{2}-\omega_{0}^{2}
x^{2}\right )+\sum_{n=1}^{N} \frac{1}{2}
\left(\dot{q}_n^{2}-\omega_{n}^{2}q_{n}^{2} \right) - \frac{1}{2}
h_{n} x q^{2}_{n} \,,
\end{equation}
which indeed is the quantum mechanical $D=1$ dimensional analogue
of the Lagrangian density in equation (\ref{action:tree2})
considered before. Here, $\omega_0$ and $\{\omega_n\}$ are the
frequencies of the oscillators as usual. The $x$ oscillator is the
system in a thermal environment of $\{q_n\}$ oscillators. We
absorb the mass in the time in our action, and the remaining
dimensionless mass dependence in the $\{q_n\}$.

\subsection{The Kadanoff-Baym Equations in Quantum Mechanics}
\label{The Kadanoff-Baym Equations in Quantum Mechanics}

The free thermal statistical and causal propagator in quantum
mechanics read:
\begin{subequations}
\label{QMStatisticalandCausalPropagator}
\begin{eqnarray}
F_{q_n}(t,t') &=&
\frac{\cos(\omega_{n}(t-t'))}{2\omega_{n}}\coth(\beta\omega_{n}/2)
\label{QMStatisticalandCausalPropagatora}\\
\Delta^{c}_{q_n}(t,t')&=&\frac{-1}{\omega_{n}}\sin(\omega_{n}(t-t'))
\label{QMStatisticalandCausalPropagatorb}\,.
\end{eqnarray}
\end{subequations}
The statistical and causal self-energies of the $x$-system at
lowest order in perturbation theory are defined by:
\begin{subequations}
\label{QMStatisticalandCausalSelfMass}
\begin{eqnarray}
M^{F}_x(t,t') & = & - \sum_{n=1}^{N} \frac{h^{2}_n}{4} \left[
\left(\imath\Delta^{+-}_{q_n}(t,t') \right )^{2}+ \left
(\imath\Delta^{-+}_{q_n}(t,t') \right)^{2}
\right] \label{QMStatisticalandCausalSelfMassa} \\
M^{c}_x(t,t') & =& - \sum_{n=1}^{N} \frac{\imath h^{2}_n}{2}
\left[ \left (\imath\Delta^{+-}_{q_n}(t,t') \right )^{2}- \left
(\imath\Delta^{-+}_{q_n}(t,t') \right)^{2} \right]
\label{QMStatisticalandCausalSelfMassb}\,,
\end{eqnarray}
\end{subequations}
and are calculated as:
\begin{subequations}
\label{QMStatisticalandCausalSelfMass2}
\begin{eqnarray}
M_{x}^{F}(t,t') & = & - \sum_{n=1}^{N}  \frac{h^{2}_n}{2} \left[(F_{q_n}(t,t'))^{2}-\frac{1}{4}(\Delta^{c}_{q_n}(t,t'))^{2} \right]\nonumber \\
& = & - \sum_{n=1}^{N} \frac{h^{2}_n}{16\omega_{n}^{2}} \left[
\left(\coth^{2} \left(\frac{\beta\omega_{n}}{2} \right )+1 \right
)\cos \left(2\omega_{n}(t-t') \right)+
\coth^{2}\left(\frac{\beta\omega_{n}}{2} \right)-1 \right]
 \label{QMStatisticalandCausalSelfMass2a}\\
M_{x}^{c}(t,t') & = & - \sum_{n=1}^{N} h^{2}_{n} F_{q_n}(t,t')
\Delta^{c}_{q_n}(t,t') =\sum_{n=1}^{N}
\frac{h^{2}_{n}}{4\omega_{n}^{2}} \sin \left(2\omega_{n}(t-t')
\right) \coth \left (\frac{\beta\omega_{n}}{2} \right
)\label{QMStatisticalandCausalSelfMass2b}\,.
\end{eqnarray}
\end{subequations}
As in our field theoretical model we neglect the backreaction from
the system on the environment. The Kadanoff-Baym equations for the
$x$-system for the statistical and causal propagators are now
given by:
\begin{subequations}
\label{QMStatisticalandCausalKBEqns}
\begin{eqnarray}
(\partial_{t}^{2} + \omega_{0}^{2}) F_{x}(t,t') + \int_{0}^{t'}
\mathrm{d} t_{1} M_{x}^{F}(t,t_{1})\Delta_{x}^{c}(t_{1},t')
-\int_{0}^{t} \mathrm{d} t_{1} M_{x}^{c}(t,t_{1})F_{x}(t_{1},t')
&=& 0 \label{QMStatisticalandCausalKBEqnsa}\\
(\partial_{t}^2+\omega_{0}^{2})\Delta_{x}^{c}(t,t') -
\int_{t'}^{t}\mathrm{d} t_{1} M_{x}^{c}(t,t_{1})
\Delta_{x}^{c}(t_{1},t') &=& 0
\label{QMStatisticalandCausalKBEqnsb}\,,
\end{eqnarray}
\end{subequations}
where $\{t,t'\} \ge t_{0}=0$. An important difference compared to
the field theoretical Kadanoff-Baym equations in
(\ref{eomcausalprop2}) and (\ref{eomnumericalc}) is that we do not
have to renormalise them. Also, we do not have to consider any
memory effects before $t_0 = 0$. We can now straightforwardly
solve the Kadanoff-Baym equations above by numerical methods to
find the statistical propagator and hence the quantum mechanical
analogue of the phase space area (\ref{deltaareainphasespace}) and
entropy (\ref{entropy}) as functions of time.

\subsection{The Master Equation in Quantum Mechanics}
\label{The Master Equation in Quantum Mechanics}

In order to derive the perturbative master equation for our model,
we follow Paz and Zurek \cite{Paz:2000le}. The perturbative master
equation is obtained straightforwardly from the Dyson series,
truncated at second order, as a solution to the von Neumann
equation and reads:
\begin{eqnarray}\label{QMMasterEquation1}
\dot{\hat{\rho}}_{\mathrm{red}}(t) &=& \frac{1}{\imath} [
\hat{H}_{S}(t),\hat{\rho}_{\mathrm{red}}(t) ] + \frac{1}{\imath}
\sum_{n=1}^{N}  \frac{h_n}{2} [ \langle \hat{q}_n^2(t)
\rangle \hat{x} , \hat{\rho}_{\mathrm{red}}(t)  ]\\
&& - \sum_{\substack{n=1 \\
m=1}}^{N} \int_0^{t} \mathrm{d}t_1 K_{nm}^{(3)}(t,t_1) [
\hat{x},[\hat{x}(t_1-t), \hat{\rho}_{\mathrm{red}}(t)  ]] +
K_{nm}^{(4)}(t,t_1) [ \hat{x},\{\hat{x}(t_1-t),
\hat{\rho}_{\mathrm{red}}(t)  \}] \,,
\end{eqnarray}
where we follow the notation of Paz and Zurek and define the
coefficients:
\begin{subequations}
\label{QMMasterEquation2}
\begin{eqnarray}
K_{nm}^{(3)}(t,t_1) &=& \frac{h_n h_m}{8} \langle \{
\hat{q}_n^2(t), \hat{q}_m^2(t_1) \}\rangle - \frac{h_n h_m}{4}
\langle
\hat{q}_n^2(t)\rangle \langle \hat{q}_m^2(t_1)\rangle  \label{QMMasterEquation2a}\\
K_{nm}^{(4)}(t,t_1) &=& \frac{h_n h_m}{8} \langle [
\hat{q}_n^2(t), \hat{q}_m^2(t_1) ]\rangle
\label{QMMasterEquation2b}\,.
\end{eqnarray}
\end{subequations}
Also, note that $\hat{x}(t) = \hat{x}\cos(\omega_0 t) + \hat{p}_x
\sin(\omega_0 t)/\omega_0$ due to changing back to the
Schr\"odinger picture from the interaction picture
\cite{Paz:2000le}. The master equation above reduces further to:
\begin{equation}\label{QMMasterEquation3}
\dot{ \hat{\rho}}_{\mathrm{red}}(t) = \frac{1}{i} [
\hat{H}_{S}(t),\hat{\rho}_{\mathrm{red}}(t) ] - \int_{0}^{t}
\mathrm{d} t_1 \nu(t_1) [\hat{x},[\hat{x}(-t_1),
\hat{\rho}_{\mathrm{red}}(t)]] - i \eta(t_1) [\hat{x}, \{
\hat{x}(-t_1), \hat{\rho}_{\mathrm{red}}(t) \}]\,.
\end{equation}
In the equation above, we dropped the linear term in equation
(\ref{QMMasterEquation1}) as a time dependent linear term will not
affect the entropy \cite{Koksma:2010zi}. The noise and dissipation
kernels $\nu(t)$ and $\eta(t)$ are straightforwardly related to
$K_{nm}^{(3)}(t,t_1)$ and $K_{nm}^{(4)}(t,t_1)$, respectively, and
read at the lowest order in perturbation theory \cite{Hu:1991di,
Hu:1993vs}:
\begin{subequations}
\label{QMMasterEquation4}
\begin{eqnarray}
\nu(t) &=& \sum_{n=1}^N \frac{h^{2}_n}{16\omega_{n}^{2}} \left[
\left(\coth^{2} \left(\frac{\beta\omega_{n}}{2} \right )+1 \right
)\cos \left(2\omega_{n}t \right)+
\coth^{2}\left(\frac{\beta\omega_{n}}{2} \right)-1 \right]
\label{QMMasterEquation4a} \\
&=&  - M_{x}^{F}(t,0)  \nonumber \\
\eta(t) &=& \sum_{n=1}^N \frac{h^{2}_{n}}{8\omega_{n}^{2}} \sin
\left(2\omega_{n}t \right) \coth \left (\frac{\beta\omega_{n}}{2}
\right ) \label{QMMasterEquation4b}\\
&=&  \frac{1}{2} M_{x}^{c}(t,0)   \nonumber \,.
\end{eqnarray}
\end{subequations}
Note that we can easily relate the noise and dissipation kernels
that appear in the master equation to the self-mass corrections in
the Kadanoff-Baym equations. This is an important identity and we
will return to it shortly. One thus finds:
\begin{equation}\label{QMMasterEquation5}
\dot{ \hat{\rho}}_{\mathrm{red}}(t) = -i [ \hat{H}_{S}(t) +
\frac{1}{2} \Omega^2(t) \hat{x}^2 ,\hat{\rho}_{\mathrm{red}}(t) ]
- i \gamma(t) [\hat{x},\{\hat{p}_x,
\hat{\rho}_{\mathrm{red}}(t)\}] - D(t) [ \hat{x},[\hat{x},
\hat{\rho}_{\mathrm{red}}(t)]] - f(t) [\hat{x}, [\hat{p}_x,
\hat{\rho}_{\mathrm{red}}(t)]],
\end{equation}
where the frequency ``renormalisation'' $\Omega(t)$, the damping
coefficient $\gamma(t)$ and the two diffusion coefficients $D(t)$
and $f(t)$ are given by:
\begin{subequations}
\label{QMMasterEquation5AA}
\begin{eqnarray}
\Omega^2(t) &=& -2 \int_0^{t} \mathrm{d} t_1 \eta(t_1)
\cos(\omega_0 t_1) \label{QMMasterEquation5a} \\
&=& \sum_{n=1}^{N} \frac{h_n^2}{4\omega_{n}^{2}(4\omega_{n}^{2} -
\omega_{0}^{2})} \coth \left (\frac{\beta\omega_{n}}{2} \right )
\left\{
-2\omega_{n}(1-\cos(\omega_{0}t)\cos(2\omega_{n}t))+\omega_{0}\sin(\omega_{0}t)\sin(2\omega_{n}t)\right\}
\nonumber \\
\gamma(t) &=&  \int_0^{t} \mathrm{d} t_1 \eta(t_1)
\frac{\sin(\omega_0 t_1)}{\omega_0} =\sum_{n=1}^{N}
\frac{h_n^2}{16\omega_{n}^{2}\omega_{0}}
\coth\left(\frac{\beta\omega_{n}}{2} \right)\left\{
\frac{\sin([\omega_{0}-2\omega_{n}]t)}{\omega_{0}-2\omega_{n}}-\frac{\sin([\omega_{0}+2\omega_{n}]t)}{\omega_{0}+2\omega_{n}}\right\}
 \label{QMMasterEquation5b}\\
D(t) &=& \int_0^{t} \mathrm{d} t_1 \nu(t_1) \cos(\omega_0 t_1)
\label{QMMasterEquation5c} \\
&=& \sum_{n=1}^{N} \frac{h_n^2}{16\omega_{n}^{2}} \Bigg[
\frac{1}{2} \left(\coth^2\left(\frac{\beta\omega_{n}}{2} \right)+1
\right) \left\{
\frac{\sin([\omega_{0}-2\omega_{n}]t)}{\omega_{0}-2\omega_{n}}+
\frac{\sin([\omega_{0}+2\omega_{n}]t)}{\omega_{0}+2\omega_{n}}\right\}
\nonumber \\
&& \qquad\qquad +\left(\coth^{2}\left(\frac{\beta\omega_{n}}{2}
\right )-1 \right
)\frac{\sin(\omega_{0}t)}{\omega_{0}} \Bigg]  \nonumber \\
f(t) &=& - \int_0^{t} \mathrm{d} t_1 \nu(t_1) \frac{\sin(\omega_0
t_1)}{\omega_0} \label{QMMasterEquation5d}  \\
&=& \sum_{n=1}^{N} \frac{-h_n^2}{16\omega_{n}^{2}\omega_{0}}
\Bigg[\left(\coth^2 \left (\frac{\beta\omega_{n}}{2} \right) +1
\right )\frac{\{\omega_{0}(1-\cos(\omega_{0}t)\cos(2\omega_{n}t))-
2\omega_{n}\sin(\omega_{0}t)\sin(2\omega_{n}t)\}}{\omega_{0}^{2}-4\omega_{n}^{2}}
\nonumber \\
&& \qquad\qquad
+\left(\coth^{2}\left(\frac{\beta\omega_{n}}{2}\right)-
1\right)\frac{1-\cos(\omega_{0}t)}{\omega_{0}}\Bigg] \nonumber \,.
\end{eqnarray}
\end{subequations}
We are now ready to solve the master equation
(\ref{QMMasterEquation5}). As we are interested in the evolution
of 2-point functions, let us make a Gaussian ansatz and project
this operator equation on the position bras and kets as follows:
\begin{equation} \label{QMreduceddensitymatrixAnsatz}
\rho_{\mathrm{red}}(x,y;t) = \langle x|
\hat{\rho}_{\mathrm{red}}(t)| y \rangle=
\tilde{\mathcal{N}}(t)\exp\left[
-\tilde{a}(t)x^2-\tilde{a}^{\ast}(t)y^2+2\tilde{c}(t)xy \right]\,.
\end{equation}
It turns out to be advantageous to directly compute the time
evolution of our three non-trivial Gaussian correlators. Analogous
methods have been used in \cite{Blume-Kohout:2003} to analyse
decoherence in an upside down simple harmonic oscillator. We can
thus derive the following set of differential equations
\cite{Koksma:2010dt}:
\begin{subequations}
\label{QMMasterEquation6Corr}
\begin{eqnarray}
\frac{\mathrm{d}\langle \hat{x}^2\rangle }{\mathrm{d}t} &=& -
\frac{ \dot{\tilde{a}}_{\mathrm{R}} -
\dot{\tilde{c}}}{4(\tilde{a}_{\mathrm{R}} - \tilde{c})^2} = 2
\left \langle \frac{1}{2} \{\hat{x},\hat{p}\}\right \rangle
\label{QMMasterEquation6Corra} \\
\frac{\mathrm{d}\langle \hat{p}^2\rangle }{\mathrm{d}t} &=& -
2(\omega_0^2+\Omega^2)\left \langle \frac{1}{2}
\{\hat{x},\hat{p}\}\right \rangle - 4 \gamma(t) \langle
\hat{p}^2\rangle + 2D(t) \label{QMMasterEquation6Corrb} \\
\frac{\mathrm{d}\left \langle \frac{1}{2}
\{\hat{x},\hat{p}\}\right \rangle }{\mathrm{d}t} &=& -
(\omega_0^2+\Omega^2)\langle \hat{x}^2 \rangle + \langle
\hat{p}^2\rangle- f(t) -2 \gamma(t) \left \langle \frac{1}{2}
\{\hat{x},\hat{p}\} \right \rangle \label{QMMasterEquation6Corrc}
\,.
\end{eqnarray}
\end{subequations}
These equations are completely equivalent to the master equation
when one is interested in the Gaussian correlators only.
Initially, we impose that the system is in a pure state:
\begin{subequations}
\label{QMMasterEquation7}
\begin{eqnarray}
\langle \hat{x}^2(t_0)\rangle &=& \frac{1}{2\omega_0}
\label{QMMasterEquation7a} \\
\langle \hat{p}^2 (t_0) \rangle  &=& \frac{\omega_0}{2}
\label{QMMasterEquation7b} \\
\left \langle \frac{1}{2} \{\hat{x},\hat{p}\}(t_0)\right \rangle
&=& 0 \,. \label{QMMasterEquation7c}
\end{eqnarray}
\end{subequations}
We can then straightforwardly find the quantum mechanical analogue
of the phase space area in equation (\ref{deltaareainphasespace})
and the von Neumann entropy for the system in equation
(\ref{entropy}).

Let us finally remark that our Kadanoff-Baym equations
(\ref{QMStatisticalandCausalKBEqns}) can also be obtained starting
from the Feynman-Vernon Gaussian path integral exponential
obtained in \cite{Hu:1993vs} that is normally used to derive the
perturbative master equation. For example, after integrating out
the environment at one loop order (which is usually a first step
in deriving a master equation), the 1PI equations of motion can be
obtained from the effective action:
\begin{eqnarray}
&& S_{S}[x^+]-S_{S}[x^-]
  - \int_{0}^{\infty}\mathrm{d}t_{1}\mathcal{T}[x^+(t_{1}))-x^-(t_{1})]
  +\imath \, \int_{0}^{\infty}\!\mathrm{d}t_{1}
\int_{0}^{t_{1}}\!\mathrm{d}t_{2}[x^+(t_{1})-x^-(t_{1})]
           \nu(t_{1}-t_{2})[x^+(t_{2})-x^-(t_{2})]
 \nonumber\\
&& \hskip 2.5cm
 +\int_{0}^{\infty}\!\mathrm{d}t_{1}\int_{0}^{t_{1}}\!\mathrm{d}t_{2}
   [x^+(t_{1})-x^-(t_{1})]
\eta(t_{1}-t_{2})[x^+(t_{2})+x^-(t_{2})]
 \label{FeynmanVernon}\\
 &&
  =\,S_{S}[x^+]-S_{S}[x^-]
 - \int_{0}^{\infty}\mathrm{d}t_{1}\mathcal{T}[x^+(t_{1})-x^-(t_{1})]
-\frac12 \sum_{a,b = \pm}ab
 \int_{0}^{\infty}\!\mathrm{d}t_{1}\int_{0}^{\infty}\!\mathrm{d}t_{2}
x^a(t_{1}) \imath M^{ab}(t_1;t_2) x^b(t_{2})
 \nonumber\,.
\end{eqnarray}
Here, $M^{ab}$ are the self-masses that can be read off from
equations (\ref{QMStatisticalandCausalSelfMass2})
and~(\ref{reduction:selfMass:MF+Mc2}), $S_{S}[x^\pm]$ is the free
action defined by equation (\ref{QMLagrangian}) and $\eta$ and
$\nu$, or, equivalently, the causal and statistical self-masses,
are given in equation (\ref{QMMasterEquation4}). In the equation
above, $\mathcal{T}$ denotes the tadpole contribution, which does
not affect the entropy \cite{Koksma:2010zi}, and reads:
\begin{equation}\label{FeynmanVernon2}
\mathcal{T} = \sum_{n=1}^N \frac{h_{n}}{4 \omega_{n}}\coth \left
(\frac{\beta\omega_{n}}{2} \right )\,,
\end{equation}
which is easily inferred from the interaction term
in~(\ref{QMLagrangian})
and~(\ref{QMStatisticalandCausalPropagatora}). The quantum
corrected equation of motion for $x(t)$ follows straightforwardly
by variation of equation (\ref{FeynmanVernon}) with respect to
$x^\pm(t)$, and setting $x^\pm(t)$ equal to $x(t)$.
 More generally, if one would introduce
non-local sources for two-point functions in the Feynman-Vernon
path integral, one would obtain the Kadanoff-Baym equations in
(\ref{QMStatisticalandCausalKBEqns}).

\subsection{Time Evolution of the Entropy in Quantum Mechanics}
\label{Time Evolution of the Entropy in Quantum Mechanics}

\begin{figure}
    \begin{minipage}[t]{.45\textwidth}
        \begin{center}
\includegraphics[width=\textwidth]{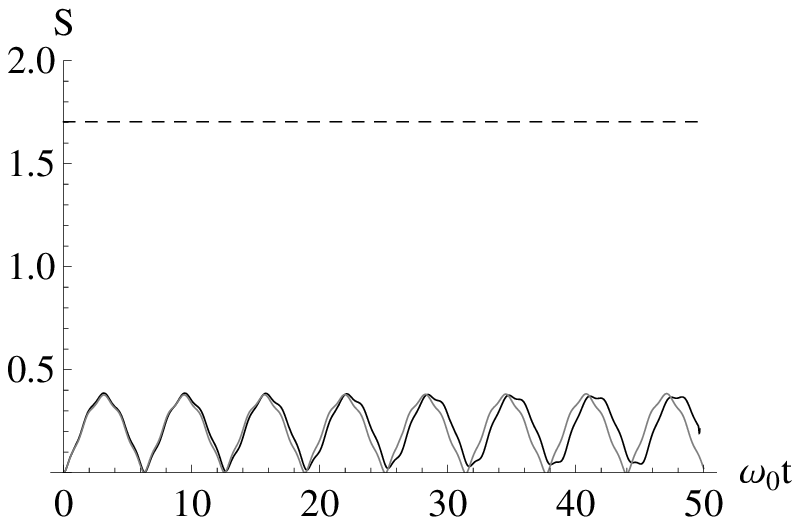}
   {\em \caption{Entropy as a function of time for $N=1$ in the
   non-resonant regime. The Gaussian von Neumann entropy (black) agrees
   with the entropy from the master equation (gray) up to the
   expected perturbative corrections. The dashed line indicates full thermalisation of
   $x$. We use
   $\omega_1/\omega_0=2$, $h/\omega_0^3=1$ and $\beta\omega_0=1/2$.
   \label{fig:QM S NRes}}}
        \end{center}
   \end{minipage}
\hfill
    \begin{minipage}[t]{.45\textwidth}
        \begin{center}
\includegraphics[width=\textwidth]{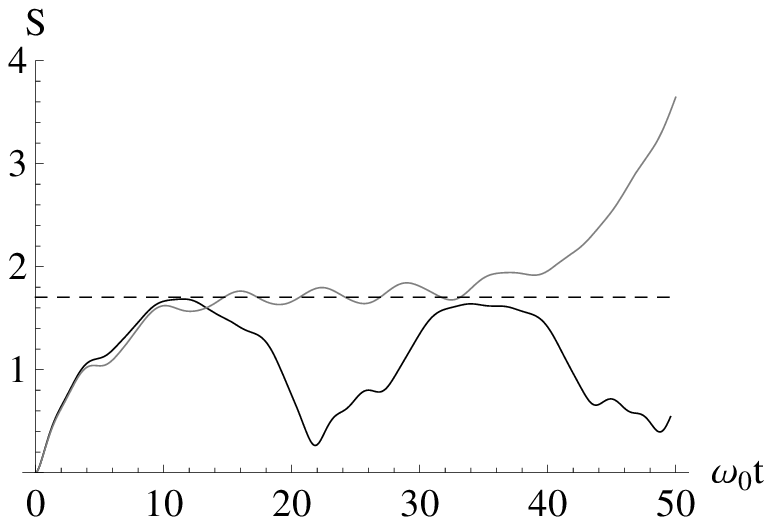}
   {\em \caption{Entropy as a function of time for $N=1$ in the
   resonant regime. The Gaussian von Neumann entropy (black)
   shows a stable behaviour, unlike the entropy from the
   perturbative master equation (gray) that reveals unphysical secular
   growth. The dashed line indicates full thermalisation of
   $x$. We use $\omega_1/\omega_0=0.53$, $h/\omega_0^3=0.1$ and $\beta\omega_0=1/2$.
   \label{fig:QM S Res}}}
        \end{center}
    \end{minipage}
\vskip 0.1cm
    \begin{minipage}[t]{.45\textwidth}
        \begin{center}
\includegraphics[width=\textwidth]{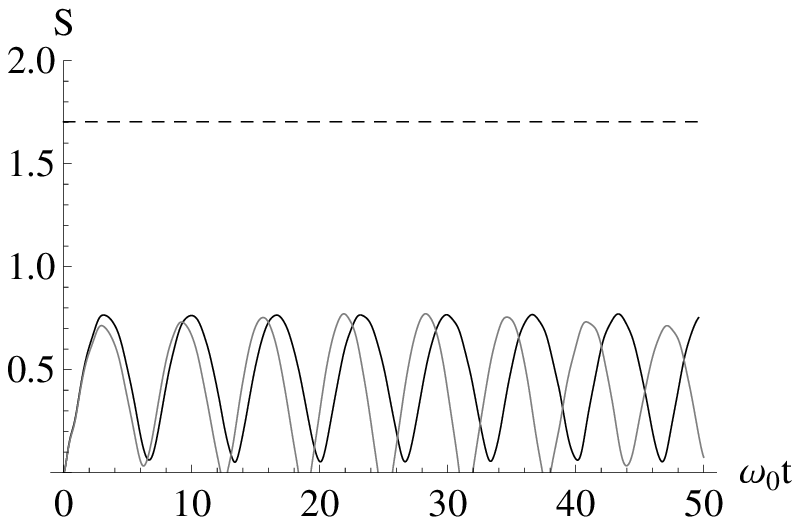}
   {\em \caption{Entropy as a function of time for $N=50$ in the
   non-resonant regime. The Gaussian von Neumann entropy (black) agrees
   with the entropy from the master equation (gray) up to the
   expected perturbative corrections. We use
   $\omega_n/\omega_0 \in [2,4]$, $1 \leq n \leq N$, $h/\omega_0^3=1/2$ and $\beta\omega_0=1/2$.
   \label{fig:QM S N NRes}}}
        \end{center}
    \end{minipage}
\hfill
    \begin{minipage}[t]{.45\textwidth}
        \begin{center}
\includegraphics[width=\textwidth]{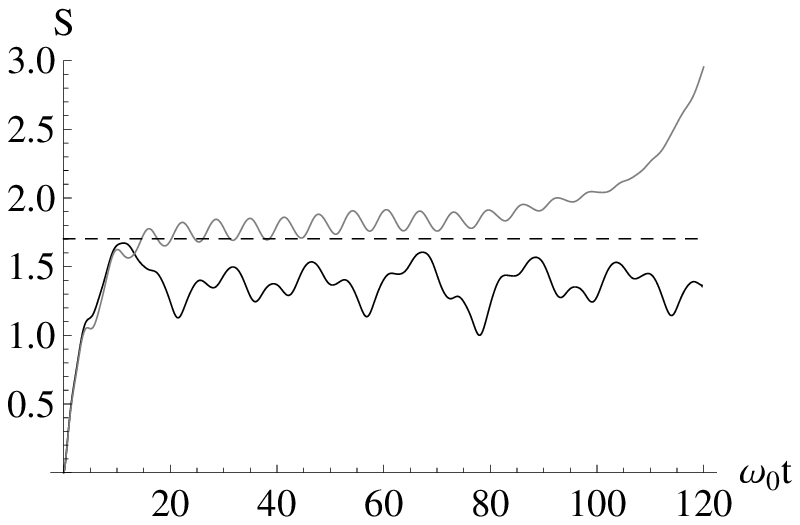}
   {\em \caption{Entropy as a function of time for $N=50$ in the
   resonant regime. The Gaussian von Neumann entropy (black)
   yields a stable behaviour in time, unlike the entropy from the
   perturbative master equation (gray) that reveals secular
   growth. We use $\omega_n/\omega_0 \in [0.5,0.6]$, $1 \leq n \leq N$, $h/\omega_0^3=0.015$ and $\beta\omega_0=1/2$.
   \label{fig:QM S N Res}}}
        \end{center}
    \end{minipage}
\end{figure}
Let us discuss our results. It is important to distinguish between
the so-called resonant regime and non-resonant regime
\cite{Koksma:2010dt}. In the former, we have that one or more
$\omega_n \simeq \omega_0/2$, with $1 \leq n \leq N$. In the
non-resonant regime all environmental frequencies differ
significantly from $\omega_0/2$ and are as a consequence
effectively decoupled from the system oscillator. If one wants to
study the efficient decoherence of such a system, the non-resonant
regime is not the relevant regime to consider.

In figure \ref{fig:QM S NRes} we show the Gaussian von Neumann
entropy resulting from the Kadanoff-Baym equations and from the
perturbative master equation as a function of time in black and
gray, respectively. At the moment, we consider just one
environmental oscillator $N=1$ in the non-resonant regime. Here,
the two entropies agree nicely up to the expected perturbative
corrections due to the inappropriate resummation scheme of the
perturbative master equation to which we will return shortly.
However, let us now consider figure \ref{fig:QM S Res} where we
study the resonant regime for $N=1$. Clearly, the entropy
resulting from the master equation breaks down and suffers from
physically unacceptable secular growth. The behaviour of the
Gaussian von Neumann entropy from the Kadanoff-Baym equations is
perfectly stable. Moreover, given the weak coupling
$h/\omega_0^3=0.1$, we do not observe perfect thermalisation
(indicated by the dashed black line).

If we consider $N=50$ environmental oscillators, the qualitative
picture does not change. In figure \ref{fig:QM S N NRes} we show
the evolution of the two entropies in the non-resonant regime, and
in figure \ref{fig:QM S N Res} in the resonant regime. The entropy
from the perturbative master equation blows up as before, whereas
the Gaussian von Neumann entropy is stable. In figure \ref{fig:QM
S N NRes} we randomly select 50 frequencies in the interval
$[2,4]$ which is what we denote by $\omega_n/\omega_0 \in [2,4]$.
In the resonant regime we use $\omega_n/\omega_0 \in [0.5,0.6]$.
The breakdown of the perturbative master equation in this regime
is generic.

Just as discussed in \cite{Koksma:2010dt}, energy is conserved in
our model such that the Poincar\'e recurrence theorem applies.
This theorem states that our system will after a sufficiently long
time return to a state arbitrary close to its initial state. The
Poincar\'e recurrence time is the amount of time this takes.
Compared to the $N=1$ case we previously considered, we observe
for $N=50$ in figure \ref{fig:QM S N Res} that the Poincar\'e's
recurrence time has increased. Thus, by including more and more
oscillators, decoherence becomes rapidly more irreversible, as one
would expect. If we extend this discussion to field theory, where
several modes couple due to the loop integrals (hence $N
\rightarrow \infty$), we conclude that clearly our Poincar\'e
recurrence time becomes infinite. Hence, the entropy increase has
become irreversible for all practical purposes and our system has
(irreversibly) decohered.

In decoherence studies, one is usually interested in extracting
two quantitative results: the decoherence rate and the total
amount of decoherence. As emphasised before, we take the point of
view that the Gaussian von Neumann entropy should be used as
\emph{the} quantitative measure of decoherence, as it is an
invariant measure of the phase space occupied by a state. Hence,
the rate of change of the phase space area (or entropy) is the
decoherence rate and the total amount of decoherence is the total
(average) amount of entropy that is generated at late times. This
is to be contrasted with most of the literature
\cite{Zurek:2003zz} where non-invariant measures of decoherence
are used. The statement regarding the decoherence rate we would
like to make here, however, is that our Gaussian von Neumann
entropy and the entropy resulting from the master equation would
give the same result as their early times evolution coincides. The
master equation does however not predict the total amount of
decoherence accurately. In the resonant regime the entropy
following from the perturbative master equation blows up at late
times and, consequently, fails to accurately predict the total
amount of decoherence that has taken place. Our correlator
approach to decoherence does not suffer from this fatal
shortcoming.

\subsection{Deriving the Master Equation from the Kadanoff-Baym Equations}
\label{Deriving the Master Equation from the Kadanoff-Baym
Equations}

The secular growth is caused by the perturbative approximations
used in deriving the master equation
(\ref{QMMasterEquation6Corr}). The coefficients appearing in the
master equation diverge when $\omega_n = \omega_0/2$ which can be
appreciated from equation (\ref{QMMasterEquation5AA}). However,
there is nothing non-perturbative about the resonant regime. Our
interaction coefficient $h$ is still very small such that the
self-mass corrections to $\omega_0^2$ are tiny.

Here we outline the perturbative approximations that cause the
master equation to fail. In order to do this, we simply derive the
master equation from the Kadanoff-Baym equations by making the
appropriate approximations. Of course, equation
(\ref{QMMasterEquation6Corra}) is trivial to prove. The
Kadanoff-Baym equations are given in equation
(\ref{QMStatisticalandCausalKBEqns}) and contain memory integrals
over the causal and statistical propagators. We make the
approximation to use the free equation of motion for the causal
propagator appearing in the memory integrals according to which:
\begin{equation}\label{QMpertapproximation3}
(\partial_t^2 + \omega_0^2 )  \Delta_x^{c, \mathrm{free}}(t,t') =
0\,.
\end{equation}
This equation is trivially solved in terms of sines and cosines.
Let us thus impose initial conditions at $t=t'$ as follows:
\begin{eqnarray}
\Delta_{x}^{c}(t,t') \simeq  \Delta_x^{c, \mathrm{free}}(t,t') &=&
\cos(\omega_0 \Delta t) \Delta_{x}^{c}(t',t') + \frac{1}{\omega_0}
\sin(\omega_0 \Delta t)
\partial_t \Delta_{x}^{c}(t,t') |_{t=t'}
\label{QMpertapproximation1}\\
&=& - \frac{1}{\omega_0} \sin(\omega_0 \Delta t) \nonumber \,.
\end{eqnarray}
Here, $\Delta t = t-t'$. We relied upon some basic properties of
the causal propagator (see e.g. equation
(\ref{boundaryconditionscausual}) in the next section). Likewise,
we approximate the statistical propagator appearing in the memory
integrals as:
\begin{eqnarray}
F_{x}(t,t') \simeq  F_{x}^{\mathrm{free}}(t,t')&=& \cos(\omega_0
\Delta t) F_x(t',t') + \frac{1}{\omega_0} \sin(\omega_0 \Delta t)
\partial_t
F_x(t,t')|_{t=t'} \label{QMpertapproximation2}\\
&=&  \cos(\omega_0 \Delta t) \langle \hat{x}^2(t')\rangle +
\frac{1}{\omega_0} \sin(\omega_0 \Delta t) \left \langle
\frac{1}{2} \{\hat{x},\hat{p}\}(t')\right\rangle \nonumber \,,
\end{eqnarray}
where we inserted how our statistical propagator can be related to
our three Gaussian correlators, from the quantum mechanical
version of equation (\ref{3 equal time correlators}). Note that
expression (\ref{QMpertapproximation2}) is not symmetric under
exchange of $t$ and $t'$, whereas the statistical propagator as
obtained from e.g. the Kadanoff-Baym equations of course respects
this symmetry.

Now, we send $t'\rightarrow t$ in the Kadanoff-Baym equations and
carefully relate the statistical propagator and derivatives
thereof to quantum mechanical expectation values. From equation
(\ref{QMStatisticalandCausalKBEqnsa}), where we change variables
to $\tau = t-t_1$, it thus follows that:
\begin{eqnarray}
\partial_{t}^{2}F_{x}(t,t')|_{t=t'} &=& -\omega_{0}^{2}\langle \hat{x}^2(t)\rangle - \int_{0}^{t} \mathrm{d}
\tau M_{x}^{F}(\tau,0) \frac{\sin(\omega_0 \tau)}{\omega_0} +
\langle \hat{x}^2(t)\rangle \int_{0}^{t} \mathrm{d} \tau
M_{x}^{c}(\tau,0)\cos(\omega_0 \tau)
\label{QMpertapproximation4}\\
&&  - \left\langle \frac{1}{2} \{\hat{x},\hat{p}\}(t)\right
\rangle \int_{0}^{t} \mathrm{d} \tau M_{x}^{c}(\tau,0)
\frac{\sin(\omega_0 \tau)}{\omega_0}\nonumber \,.
\end{eqnarray}
Using equations (\ref{QMMasterEquation4}) and
(\ref{QMMasterEquation5AA}), equation (\ref{QMpertapproximation4})
above reduces to (\ref{QMMasterEquation6Corrc}):
\begin{equation}
\frac{\mathrm{d}\left \langle \frac{1}{2}
\{\hat{x},\hat{p}\}\right \rangle }{\mathrm{d}t} = -
(\omega_0^2+\Omega^2)\langle \hat{x}^2 \rangle + \langle
\hat{p}^2\rangle- f(t) -2 \gamma(t) \left \langle \frac{1}{2}
\{\hat{x},\hat{p}\} \right \rangle\,. \nonumber
\end{equation}
Here, we used the identities derived in equation
(\ref{QMMasterEquation4}) that relate the noise and dissipation
kernels of the master equation to our causal and statistical
self-mass. In order to derive the final master equation for the
correlator $\langle \hat{p}^2\rangle$, we have to use the
following subtle argument:
\begin{equation}\label{QMpertapproximation5}
\partial_t^2\partial_{t'} F(t,t')|_{t=t'} = \frac{1}{2}
\frac{\mathrm{d}}{\mathrm{d}t} \langle \hat{p}^2(t)\rangle \,.
\end{equation}
In order to derive its corresponding differential equation, we
thus have to act with $\partial_{t'}$ on equation
(\ref{QMStatisticalandCausalKBEqnsa}) and then send $t'\rightarrow
t$. As an intermediate step, we can present:
\begin{eqnarray}
 \frac{1}{2} \frac{\mathrm{d}}{\mathrm{d}t}
\langle \hat{p}^2(t)\rangle &=& -\omega_{0}^{2} \left\langle
\frac{1}{2} \{\hat{x},\hat{p}\}(t)\right \rangle - \int_{0}^{t}
\mathrm{d} \tau M_{x}^{F}(\tau,0) \cos(\omega_0 \tau)
\label{QMpertapproximation6}\\
&& + \int_{0}^{t} \mathrm{d} \tau M_{x}^{c}(\tau,0) \left[
-\omega_0 \sin(\omega_0 \tau)\langle \hat{x}^2(t')\rangle+
\cos(\omega_0 \tau) \left\langle \frac{1}{2}
\{\hat{x},\hat{p}\}(t')\right \rangle -
\frac{\sin(\omega_0\tau)}{\omega_0}
\partial_{t'}\left\{\partial_{t}F_x(t,t')|_{t=t'}\right\}\right]\nonumber
\,,
\end{eqnarray}
where we still have to send $t'\rightarrow t$ on the second line.
Now, one can use:
\begin{equation}\label{QMpertapproximation7}
\partial_{t'}\left\{\partial_{t}F_x(t,t')|_{t=t'}\right\} = \langle \hat{p}^2(t')\rangle -
\omega_0^2 \langle \hat{x}^2(t')\rangle \,.
\end{equation}
In the light of equations (\ref{QMMasterEquation4}) and
(\ref{QMMasterEquation5AA}), equation (\ref{QMpertapproximation6})
simplifies to equation (\ref{QMMasterEquation6Corrb}):
\begin{equation}
\frac{\mathrm{d}\langle \hat{p}^2\rangle }{\mathrm{d}t}= -
2(\omega_0^2+\Omega^2)\left \langle \frac{1}{2}
\{\hat{x},\hat{p}\}\right \rangle - 4 \gamma(t) \langle
\hat{p}^2\rangle + 2D(t)\nonumber \,.
\end{equation}
We thus conclude that we can derive the master equation for the
correlators from the Kadanoff-Baym equations using the
perturbative approximation in equations
(\ref{QMpertapproximation1}) and (\ref{QMpertapproximation2}).
Clearly, this approximation invalidates the intricate resummation
techniques of the quantum field theoretical 2PI scheme. In the 2PI
framework, one resums an infinite number of Feynman diagrams in
order to obtain a stable and thermalised late time evolution. By
approximating the memory integrals in the Kadanoff-Baym equations,
the master equation spoils this beautiful property.

The derivation presented here can be generalised to quantum field
theory. By using similar approximations, one can thus derive the
renormalised correlator equations that would follow from the
perturbative master equation.

\section{Results: Entropy Generation in Quantum Field Theory}
\label{Results: Entropy Generation in Quantum Field Theory}

Let us now return to field theory and solve for the statistical
propagator and hence fix the Gaussian von Neumann entropy of our
system. For completeness, let us here just once more recall
equation (\ref{eomcausalprop2}) and (\ref{eomnumericalc}) for the
causal and statistical propagator:
\begin{subequations}
\label{KBEqn}
\begin{eqnarray}
(\partial_{t}^{2}+k^{2}+m^{2}_{\phi}) \Delta^{c}_{\phi}(k,t,t') &-&
\left(\partial_{t}^{2}+k^{2}\right) \int_{t'}^{t} \mathrm{d}t_{1}
Z^{c}_{\phi}(k,t,t_{1}) \Delta^{c}_{\phi}(k,t_{1},t') -
\int_{t'}^{t} \mathrm{d}t_{1} M^{c}_{\phi,\mathrm{th}}(k,t,t_{1})
\Delta^{c}_{\phi}(k,t_{1},t') =0 \label{KBEqnCausalProp} \\
(\partial_{t}^{2}+k^{2}+m^{2}_{\phi}) F_{\phi}(k,t,t') &-&
\left(\partial_{t}^{2}+k^{2}\right) \Bigg[ \int_{-\infty}^{t_{0}}
\mathrm{d}t_{1} Z^{c}_{\phi}(k,t,t_{1})
F_{\phi}^{\mathrm{free}}(k,t_{1},t') + \int_{t_{0}}^{t}
\mathrm{d}t_{1} Z^{c}_{\phi}(k,t,t_{1}) F_{\phi}(k,t_{1},t')
\label{KBEqnStatProp}\\
&& \qquad\qquad - \int_{-\infty}^{t_{0}} \mathrm{d}t_{1}
Z^{F}_{\phi}(k,t,t_{1})
\Delta^{c,\mathrm{free}}_{\phi}(k,t_{1},t') - \int_{t_{0}}^{t'}
\mathrm{d}t_{1} Z^{F}_{\phi}(k,t,t_{1})
\Delta^{c}_{\phi}(k,t_{1},t') \Bigg] \nonumber \\
&& - \int_{-\infty}^{t_{0}} \mathrm{d}t_{1}
M^{c}_{\phi,\mathrm{th}}(k,t,t_{1})
F_{\phi}^{\mathrm{free}}(k,t_{1},t') - \int_{t_{0}}^{t}
\mathrm{d}t_{1} M^{c}_{\phi,\mathrm{th}}(k,t,t_{1})
F_{\phi}(k,t_{1},t') \nonumber\\
&& + \int_{-\infty}^{t_{0}} \mathrm{d}t_{1}
M^{F}_{\phi,\mathrm{th}}(k,t,t_{1})
\Delta^{c,\mathrm{free}}_{\phi}(k,t_{1},t') + \int_{t_{0}}^{t'}
\mathrm{d}t_{1} M^{F}_{\phi,\mathrm{th}}(k,t,t_{1})
\Delta^{c}_{\phi}(k,t_{1},t') = 0 \nonumber \,.
\end{eqnarray}
\end{subequations}
We use all self-masses calculated previously: we need the vacuum
self-masses in equation (\ref{selfmasses}), one of the two
following infinite past memory kernels in equation
(\ref{MemoryKernel}) or (\ref{MemoryKernelvac}) depending on the
initial conditions chosen, the thermal causal self-mass in
(\ref{selfMass:causal3}), the vacuum-thermal contribution to the
statistical self-mass in equation (\ref{selfMass:statistical6})
and finally the high temperature or low temperature contribution
to the thermal-thermal statistical self-mass in equation
(\ref{MFphi:thth:hiT:FinalResult}) or (\ref{MFphi:thth:4}). We are
primarily interested in two cases, a constant mass for our system
field and a changing one:
\begin{subequations}
\label{massbehaviour}
\begin{eqnarray}
m_{\phi}(t) &=& m_{0} = \mathrm{const} \label{massbehavioura} \\
m^{2}_{\phi}(t) &=& A+B\tanh(\rho \{t-t_{\mathrm{m}}\})
\label{massbehaviourb} \,,
\end{eqnarray}
\end{subequations}
where we let $A$ and $B$ take different values. Also,
$t_{\mathrm{m}}$ is the time at which the mass changes, which we
take to be $\rho t_{\mathrm{m}}=30$. Let us outline our numerical
approach. In the code, we take $t_{0}=0$ and we let $\rho t$ and
$\rho t'$ run between 0 and 100 for example. As in the vacuum
case, we first need to determine the causal propagator, as it
enters the equation of motion of the statistical propagator. The
boundary conditions for determining the causal propagator are as
follows:
\begin{subequations}
\label{boundaryconditionscausual}
\begin{eqnarray}
\Delta^{c}_{\phi}(t,t) &=& 0 \label{boundaryconditionscausuala} \\
\partial_{t} \Delta^{c}_{\phi}(t,t')|_{t=t'} &=& -1
\label{boundaryconditionscausualb} \,,
\end{eqnarray}
\end{subequations}
Condition (\ref{boundaryconditionscausuala}) has to be satisfied
by definition and condition (\ref{boundaryconditionscausualb})
follows from the commutation relations.

Once we have solved for the causal propagator, we can consider
evaluating the statistical propagator. As in the $T=0$ case, the
generated entropy is a constant which can be appreciated from a
rather simple argument \cite{Koksma:2009wa}. When
$m_{\phi}=\mathrm{const}$, we have $F_{\phi}(k,t,t') =
F_{\phi}(k,t-t')$ such that quantities like:
\begin{subequations}
\label{Fconstantmass2}
\begin{eqnarray}
F_{\phi}(k,0) &=& \int_{-\infty}^{\infty} \frac{\mathrm{d}k^{0}}{2\pi} F_{\phi}(k^{\mu}) \label{Fconstantmass2a} \\
\left.\partial_{t} F_{\phi}(k, \Delta t) \right|_{\Delta t =0} &=&
- \imath \int_{-\infty}^{\infty} \frac{\mathrm{d}k^{0}}{2\pi}
k^{0}
F_{\phi}(k^{\mu}) \label{Fconstantmass2b} \\
\left.\partial_{t'}\partial_{t} F_{\phi}(k, \Delta t)
\right|_{\Delta t =0} &=& \int_{-\infty}^{\infty}
\frac{\mathrm{d}k^{0}}{2\pi} k_{0}^{2} F_{\phi}(k^{\mu})
\label{Fconstantmass2c} \,,
\end{eqnarray}
\end{subequations}
are time independent. Consequently, the phase space area
$\Delta_{k}$ is constant, and so is the generated entropy. If our
initial conditions differ from these values, we expect to observe
some transient dependence. This entropy is thus the interacting
thermal entropy. The total amount of generated entropy measures
the total amount of decoherence that has occurred. Given a
temperature $T$, the thermal entropy provides a good estimate of
the maximal amount of entropy that can be generated (perfect
decoherence), however depending on the particular parameters in
the theory this maximal amount of entropy need not always be
reached (imperfect decoherence). Effectively, the interaction
opens up phase space for the system field implying that less
information about the system field is accessible to us and hence
we observe an increase in entropy. In order to evaluate the
integrals above, we need the statistical propagator in Fourier
space:
\begin{eqnarray}
F_{\phi}(k^{\mu}) &=& \frac{1}{2} \frac{ \imath M^{+-}(k^{\mu}) +
\imath M^{-+}(k^{\mu})}{ \imath M^{+-}(k^{\mu}) - \imath
M^{-+}(k^{\mu})}\left[ \frac{\imath}{k_\mu k^\mu + m_{\phi}^{2} +
\imath M^{\mathrm{r}}(k^\mu)} - \frac{\imath}{k_\mu k^\mu +
m_{\phi}^{2} + \imath M^{\mathrm{a}}(k^\mu)}\right]
\label{FourierStatisticalA} \,.
\end{eqnarray}
Here, $\imath M^{\mathrm{r}}$ and $\imath M^{\mathrm{a}}$ are the
retarded and advanced self-masses, respectively. All the
self-masses in Fourier space in this expression are derived in
appendix \ref{AppendixB}. The discussion above is important for
understanding how to impose boundary conditions for the
statistical propagator at $t_0$. We impose either so-called ``pure
state initial conditions'' or ``mixed state initial conditions''.
If we constrain the statistical propagator to occupy the minimal
allowed phase space area initially, we impose pure state initial
conditions and set:
\begin{subequations}
\label{PSboundaryconditionsstat}
\begin{eqnarray}
F_{\phi}(t_{0},t_{0}) &=& \frac{1}{2 \omega_{\mathrm{in}}} \label{PSboundaryconditionsstata} \\
\partial_{t} F_{\phi}(t,t_{0})|_{t=t_{0}} &=& 0
\label{PSboundaryconditionsstatb} \\
\partial_{t'}\partial_{t} F_{\phi}(t,t')|_{t=t'=t_{0}} &=&
\frac{\omega_{\mathrm{in}}}{2} \label{PSboundaryconditionsstatc}
\,,
\end{eqnarray}
\end{subequations}
where $\omega_{\mathrm{in}}$ refers to the initial mass
$m_{\phi}(t_0)$ of the field if the mass changes throughout the
evolution. This yields $\Delta_{k}(t_{0}) = 1$ such that:
\begin{equation}
S_{k}(t_{0})=0 \label{PSboundaryconditionsstatd} \,,
\end{equation}
Initially, we thus force the field to occupy the minimal area in
phase space. Clearly, if we constrain our field to be in such an
out-of-equilibrium state initially, we should definitely not
include all memory kernels pretending that our field has already
been interacting from negative infinity to $t_0$. Otherwise, our
field would have thermalised long before $t_0$ and could have
never began the evolution in its vacuum state. If we thus impose
pure state initial conditions, we must drop the ''thermal memory
kernels'':
\begin{equation}\label{ThMemoryDrop}
\int_{-\infty}^{t_{0}} \mathrm{d}t_{1}
M^{c}_{\phi,\mathrm{th}}(k,t,t_{1})
F_{\phi}^{\mathrm{free}}(k,t_{1},t') \qquad \mathrm{and} \qquad \int_{-\infty}^{t_{0}} \mathrm{d}t_{1}
M^{F}_{\phi,\mathrm{th}}(k,t,t_{1})
\Delta^{c,\mathrm{free}}_{\phi}(k,t_{1},t') \,,
\end{equation}
but rather keep the ''vacuum memory kernels'' in equation
(\ref{KBEqnStatProp}), which are the other two memory kernels
involving free propagators. We evaluated the relevant integrals in
closed form in equation (\ref{MemoryKernelvac}). This setup
roughly corresponds to switching on the coupling $h$ adiabatically
slowly at times before $t_0$. At $t_0$, the temperature of the
environment is suddenly switched on such that the system responds
to this change from $t_0$ onwards. Note that if we would not
include any memory effects and switch on the coupling $h$
non-adiabatically at $t_0$, the pure state initial conditions
would correspond to the physically natural choice. This would
however also instantaneously change the vacuum of our theory, and
we would thus need to renormalise our theory both before and after
$t_0$ separately. Including the vacuum memory kernels is thus
essential, as it ensures that our evolution is completely finite
at all times without the need for time dependent
counterterms\footnote{In \cite{Baacke:1998di, Baacke:1999nq} the
renormalisation of fermions in an expanding Universe is
investigated where a similar singularity at the initial time $t_0$
is encountered. It could in their case however be removed by a
suitably chosen Bogoliubov transformation.}.

Secondly, we can impose mixed state boundary conditions, where we
use the numerical values for the statistical propagator and its
derivatives calculated from equations (\ref{Fconstantmass2}) and
(\ref{FourierStatisticalA}), such that we have $\Delta_{k}(t_{0})
= \Delta_{\mathrm{ms}}=\mathrm{const}$ and:
\begin{equation}
S_{k}(t_{0}) =  S_{\mathrm{ms}}>0
\label{MSboundaryconditionsstatd} \,,
\end{equation}
where we use the subscript ``ms'' to denote ``mixed state''. In
other words, we constrain our system initially to be in the
interacting thermal state and $ S_{\mathrm{ms}}$ is the value of
the interacting thermal entropy. The integrals in equation
(\ref{Fconstantmass2}) can now be evaluated numerically to yield
the appropriate initial conditions. For example when $\beta
\rho=1/2$, $k/\rho=1$, $m_{\phi}/\rho=1$ and $h/\rho=3$, we find:
\begin{subequations}
\label{Fconstantmass3}
\begin{eqnarray}
\left. F_{\phi}(k/\rho=1,\Delta t)\right|_{\Delta t =0} &=&
1.89885
\label{Fconstantmass3a} \\
\left.\partial_{t} F_{\phi}(k/\rho=1, \Delta t) \right|_{\Delta t
=0} &=& 0 \label{Fconstantmass3b} \\
\left.\partial_{t'}\partial_{t} F_{\phi}(k/\rho=1, \Delta t)
\right|_{\Delta t =0} &=& 2.08941 \label{Fconstantmass3c} \,.
\end{eqnarray}
\end{subequations}
Clearly, equation (\ref{Fconstantmass3b}) always vanishes as the
integrand is an odd function of $k^{0}$. The numerical value of
the phase space area in this case follows from equations
(\ref{Fconstantmass3}) and (\ref{deltaareainphasespace}) as:
\begin{equation}
\Delta_{\mathrm{ms}} = 3.98371 \label{DeltaconstantMass} \,.
\end{equation}
The interacting thermal entropy hence reads:
\begin{equation}
S_{\mathrm{ms}} = 1.67836 \label{SconstantMass} \,.
\end{equation}
The mixed state initial condition basically assumes that our
system field has already equilibrated before $t_0$ such that the
entropy has settled to the constant mixed state value. In this
case, we include of course both the vacuum memory kernels and the
thermal memory kernels\footnote{Let us make an interesting
theoretical observation that to our knowledge would apply for any
interacting system in quantum field theory. Suppose our coupling
$h$ would be time independent. Suppose also that the system field
$\phi$ and the environment field $\chi$ form a closed system
together. Now, imagine that we are interested in the time
evolution of the entropy at some finite time $t_0$. Our system
field has then already been interacting with the environment at
times before $t_0$ such that one can expect that our system has
equilibrated at $t_0$. Hence, to allow out-of-equilibrium initial
conditions, one must always change the theory slightly. The
possibility that we advocate is to drop those memory kernels that
do not match the chosen initial condition. In this way, the
evolution history of our field is consistent.}.

A few more words on the memory kernels for the mixed state
boundary conditions are in order. For the vacuum memory kernels,
we of course use equation (\ref{MemoryKernel}). It is
unfortunately not possible to evaluate the thermal memory kernels
in closed form too. The two integrals in equation
(\ref{ThMemoryDrop}) have to be evaluated numerically as a
consequence. One can numerically verify that the integrands are
highly oscillatory and do not settle quickly to some constant
value for each $t$ and $t'$ due to the competing frequencies
$\omega$ and $k$. We chose to integrate from -300 to $t_0 \rho=0$
and smooth out the remaining oscillations of the integral by
defining a suitable average over half of the period of the
oscillations.

Finally, let us outline the numerical implementation of the
Kadanoff-Baym equations (\ref{KBEqn}). Solving for the causal
propagator is straightforward as equation
(\ref{boundaryconditionscausual}) provides us for each $t'$ with
two initial conditions at $t=t'$ and at $t=t'+ \Delta t$, where
$\Delta t$ is the numerical step size. We can thus solve the
causal propagator as a function of $t$ for each fixed $t'$.
Solving for the statistical propagator is somewhat more subtle.
The initial conditions, e.g. in equation
(\ref{PSboundaryconditionsstat}), for a given choice of parameters
only fix $F_{\phi}(t_0,t_0)$, $F_{\phi}(t_0+\Delta t,t_0) =
F_{\phi}(t_0,t_0+\Delta t)$ and $F_{\phi}(t_0+\Delta t,t_0+\Delta
t)$. This is sufficient to solve for $F_{\phi}(t,t_0)$ and
$F_{\phi}(t,t_0+\Delta t)$ as functions of time for fixed $t'=t_0$
and $t'=t_0+\Delta t$. Now, we can use the symmetry relation
$F_{\phi}(t,t')=F_{\phi}(t',t)$ such that we can also find
$F_{\phi}(t_0,t')$ and $F_{\phi}(t_0+\Delta t,t')$ as functions of
$t'$ for fixed $t=t_0$ and $t=t_0+\Delta t$. The latter step
provides us with the initial data that is sufficient to find
$F_{\phi}(t,t')$ as a function of $t$ for each fixed $t'$.

Once we have solved for the statistical propagator, our life
becomes much easier as we can immediately find the phase space
area via relation (\ref{deltaareainphasespace}). The phase space
area fixes the entropy.

\subsection{Evolution of the Entropy: Constant Mass}
\label{Constant Mass}

\begin{figure}
    \begin{minipage}[t]{.45\textwidth}
        \begin{center}
\includegraphics[width=\textwidth]{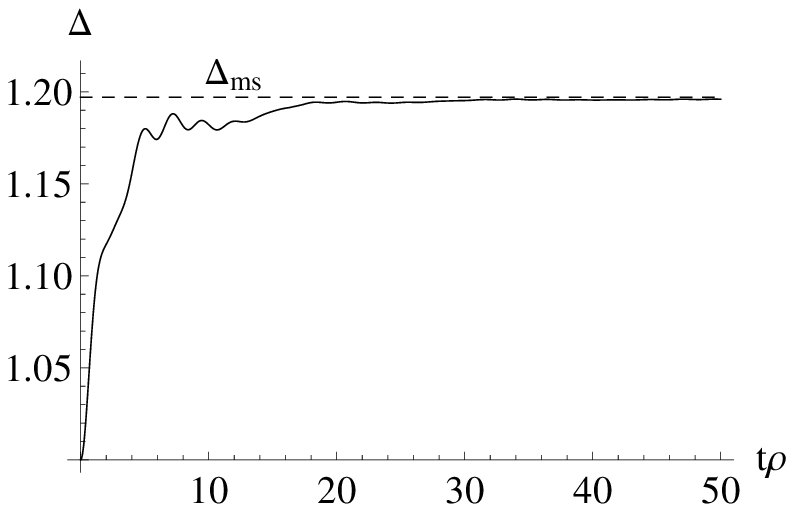}
   {\em \caption{Phase space area as a function of time. It
   settles nicely to $\Delta_{\mathrm{ms}}$, indicated by the
   dashed black line. We use $\beta \rho = 2$, $k/\rho=1$, $m_\phi / \rho = 1$,
   $h/\rho=4$ and a total number of steps $N=2000$ up to $t\rho=100$.
   \label{fig:PSA_lowT_1}}}
        \end{center}
   \end{minipage}
\hfill
    \begin{minipage}[t]{.45\textwidth}
        \begin{center}
\includegraphics[width=\textwidth]{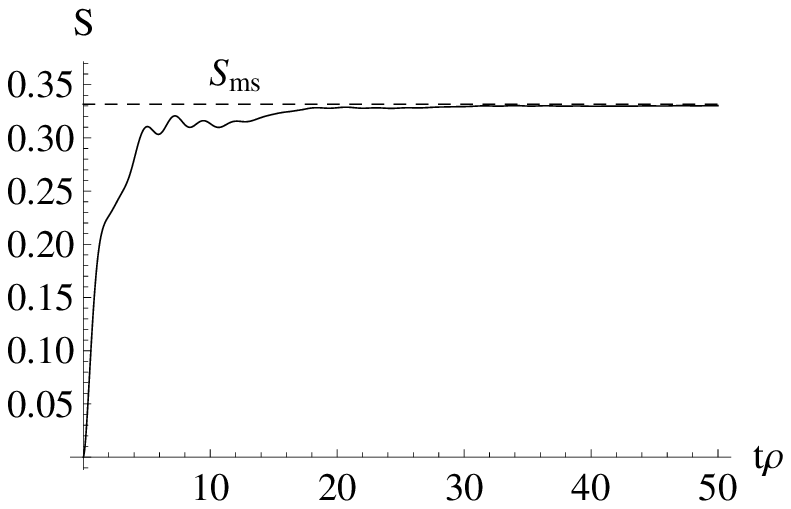}
   {\em \caption{Entropy as a function of time. The evolution of
   the entropy is obtained from the phase space area in figure
   \ref{fig:PSA_lowT_1}. For $50< t\rho < 100$, the entropy
   continues to coincide with $S_{\mathrm{ms}}$.
   \label{fig:S_lowT_1}}}
        \end{center}
    \end{minipage}
\vskip 0.1cm
    \begin{minipage}[t]{.45\textwidth}
        \begin{center}
\includegraphics[width=\textwidth]{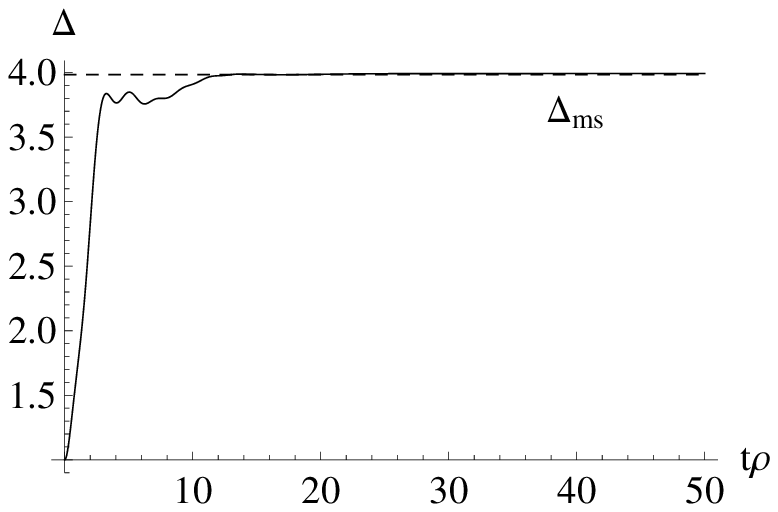}
   {\em \caption{Phase space area as a function of time. At high
   temperatures, we see that the phase space area settles quickly
   again to $\Delta_{\mathrm{ms}}$. We use $\beta \rho = 1/2$, $k/\rho=1$, $m_\phi / \rho = 1$,
   $h/\rho=3$ and $N=2000$ up to $t\rho=100$.
   \label{fig:PSA_highT_1}}}
        \end{center}
    \end{minipage}
\hfill
    \begin{minipage}[t]{.45\textwidth}
        \begin{center}
\includegraphics[width=\textwidth]{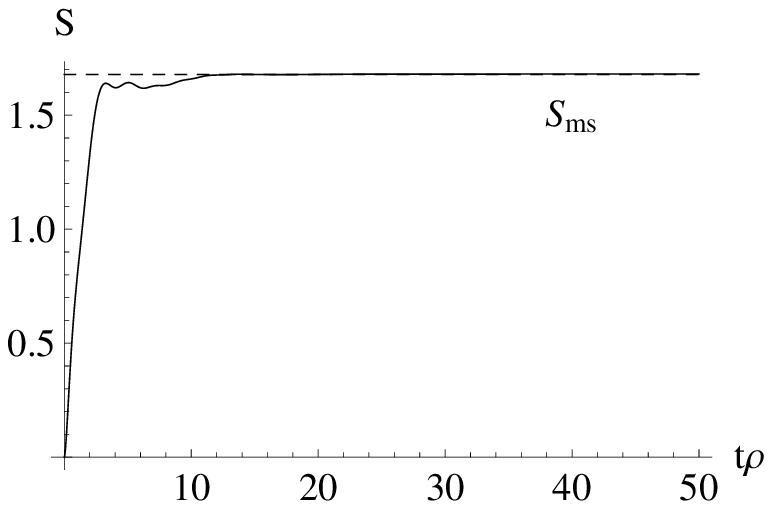}
   {\em \caption{Entropy as a function of time, which follows
   again from the evolution of the phase space area as a function
   of time as depicted in figure \ref{fig:PSA_highT_1}. Our pure
   state quickly appears to our observer as a mixed state with a
   large positive entropy $S_{\mathrm{ms}}$.
   \label{fig:S_highT_1}}}
        \end{center}
    \end{minipage}
\vskip 0.1cm
    \begin{minipage}[t]{.45\textwidth}
        \begin{center}
\includegraphics[width=\textwidth]{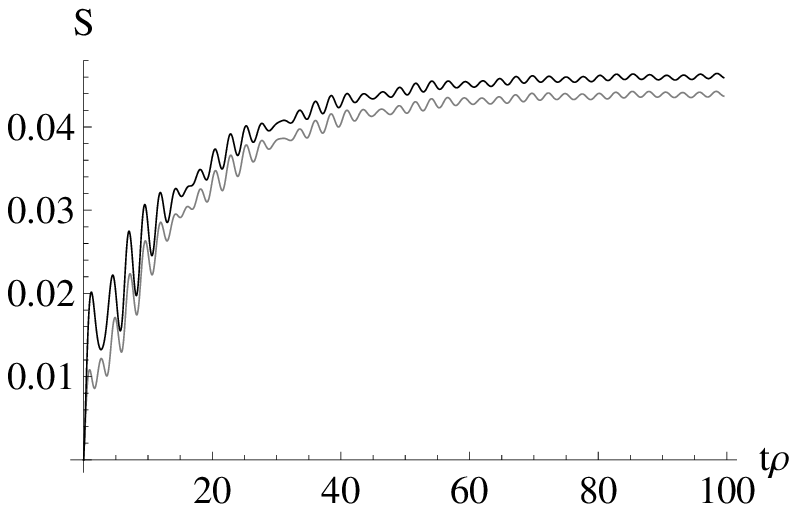}
   {\em \caption{Entropy as a function of time. The evolution at
   very low temperatures $\beta \rho = 10$ (black) resembles the vacuum
   evolution $\beta \rho = \infty$ (gray) from \cite{Koksma:2009wa} as
   one would intuitively expect. We furthermore use $k/\rho=1$, $m_\phi / \rho = 1$,
   $h/\rho=4$ and $N=2000$ up to $t\rho=100$.
   \label{fig:PSA_comparewithvacuum}}}
        \end{center}
    \end{minipage}
\hfill
    \begin{minipage}[t]{.45\textwidth}
        \begin{center}
\includegraphics[width=\textwidth]{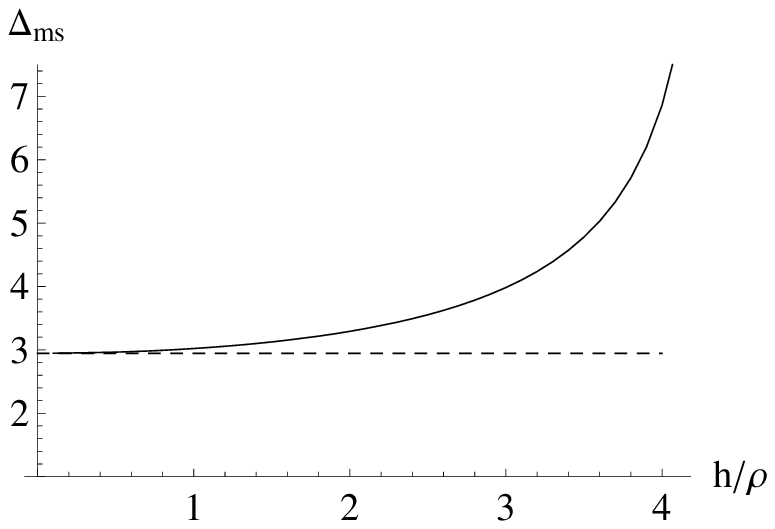}
   {\em \caption{We show $\Delta_{\mathrm{ms}}$, the interacting thermal phase space area, as a function of
   $h/\rho$. For $h/\rho \ll 1$, we see that
   $\Delta_{\mathrm{ms}}$ is almost equal to the free phase
   space area $\Delta_{\mathrm{free}} = \coth(\beta\omega/2)$
   indicated by the dashed black line. For larger values of $h/\rho$ we approach the
   non-perturbative regime. We use $\beta \rho = 1/2$, $k/\rho=1$ and $m_\phi / \rho =
   1$.
   \label{fig:nonperturbativeregime}}}
        \end{center}
    \end{minipage}
\end{figure}
Let us firstly turn our attention to figure \ref{fig:PSA_lowT_1}.
This plot shows the phase space area as a function of time at a
fairly low temperature $\beta \rho = 2$. Starting at
$\Delta_k(t_0)=1$, its evolution settles precisely to
$\Delta_{\mathrm{ms}}$, indicated by the dashed black line, as one
would expect. From the evolution of the phase space area, one
readily finds the evolution of the entropy as a function of time
in figure \ref{fig:S_lowT_1}.

At a higher temperature, $\beta \rho = 1/2$, we observe in figures
\ref{fig:PSA_highT_1} and \ref{fig:S_highT_1} that the generated
phase space area and entropy as a function of time is larger. This
can easily be understood by realising that the thermal value of
the entropy, set by the environment, provides us with a good
estimate of the maximal amount of decoherence that our system can
experience. Again we observe an excellent agreement between
$\Delta_{\mathrm{ms}}$ or $S_{\mathrm{ms}}$ and the corresponding
numerical evolution.

Let us now discuss figure \ref{fig:PSA_comparewithvacuum}. Here,
we show two separate cases for the evolution of the entropy: one
at a very low temperature $\beta \rho = 10$ (in black) and one
vacuum evolution $\beta \rho = \infty$ (in gray) which we already
calculated in \cite{Koksma:2009wa}. As we would intuitively
expect, we see that the former case settles to an entropy
$S_{\mathrm{ms}}=0.04551$ that is slightly above the vacuum
asymptote $S_{\mathrm{ms}}=0.04326$.

Finally, in figure \ref{fig:nonperturbativeregime} we show the
interacting phase space area $\Delta_{\mathrm{ms}}$ as a function
of the coupling $h$. For $h/\rho \ll 1$, we see that
$\Delta_{\mathrm{ms}}$ approaches the free thermal phase space
area $\Delta_{\mathrm{free}} = \coth(\beta\omega/2)$. For larger
values of the coupling, we see that $\Delta_{\mathrm{ms}} >
\Delta_{\mathrm{free}}$. If these two differ significantly, we
enter the non-perturbative regime. In the perturbative regime,
this plot substantiates our earlier statement that the free
thermal entropy $\Delta_{\mathrm{free}}$ provides us with a good
estimate of the total amount of decoherence that our system can
experience. Our system however thermalises to
$\Delta_{\mathrm{ms}}$, and not to $\Delta_{\mathrm{free}}$ as the
interaction changes the nature of the free thermal state.

The most important point of the results shown here is that,
although a pure state with vanishing entropy $S_k = 0$ remains
pure under unitary evolution, we perceive this state over time as
a mixed state with positive entropy $S_{\mathrm{ms}}>0$ as
non-Gaussianities are generated by the evolution (both in the
correlation between the system and environment as well as higher
order correlations in the system itself) and subsequently
neglected in our definition of the Gaussian von Neumann entropy.
The total amount of decoherence corresponds to the interacting
thermal entropy $S_{\mathrm{ms}}$.

\subsection{Decoherence Rates}
\label{Decoherence Rates}

As the Gaussian von Neumann entropy in equation (\ref{entropy}) is
the only invariant measure of the entropy of a Gaussian state, we
take the point of view that this quantity, or equivalently the
phase space area in equation (\ref{deltaareainphasespace}), should
be taken as the quantitative measure for decoherence. This agrees
with the general view on decoherence according to which the
decoherence rate is the rate at which a system in a pure state
evolves into a mixed state due to its interaction with an
environment. This is to be contrasted with some of the literature
where different, non-invariant measures are proposed
\cite{Zurek:2003zz, Giraud:2009tn}. For example in
\cite{Zurek:2003zz, Zurek:1991vd}, the superposition of two
minimum uncertainty Gaussian states located at positions $x$ and
$x'$ is considered. The decoherence rate is defined differently,
i.e.: it is the characteristic timescale at which the off-diagonal
contributions in the total density matrix decay and coincides with
the timescale at which the interference pattern in the Wigner
function decays. It is given by:
\begin{equation}\label{timescaleZurek1}
\tau_{\mathrm{D}}^{-1} = \gamma
\left(\frac{x-x'}{\lambda_{T}}\right)^{2} \,,
\end{equation}
where the thermal de Broglie wavelength is given by
$\lambda_{T}=(2mk_{B}T)^{-1/2}$. In other words, according to
\cite{Zurek:2003zz}, the decoherence rate depends on the spatial
separation $x-x'$ of the two Gaussians. Note that in quantum field
theory the expression would generalise to $\tau_{\mathrm{D}}^{-1}
\propto (\phi-\phi')^2$. This is just one example, one can find
other definitions of decoherence in the literature.

The main difference is that our decoherence rate does not depend
on the configuration space variables $x$ or $\phi$ but is an
intrinsic property of the state. In other words, we do not look at
different spatial regions of the state, but rather to the state as
a whole from which we extract one decoherence rate. As we outlined
in \cite{Koksma:2010zi}, a nice intuitive way to visualise the
process of decoherence is in Wigner space. The Wigner transform of
a density matrix coincides with the Fourier transform with respect
to its off-diagonal entries. As discussed previously, the phase
space area measures the area the state occupies in Wigner space in
units of the minimum phase space area $\hbar/2$, which we refer to
as the statistical particle number $n$. The pure state considered
in the previous subsection decoheres and its phase space area
increases to approximately its thermal value. When $\Delta \gg 1$
($n\gg 1$), different regions in phase space of area $\hbar/2$
are, to a good approximation, not correlated and thus evolve
independently. As we have considered Gaussian states only and not
the superposition of two spatially separated Gaussians, which when
considered together is in fact a highly non-Gaussian state, a
direct comparison is not straightforward.

Let us extract the decoherence rate from the evolution of the
entropy. We define the decoherence time scale to be the
characteristic time it takes for the phase space area
$\Delta_{k}(t)$ to settle to its constant mixed state value
$\Delta_{\mathrm{ms}}$. The phase space area approaches the
constant asymptotic value in an exponential manner:
\begin{equation}\label{timescale1}
\frac{\mathrm{d}}{\mathrm{d}t}\delta\Delta_{k}(t) +
\Gamma_{\mathrm{dec}} \delta\Delta_{k}(t) =0 \,,
\end{equation}
where $\delta\Delta_{k}(t)=\Delta_{\mathrm{ms}} - \Delta_{k}(t)$
and where $ \Gamma_{\mathrm{dec}}$ is the decoherence rate. This
equation is equivalent to $\dot{n}_{k}= - \Gamma_{\mathrm{dec}}(
n_{k}-n_{\mathrm{ms}})$, where $n_{k}$ is defined in equation
(\ref{particlenumber}) and $n_{\mathrm{ms}}$ is the stationary $n$
corresponding to $\Delta_{\mathrm{ms}}$. As in the vacuum case
\cite{Koksma:2009wa}, we anticipate that the decoherence rate is
given by the single particle decay rate of the interaction $\phi
\rightarrow \chi^2$. The single particle decay rate
reads\footnote{For cases where $m_{\chi} \neq 0$, see
\cite{Boyanovsky:2004dj}.}:
\begin{equation}\label{singleparticledecayrate}
\Gamma_{\phi \rightarrow \chi \chi} = - \left.
\frac{\mathrm{Im}(\imath
M^{\mathrm{r}}_{\phi})}{\omega_{\phi}}\right|_{k^{0}=\omega_{\phi}}
= \frac{h^2}{32 \pi \omega_{\phi}} +\frac{h^2}{16\pi k \beta
\omega_{\phi}} \log\left(
\frac{1-e^{-\frac{\beta}{2}(\omega_{\phi}+k)}}
{1-e^{-\frac{\beta}{2}(\omega_{\phi}-k)}}\right) \,,
\end{equation}
where we used the retarded self-mass in Fourier space in equation
(\ref{FourierSelfMassRelationa}) and several relevant self-masses
in appendix~\ref{AppendixB}. Let us briefly outline the steps
needed to derive the result above. In order to calculate $\imath
M^{\mathrm{r}}_{\phi}(k^{\mu})$, we use $\imath
M_{\phi,\mathrm{vac}}^{++}(k^{\mu})$ in equation
(\ref{SelfMassFourier3a}) and $\imath
M_{\phi,\mathrm{vac}}^{+-}(k^{\mu})$ in equation
(\ref{SelfMassFourier9}). There are no thermal-thermal
contributions to $\imath M^{\mathrm{r}}_{\phi}(k^{\mu})$ which can
be appreciated from equation (\ref{SelfMassFourier7}). Finally, in
order to derive the vacuum-thermal contribution, let us recall
equation (\ref{reduction:selfMass:MF+Mc2c}) given by:
$M^{++}_{\phi}(k,t,t') = M^{F}_{\phi}(k,t,t') + {\rm
sgn}(t-t^\prime)\imath M^{c}_{\phi}(k,t,t')/2$. We clearly need
the vacuum-thermal contribution to $M^{F}_{\phi}(k^{\mu})$ which
is given in equation (\ref{SelfMassFourier11}). The imaginary part
of the second term vanishes, which can be seen by making use of an
inverse Fourier transform, just as in the first lines of equations
(\ref{DecomposeiM++}) and (\ref{DecomposeiM++2}). This fixes
$\imath M^{\mathrm{r}}_{\phi}(k^{\mu})$ completely.

\begin{figure}[t!]
    \begin{minipage}[t]{.43\textwidth}
        \begin{center}
\includegraphics[width=\textwidth]{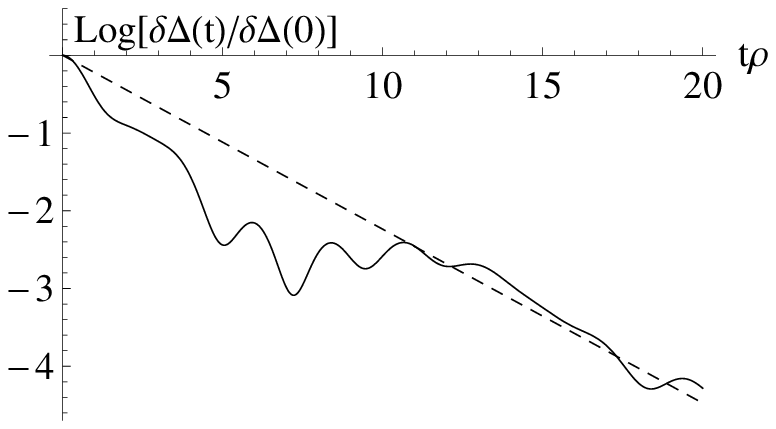}
   {\em \caption{Decoherence rate at low temperatures. We show the
   exponential approach to $\Delta_{\mathrm{ms}}$ in solid black
   and the corresponding decoherence rate given in equation
   (\ref{decoherencerate}) (dashed line). We use the phase space
   area from figure~\ref{fig:PSA_lowT_1}.
    \label{fig:DecoherenceRatelowT}}}
        \end{center}
    \end{minipage}
\hfill
    \begin{minipage}[t]{.43\textwidth}
        \begin{center}
\includegraphics[width=\textwidth]{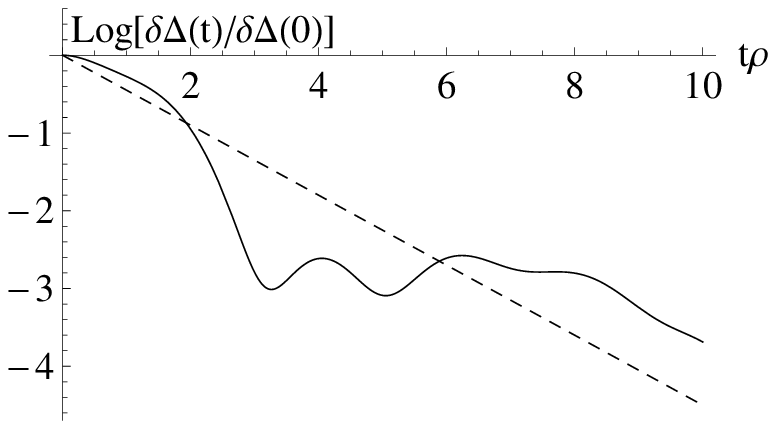}
   {\em \caption{Decoherence rate at high temperatures. We show the
   exponential approach to $\Delta_{\mathrm{ms}}$ in solid black
   and the corresponding decoherence rate given in equation
   (\ref{decoherencerate}) (dashed line). We use the phase space
   area from figure~\ref{fig:PSA_highT_1}.
   \label{fig:DecoherenceRatehighT}}}
        \end{center}
    \end{minipage}
\end{figure}

One should calculate the imaginary part of the retarded self-mass
as it characterises our decay process, which follows from equation
(\ref{FourierStatisticalA}). In order to calculate the decay rate,
we have to project the retarded self-mass on the quasi particle
shell $k^{0} = \omega_{\phi}$. Of course, one should really take
the perturbative correction to the dispersion relation of order
$\mathcal{O}(h^2/\omega_{\phi}^{2})$ into account but this effect
is rather small. Alternatively, we can project the advanced
self-mass in Fourier space on $k^{0} = - \omega_{\phi}$. We thus
expect:
\begin{equation}\label{decoherencerate}
\Gamma_{\mathrm{dec}} \simeq \Gamma_{\phi \rightarrow \chi \chi}
\,.
\end{equation}
Let us examine figures \ref{fig:DecoherenceRatelowT} and
\ref{fig:DecoherenceRatehighT}. From our numerical calculation, we
can thus easily find $\delta\Delta_{k}(t)=\Delta_{\mathrm{ms}} -
\Delta_{k}(t)$ which we show in solid black. We can now compare
with the single particle decay rate in equation
(\ref{singleparticledecayrate}) and plot $- \Gamma_{\phi
\rightarrow \chi \chi} t$. We conclude that the decoherence rate
can be well described by the single particle decay rate in our
model, thus confirming equation (\ref{decoherencerate}) above.

\subsection{The Emerging $k^0=0$ Shell}
\label{The Emerging $k^0=0$ Shell}

\begin{figure}[t]
    \begin{minipage}[t]{.45\textwidth}
        \begin{center}
\includegraphics[width=\textwidth]{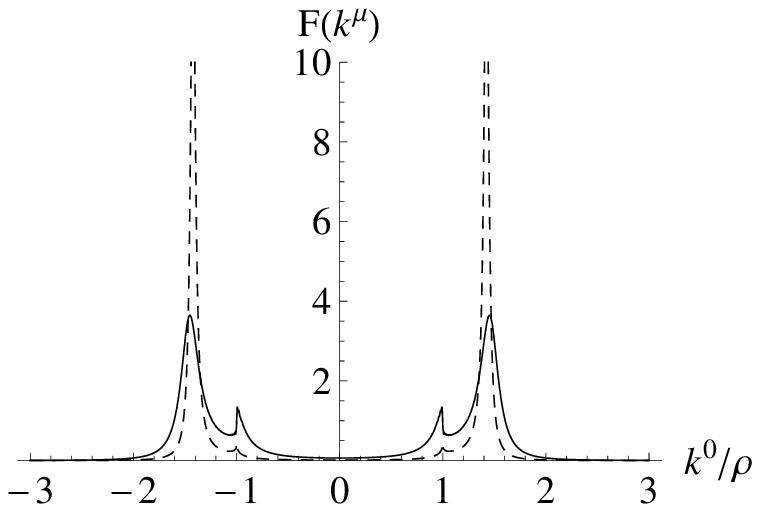}
   {\em \caption{The statistical propagator at low temperatures.
   Apart from the original quasi particle peaks, two more peaks emerge
   at $|k^0| \leq k$. We use  $\beta\rho = 2$, $k/\rho = 1$, $m_{\phi}/\rho=1$ and
   $h/\rho=4$ (solid black), $h/\rho=2$ (dashed). In the latter
   case, we do not show the entire original quasi particle peak for
   illustrative reasons.
   \label{fig:FlowT}}}
        \end{center}
   \end{minipage}
\hfill
    \begin{minipage}[t]{.45\textwidth}
        \begin{center}
\includegraphics[width=\textwidth]{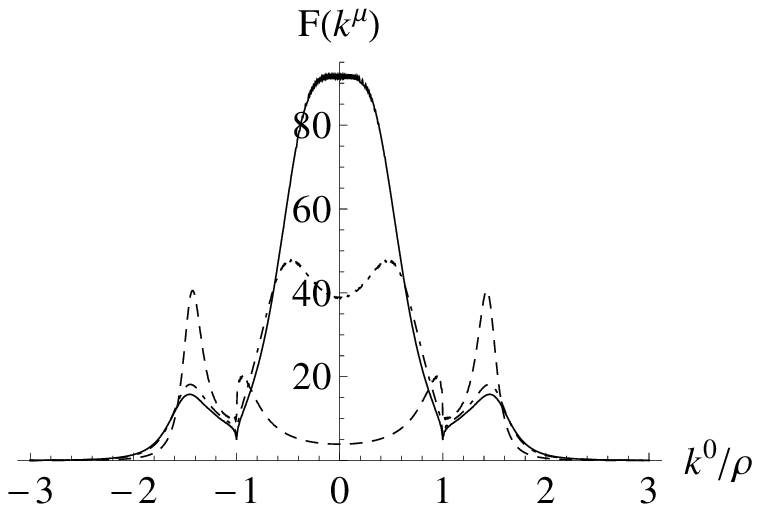}
   {\em \caption{The statistical propagator at high temperatures.
   For larger and larger coupling, we observe that the newly
   emerging quasi particle peaks move closer to $k^0=0$, where they
   eventually almost completely overlap. We use
   $\beta\rho = 0.1$, $k/\rho = 1$, $m_{\phi}/\rho=1$ and
   $h/\rho=1.61$ (solid black), $h/\rho=1.5$ (dot-dashed)
   and $h/\rho=1$ (dashed).
   \label{fig:FhighT}}}
        \end{center}
    \end{minipage}
\end{figure}

To develop some intuition, we depict $F_{\phi}(k^{\mu})$ as a
function of $k^{0}$ keeping various other parameters fixed. In the
vacuum $\beta\rho=\infty$, it is clear from the analytic form of
the statistical propagator that a $k^0=0$ shell does not exist. In
the vacuum, we have that $F_{\phi}(k^{\mu})=0$ for $|k^0| \leq k$.
At low temperatures, $\beta\rho =2$, we observe in figure
\ref{fig:FlowT} that two more quasi particle peaks emerge where
$|k^0| \leq k$. The original quasi particle peaks at $|k^0| \simeq
\omega_{\phi}$ however still dominate. At high temperatures,
$\beta\rho=0.1$, we observe in figure \ref{fig:FhighT} that the
two additional quasi particle peaks already present at lower
temperatures increase in size and move closer to $k^0=0$, where
they overlap. The original quasi particle peaks located at $|k^0|
\simeq \omega_{\phi}$ broaden as the interaction strength $h$
increases. Moreover, for increasing $h$, the original quasi
particle peaks get dwarfed by the new quasi particle peaks at
$|k^0| \leq k$ that by now almost completely overlap at $k^0=0$.

What we observe here is related to the coherence shell at $k^0=0$
first introduced by Herranen, Kainulainen and Rahkila
\cite{Herranen:2008di, Herranen:2010mh} to study quantum
mechanical reflection and quantum particle creation in a thermal
field theoretical setting (and for a discussion of fermions see
\cite{Herranen:2008hi, Herranen:2008hu}). They interpret this new
spectral solution of the statistical two point function as a
manifestation of non-local quantum coherence. As we have just
seen, the statistical propagator at late times will basically
evolve to equation (\ref{FourierStatisticalA}). We conclude that
the emerging $k^0 = 0$ shell translates to large entropy
generation at high temperatures. It is also clear that the naive
quasi particle picture of free thermal states breaks down in the
high temperature regime.

\subsection{Evolution of the Entropy: Changing Mass}
\label{Changing Mass}

Let us now study the evolution of the entropy where the mass of
the system field changes according to equation
(\ref{massbehaviourb}). For a constant mass $m_{\phi}$, the
statistical propagator depends only on the time difference of its
arguments $F_{\phi}(k,t,t')=F_{\phi}(k,t-t')$ due to time
translation invariance. This observation allowed us to find the
asymptotic value of the phase space area by means of another
Fourier transformation with respect to $t-t'$. When the mass of
the system field changes, however, we introduce a genuine time
dependence in the problem and we can only asymptotically compare
the entropy to the stationary values well before and after the
mass change. It is important to appreciate that the counterterms
introduced to renormalise the theory do not depend on $m_{\phi}$
so we do not have to consider renormalisation again
\cite{Koksma:2009wa}.

Depending on the size of the mass change, we can identify the
following two regimes:
\begin{subequations}
\label{adiabatic}
\begin{eqnarray}
|\beta_{k}|^{2} \ll 1 \qquad && \mathrm{adiabatic \,\, regime} \label{adiabatica}\\
|\beta_{k}|^{2} \gg 1 \qquad && \mathrm{non-adiabatic \,\, regime}
\label{adiabatib}\,,
\end{eqnarray}
\end{subequations}
where $\beta_{k}$ is one of the coefficients of the Bogoliubov
transformation that relates the initial (in) vacuum to the final
(out) vacuum state. As a consequence of the mass change, the state
gets squeezed \cite{Koksma:2010zi}. If $\beta_{k}=0$, the in and
out vacuum state are equal such that $|\beta_{k}|^{2}$ quantifies
the amount of particle creation and reads \cite{Birrell:1982ix}:
\begin{equation}\label{betaparticlecreation}
|\beta_{k}|^2 =
\frac{\sinh^2\left(\frac{\pi\omega_{-}}{\rho}\right)}{\sinh\left(\frac{\pi
\omega_{\mathrm{in}}}{\rho}\right) \sinh\left(\frac{\pi
\omega_{\mathrm{out}}}{\rho}\right)} \,\,\, \stackrel{
\substack{\omega_{\mathrm{in}} \ll \rho \\
\omega_{\mathrm{out}} \gg \rho}}{\longrightarrow} \,\,\,
\frac{\rho}{2\pi
\omega_{\mathrm{in}}}\left(1-\frac{\pi\omega_{\mathrm{in}}}{2\rho}\right)^2
\,.
\end{equation}
Here, $\omega_{\mathrm{in}}^2=m_{\phi, \mathrm{in}}^2+k^2$ and
$\omega_{\mathrm{out}}^2=m_{\phi, \mathrm{out}}^2+k^2$ are the
initial and final frequencies. Also, $m^2_{\phi, \mathrm{in}} =
A-B$ and $m^2_{\phi, \mathrm{out}} = A + B$ where we made use of
equation (\ref{massbehaviourb}). Finally, we defined $\omega_{\pm}
= \frac{1}{2}( \omega_{\mathrm{out}} \pm \omega_{\mathrm{in}})$.
The word ``particle'' in particle creation is not to be confused
with the statistical particle number defined by means of the phase
space area in equation (\ref{particlenumber}). Whereas the latter
counts the phase space occupied by a state in units of the minimal
uncertainty wave packet, the former corresponds to the
conventional notion of particles in curved spacetimes where one
plane wave field excitation $\hat{a}_{\vec{k}}^{\dag}|0\rangle = |
\vec{k}\rangle$ is referred to as one particle (for a discussion
on wave packets in quantum field theory, see \cite{Koksma:2010zy,
Westra:2010zx}). When we consider a changing mass in the absence
of any interaction terms, $|\beta_{k}|^{2}$ increases whereas the
phase space area remains constant. For the parameters we consider
in this paper $|\beta_{k}|^{2} \simeq \mathcal{O}(10^{-4})$ such
that we are in the adiabatic regime.

\begin{figure}
    \begin{minipage}[t]{.45\textwidth}
        \begin{center}
\includegraphics[width=\textwidth]{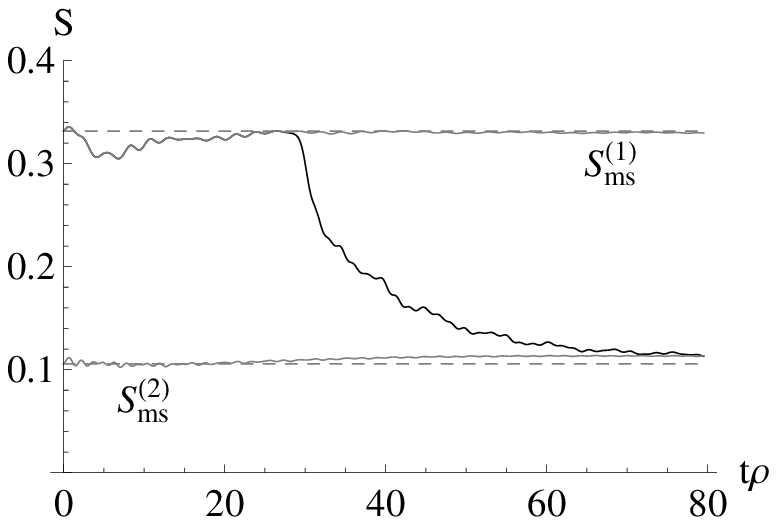}
   {\em \caption{Entropy as a function of time for a mass increase
   from $m_{\phi}/\rho=1$ to $m_{\phi}/\rho=2$, giving rise to
   the constant interacting thermal entropies $S_{\mathrm{ms}}^{(1)}$ and
   $S_{\mathrm{ms}}^{(2)}$, respectively. The mass changes
   rapidly at $t \rho=30$. We use $\beta \rho = 2$, $k/\rho=1$,
   $h/\rho=4$ and $N=1600$. The gray lines are the
   corresponding constant mass entropy functions where we use mixed state
   initial conditions.
   \label{fig:SMassIncreaseLowT}}}
        \end{center}
   \end{minipage}
\hfill
    \begin{minipage}[t]{.45\textwidth}
        \begin{center}
\includegraphics[width=\textwidth]{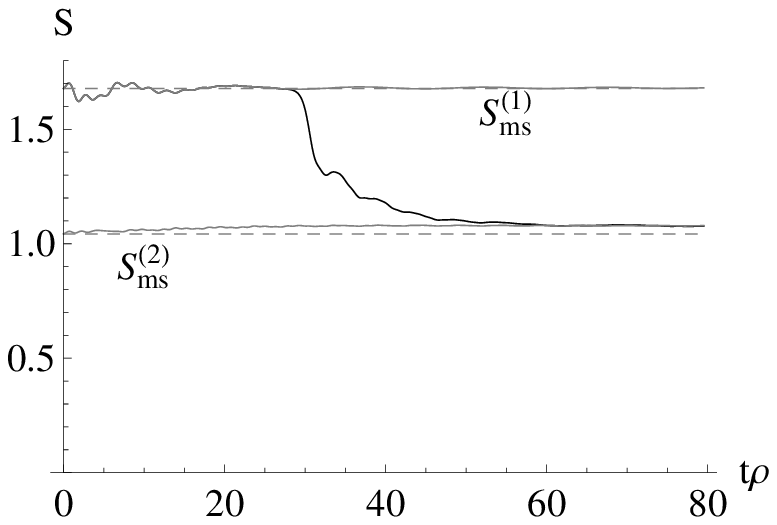}
   {\em \caption{Entropy as a function of time for a mass increase
   from $m_{\phi}/\rho=1$ to $m_{\phi}/\rho=2$, giving rise to
   the constant interacting thermal entropies $S_{\mathrm{ms}}^{(1)}$ and
   $S_{\mathrm{ms}}^{(2)}$, respectively. The mass changes
   rapidly at $t \rho=30$. We use $\beta \rho = 1/2$, $k/\rho=1$,
   $h/\rho=3$ and $N=1600$. The gray lines are the
   corresponding constant mass entropy functions where we used mixed state
   initial conditions.
   \label{fig:SMassIncreaseHighT}}}
        \end{center}
    \end{minipage}
\vskip 0.1cm
    \begin{minipage}[t]{.45\textwidth}
        \begin{center}
\includegraphics[width=\textwidth]{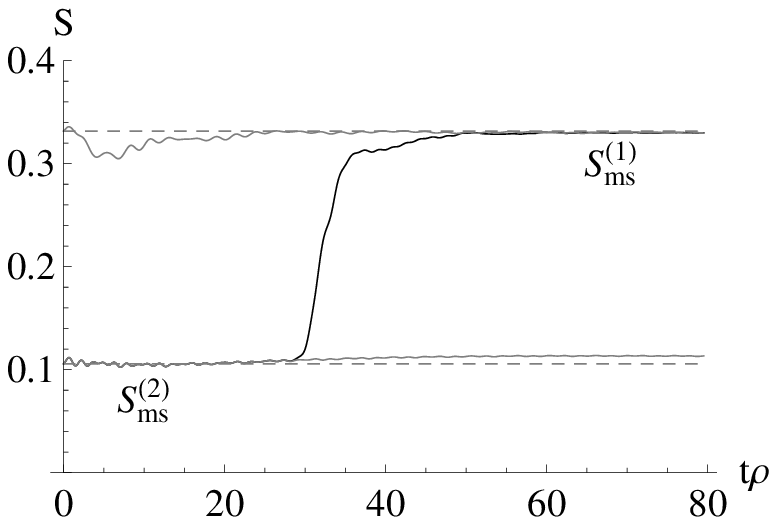}
   {\em \caption{Entropy as a function of time for a mass decrease
   where we used the same parameters as in figure
   \ref{fig:SMassIncreaseLowT}.
   \label{fig:SMassDecreaseLowT}}}
        \end{center}
    \end{minipage}
\hfill
    \begin{minipage}[t]{.45\textwidth}
        \begin{center}
\includegraphics[width=\textwidth]{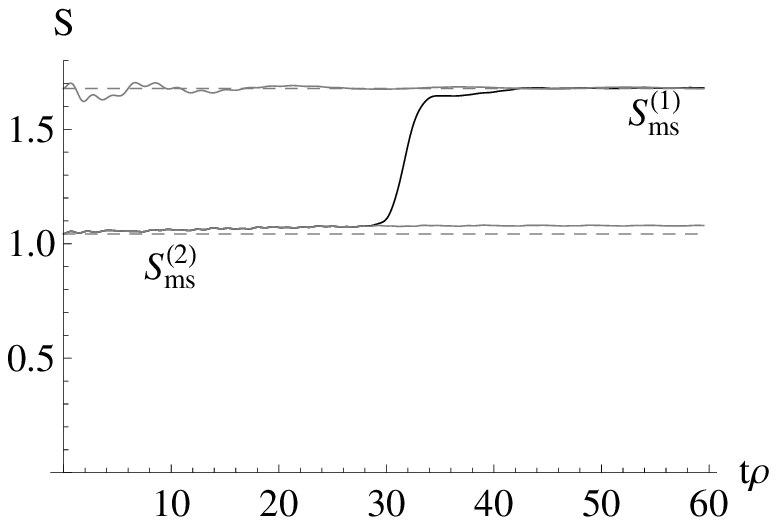}
   {\em \caption{Entropy as a function of time for a mass decrease
   where we used the same parameters as in figure
   \ref{fig:SMassIncreaseHighT}.
   \label{fig:SMassDecreaseHighT}}}
        \end{center}
    \end{minipage}
\end{figure}

\begin{figure}[t]
    \begin{minipage}[t]{.45\textwidth}
        \begin{center}
\includegraphics[width=\textwidth]{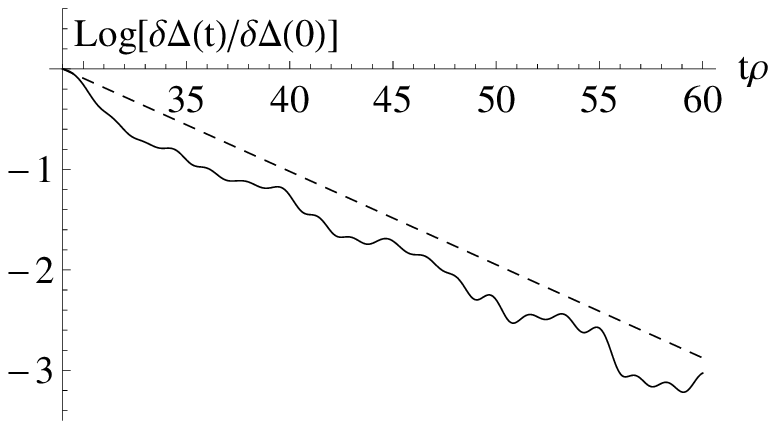}
   {\em \caption{Decoherence rate at low temperatures. We show the
   exponential approach to $\Delta_{\mathrm{ms}}^{(2)}$ in solid black
   and the corresponding decoherence rate given in equation
   (\ref{decoherencerate}) (dashed line). We use the phase space
   area from figure~\ref{fig:SMassIncreaseLowT}.
   \label{fig:DecoherenceRateChangingMasslowT}}}
        \end{center}
   \end{minipage}
\hfill
    \begin{minipage}[t]{.45\textwidth}
        \begin{center}
\includegraphics[width=\textwidth]{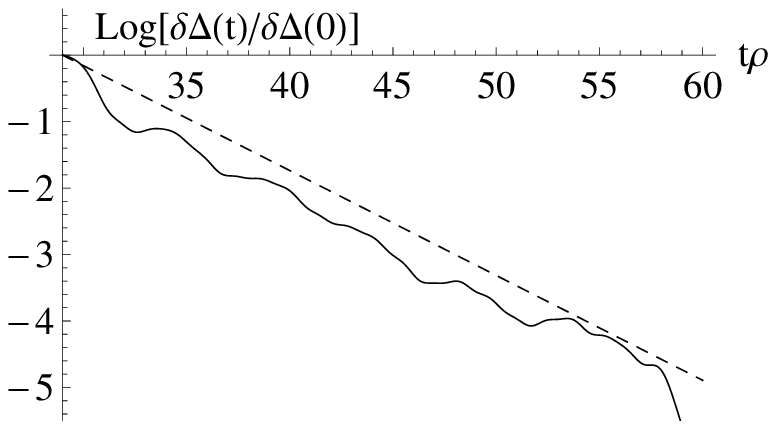}
   {\em \caption{Decoherence rate at high temperatures. We show the
   exponential approach to $\Delta_{\mathrm{ms}}^{(2)}$ in solid black
   and the corresponding decoherence rate given in equation
   (\ref{decoherencerate}) (dashed line). We use the phase space
   area from figure~\ref{fig:SMassIncreaseHighT}.
   \label{fig:DecoherenceRateChangingMasshighT}}}
        \end{center}
    \end{minipage}
\end{figure}

Let us consider the coherence effects due to a mass increase and
decrease in figures \ref{fig:SMassIncreaseLowT},
\ref{fig:SMassIncreaseHighT}, \ref{fig:SMassDecreaseLowT} and
\ref{fig:SMassDecreaseHighT}. Here, we take $m_{\phi}/\rho=1$ and
$m_{\phi}/\rho=2$ giving rise to the constant interacting thermal
entropies $S_{\mathrm{ms}}^{(1)}$ and $S_{\mathrm{ms}}^{(2)}$,
respectively. The numerical value of these asymptotic entropies is
calculated just as in the constant mass case such that we find
$S_{\mathrm{ms}}^{(1)} > S_{\mathrm{ms}}^{(2)}$. We use mixed
state initial conditions as outlined in equation
(\ref{MSboundaryconditionsstatd}) and moreover we insert the
initial mass in the memory kernels.
In figure \ref{fig:SMassIncreaseLowT} we show the effects on the
entropy for a mass increase at fairly low temperatures
$\beta\rho=2$. In gray we depict the two corresponding constant
mass entropy functions to compare the asymptotic behaviour. In
order to calculate the latter, we also use mixed state boundary
conditions. Clearly, well before and after the mass increase, the
entropy is equal to the constant interacting thermal entropy,
$S_{\mathrm{ms}}^{(1)}$ and $S_{\mathrm{ms}}^{(2)}$, respectively.
The small difference between the numerical value of the
interacting thermal entropy $S_{\mathrm{ms}}^{(2)}$ (in dashed
gray) and the corresponding $m_{\phi}/\rho=2$ constant mass
evolution is just due to numerical accuracy. It is interesting to
observe that the new interacting thermal entropy is reached on a
different time scale than $\rho^{-1}$, the one at which the
system's mass has changed. Again, we verify that the rate at which
the phase space area changes, defined analogously to equation
(\ref{timescale1}), can be well described by the single particle
decay rate (\ref{singleparticledecayrate}). Given the fact that
the mass changes so rapidly in our case, one should use the final
mass $m_{\phi, \mathrm{out}}$ in equation
(\ref{singleparticledecayrate}). In figure
\ref{fig:DecoherenceRateChangingMasslowT} we show both the
exponential approach towards the constant interacting phase space
area $\Delta_{\mathrm{ms}}^{(2)}$ and the decay rate
(\ref{singleparticledecayrate}). In order to produce figure
\ref{fig:DecoherenceRateChangingMasslowT}, we subtract the
constant mass evolution of the phase space area using mixed state
initial conditions rather than $\Delta_{\mathrm{ms}}^{(2)}$ to
find $\delta\Delta_{k}(t)$ in equation (\ref{timescale1}).

This qualitative picture does not change when we consider the same
mass increase only now at higher temperatures $\beta\rho=1/2$ in
figure \ref{fig:SMassIncreaseHighT}. The interacting thermal
entropies in this case are larger due to the fact that the
temperature is higher. Again we observe a small difference between
$S_{\mathrm{ms}}^{(2)}$ and the $m_{\phi}/\rho=2$ constant mass
evolution due to numerical accuracy. Also, the decoherence rate
can be well described by the single particle decay rate which we
depict in figures (\ref{fig:DecoherenceRateChangingMasslowT}) and
(\ref{fig:DecoherenceRateChangingMasshighT}).

When we consider the ``time reversed process'', i.e.: a mass
decrease from $m_{\phi}/\rho=2$ to $m_{\phi}/\rho=1$, we observe
an entropy increase. We show the resulting evolution of the
entropy in figures \ref{fig:SMassDecreaseLowT} and
\ref{fig:SMassDecreaseHighT} for $\beta\rho=2$ and
$\beta\rho=1/2$, respectively. The evolution of the entropy
reveals no further surprises and corresponds to the time reversed
picture of figures \ref{fig:SMassIncreaseLowT} and
\ref{fig:SMassIncreaseHighT}. The decoherence rate for a mass
decrease can again be well described by the single particle decay
rate in equation (\ref{singleparticledecayrate}).

We observe that the rate at which the mass changes is much larger
than the decoherence rate. As long as this condition is satisfied,
coherence effects continue to be important. Eventually though, the
Gaussian von Neumann entropy settles to its new constant value and
no particle creation remains as our state thermalises again. In
the context of baryogenesis, we thus expect that quantum coherence
effects remain important as long as this condition persists too.
Of course, one would have to generalise our model to a CP
violating model in which the effects that are of relevance for
coherent baryogenesis scenarios are captured.

\subsection{Squeezed States}
\label{Squeezed States}

The effect of a large non-adiabatic mass change on the quantum
state is a rapid squeezing of the state which can neatly be
visualised in Wigner space. Although it is numerically challenging
to implement a case where the mass changes non-adiabatically fast,
we can probe its most important effect on the state by considering
a state that is significantly squeezed initially. A pure and
squeezed state is characterised by the following initial
conditions:
\begin{subequations}
\label{SquuezedState1}
\begin{eqnarray}
F_{\phi}(k,t_0,t_0) &=& \frac{1}{2\omega_{\phi}}\left[
\cosh(2r)-\sinh(2r)\cos(2\varphi)\right]
\label{SquuezedState1a}\\
\left. \partial_t \partial_{t'} F_{\phi}(k,t,t')
\right|_{t=t'=t_0} &=& \frac{\omega_{\phi}}{2} \left[
\cosh(2r)+\sinh(2r)\cos(2\varphi)\right]
\label{SquuezedState1b}\\
\left. \partial_{t} F_{\phi}(k,t,t_0)\right|_{t=t_0} &=&
\frac{1}{2} \sinh(2r)\sin(2\varphi) \,.\label{SquuezedState1c}
\end{eqnarray}
\end{subequations}
Here, $\varphi$ characterises the angle along which the state is
squeezed and $r$ indicates the amount of squeezing. As a squeezed
state is pure, we have $\Delta_k(t_0)=1$ initially. A mixed
initial squeezed state condition can be achieved by multiplying
equation (\ref{SquuezedState1}) by a factor.

\begin{figure}[t!]
    \begin{minipage}[t]{.45\textwidth}
        \begin{center}
\includegraphics[width=\textwidth]{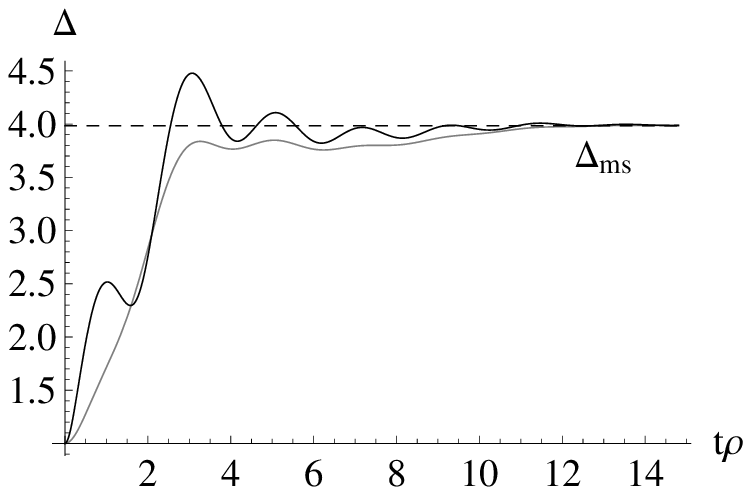}
   {\em \caption{Phase space area as a function of time for a
   squeezed initial state. We use $\varphi=0$, $e^{2r}=1/5$, $\beta \rho = 0.5$, $k/\rho=1$, $m_{\phi}/\rho=1$,
   $h/\rho=3$ and $N=300$ up to $t\rho=15$. The gray line
   indicates the pure state evolution previously considered
   in figure \ref{fig:PSA_highT_1}.
   \label{fig:DeltaSqueezedState1}}}
        \end{center}
   \end{minipage}
\hfill
    \begin{minipage}[t]{.45\textwidth}
        \begin{center}
\includegraphics[width=\textwidth]{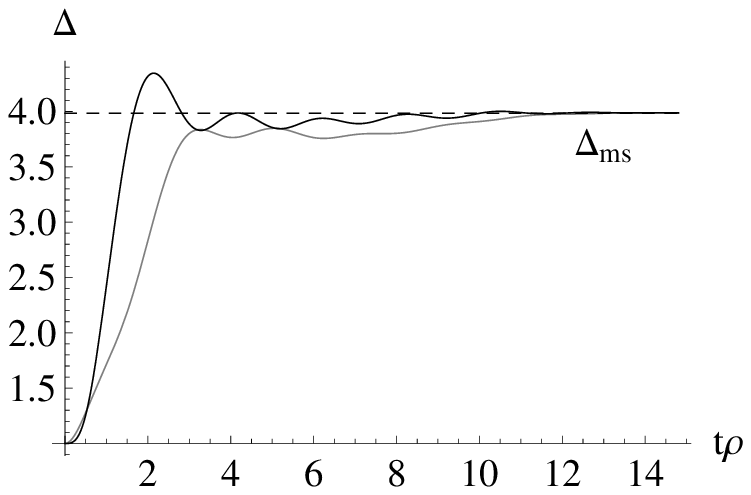}
   {\em \caption{Phase space area as a function of time for a
   squeezed initial state. On top of the usual exponential
   approach towards $\Delta_{\mathrm{ms}}$, we observe oscillatory
   behaviour. The amplitude of the oscillations decays with the
   single particle decay rate as well. We use $\varphi=0$,
   $e^{2r}=5$ and the other parameters are given in figure
   \ref{fig:DeltaSqueezedState1}.
   \label{fig:DeltaSqueezedState2}}}
        \end{center}
    \end{minipage}
\end{figure}

We show the corresponding evolution for the phase space area in
two cases in figures \ref{fig:DeltaSqueezedState1} and
\ref{fig:DeltaSqueezedState2}. As the squeezed state thermalises,
we observe two effects. Firstly, there is the usual exponential
approach towards the thermal interacting value
$\Delta_{\mathrm{ms}}$ we observed before. As we showed
previously, this process is characterised by the single particle
decay rate in equation (\ref{singleparticledecayrate}). Secondly,
superimposed to that behaviour, we observe damped oscillatory
behaviour of the phase space area as a function of time that is
induced by the initial squeezing.

The latter process in principle introduces a second decay rate in
the evolution: one can associate a characteristic time scale at
which the amplitude of the oscillations decay (superimposed on the
exponential approach towards $\Delta_{\mathrm{ms}}$). One can read
off from figures \ref{fig:DeltaSqueezedState1} and
\ref{fig:DeltaSqueezedState2} that the exponential decay of the
envelope of the oscillations can also be well described by the
single particle decay rate in equation
(\ref{singleparticledecayrate}). We thus observe only one relevant
time scale of the process of decoherence in our scalar field
model: the single particle decay rate. We thus conclude that in
the case of a non-adiabatic mass change, the decay of the
amplitude of the resulting oscillations will be in agreement with
the single particle decay rate too.

\section{Conclusion} \label{Conclusion}

We study the decoherence of a quantum field theoretical system in
a renormalised and perturbative 2PI scheme. As most of the
non-Gaussian information about a system is experimentally hard to
access, we argue in our ``correlator approach'' to decoherence
that neglecting this information and, consequently, keeping only
the information stored in Gaussian correlators, leads to an
increase of the Gaussian von Neumann entropy of the system. We
argue that the Gaussian von Neumann entropy should be used as
\emph{the} quantitative measure for decoherence.

The most important result in this paper is shown in figure
\ref{fig:S_highT_1}, where we depict the time evolution of the
Gaussian von Neumann entropy for a pure state at a high
temperature. Although a pure state with vanishing entropy $S_k =
0$ remains pure under unitary evolution, the observer perceives
this state over time as a mixed state with positive entropy
$S_{\mathrm{ms}}>0$. The reason is that non-Gaussianities are
generated by the unitary evolution (both in the correlation
between the system and environment as well as in higher order
correlations in the system itself) and subsequently neglected in
our Gaussian von Neumann entropy.

We have extracted two relevant quantitative measures of
decoherence: the maximal amount of decoherence $S_{\mathrm{ms}}$
and the decoherence rate $\Gamma_{\mathrm{dec}}$. The total amount
of decoherence corresponds to the interacting thermal entropy
$S_{\mathrm{ms}}$ and is slightly larger than the free thermal
entropy, depending on the strength of the interaction $h$. The
decoherence rate can be well described by the single particle
decay rate of our interaction $\Gamma_{\phi \rightarrow \chi
\chi}$.

This study builds the quantum field theoretical framework for
other decoherence studies in various relevant situations where
different types of fields and interactions can be involved. In
cosmology for example, the decoherence of scalar gravitational
perturbations can be induced by e.g. fluctuating tensor modes
(gravitons) \cite{Prokopec:1992ia}, isocurvature modes
\cite{Prokopec:2006fc} or even gauge fields. In quantum
information physics it is very likely that future quantum
computers will involve coherent light beams that interact with
other parts of the quantum computer as well as with an environment
\cite{QuantumComputing, Knill}. For a complete understanding of
decoherence in such complex systems it is clear that a quantum
field theoretical framework such as developed here is necessary.

We also studied the effects on the Gaussian von Neumann entropy of
a changing mass. The Gaussian von Neumann entropy changes to the
new interacting thermal entropy after the mass change on a time
scale that is again well described by the single particle decay
rate in our model. It is the same decay rate that describes the
decay of the amplitude of the oscillations for a squeezed initial
state. One can view our model as a toy model relevant for
electroweak baryogenesis scenarios. It is thus interesting to
observe that the coherence time scale (the time scale at which the
entropy changes) is much larger than the time scale $\rho^{-1}$ at
which the mass of the system field changes. We conclude that the
coherent effect of a non-adiabatic mass change (squeezing) does
not get immediately destroyed by the process of decoherence and
thermalisation.

Finally, we compared our correlator approach to decoherence to the
conventional approach relying on the perturbative master equation.
It is unsatisfactory that the reduced density matrix evolves
non-unitarily while the underlying quantum theory is unitary. We
are not against non-unitary equations or approximations in
principle, however, one should make sure that the essential
physical features of the system one is describing are kept. The
perturbative master equations does not break unitarity correctly,
as we have shown in this paper. On the practical side, the master
equation is so complex that field theoretical questions have
barely been addressed: there does not exist a treatment to take
perturbative interactions properly into account, nor has any
reduced density matrix ever been renormalised. This is the reason
for our quantum mechanical comparison, rather than a proper field
theoretical study of the reduced density matrix. In section
\ref{Deriving the Master Equation from the Kadanoff-Baym
Equations} however, we outline the perturbative approximations
used to derive the master equation from the Kadanoff-Baym
equations, i.e.: in the memory kernels of the Kadanoff-Baym
equations we insert free propagators with appropriate initial
conditions. A proper generalisation to derive the renormalised
perturbative master equation in quantum field theory from the
Kadanoff-Baym equations should be straightforward. In the simple
quantum mechanical situation, we show that the entropy following
from the perturbative master equation generically suffers from
physically unacceptable secular growth at late times in the
resonant regime. This leads to an incorrect prediction of the
total amount of decoherence that has occurred. We show that the
time evolution of the Gaussian von Neumann entropy behaves well in
both the resonant and in the non-resonant regime.

\

\noindent \textbf{Acknowledgements}

\noindent JFK thanks Jeroen Diederix for many useful suggestions.
JFK and TP acknowledge financial support from FOM grant 07PR2522
and by Utrecht University. JFK also gratefully acknowledges the
hospitality of the MIT Kavli Institute for Astrophysics and Space
Research (MKI) during his stay in Cambridge, MA.

\appendix

\section{Derivation of $M^{F}_{\phi,\mathrm{th-th}}(k,\Delta t)$}
\label{AppendixA}

Only the high and low temperature limits of
$M^{F}_{\phi,\mathrm{th-th}}(k,\Delta t)$ can be evaluated in
closed form. We derive these expressions in this appendix.

\subsection{Low Temperature Contribution}

Let us recall equation (\ref{selfMass:statistical3b}) where we can
perform the $\omega$-integral by making use of equation
(\ref{angular integral:D-2}) and $1/({\rm e}^{\beta
\omega}-1)=\sum_{n=1}^\infty {\rm e}^{-\beta n \omega}$:
\begin{eqnarray}
M^{F}_{\phi,\mathrm{th-th}}(k,\Delta t) &=& -\frac{h^2}{8\pi^2 k}
\int_0^\infty \mathrm{d} k_1 \frac{\cos(k_{1}\Delta t)}{{\rm
e}^{\beta k_1}-1}\sum_{n=1}^\infty \frac{{\rm
e}^{-n\beta\omega}}{(\Delta t)^2 + (n\beta)^2}
\left[-n\beta\cos(\omega\Delta t)+\Delta t\sin(\omega\Delta t)
\right]
      \Bigl|_{\omega=\omega_-}^{\omega=\omega_+}\Bigr.  \,,
\label{MFphi:thth:1}
\end{eqnarray}
where $\omega_\pm^2= (k\pm k_1)^2+m_\chi^2$. We now prepare this
expression for $k_1$ integration by making use of $1/({\rm
e}^{\beta k_1}-1)=\sum_{m=1}^\infty {\rm e}^{-\beta m k_1}$ and
some familiar trigonometric identities:
\begin{eqnarray}
&& M^{F}_{\phi,\mathrm{th-th}}(k,\Delta t) = -\frac{
h^2}{16\pi^2k}\sum_{m,n=1}^\infty  \frac{1}{(\Delta t)^2 +
(n\beta)^2} \int _0^\infty \mathrm{d}k_1 \, {\rm e}^{-\beta m k_1}
\label{MFphi:thth:2}
\\
&& \,\,\, \times \Bigg\{ - \beta n\,{\rm e}^{-\beta n(k+k_1)}
   \Big[\cos[(2k_1\!+\!k)\Delta t] + \cos(k\Delta t)\Big]
 +\Delta t \, {\rm e}^{-\beta n(k+k_1)}
   \Big[\sin[(2k_1\!+\!k)\Delta t] + \sin(k\Delta t)\Big]
\nonumber\\
&& \,\quad + \theta(k\!-\!k_1)\beta n \, {\rm e}^{-\beta n(k-k_1)}
   \Big[\cos[(2k_1\!-\!k)\Delta t] + \cos(k\Delta t)\Big]
 \!+\!\theta(k\!-\!k_1)\Delta t \,{\rm e}^{-\beta n(k-k_1)}
   \Big[\sin[(2k_1\!-\!k)\Delta t] - \sin(k\Delta t)\Big]
\nonumber\\
&&  \,\quad + \theta(k_1\!-\!k)\beta n \,{\rm e}^{-\beta n(k_1-k)}
   \Big[\cos[(2k_1\!-\!k)\Delta t] + \cos(k\Delta t)\Big]
 \!-\!\theta(k_1\!-\!k)\Delta t \, {\rm e}^{-\beta n(k_1-k)}
   \Big[\sin[(2k_1\!-\!k)\Delta t] - \sin(k\Delta t)\Big]
\! \Bigg\}\!. \nonumber
\end{eqnarray}
Upon integrating over $k_1$ and rearranging the terms we obtain:
\begin{eqnarray}
&& M^{F}_{\phi,\mathrm{th-th}}(k,\Delta t) = -\frac{h^2}{16\pi^2
k} \sum_{m,n=1}^\infty \frac{1}{(\Delta t)^2 + (n\beta)^2} \Bigg\{
\sin(k\Delta t)\Big({\rm e}^{-\beta nk}-{\rm e}^{-\beta mk}\Big)
\Bigg[\frac{\beta\Delta t(m+3n)}{[\beta(m+n)]^2+(2\Delta t)^2} -
\frac{\Delta t}{\beta(m-n)} \Bigg]\nonumber
\\
&& \qquad\qquad + \sin(k\Delta t)\Big({\rm e}^{-\beta nk}+{\rm
e}^{-\beta mk}\Big)\Bigg[ \frac{-\beta \Delta
t(m-3n)}{[\beta(m-n)]^2+(2\Delta t)^2} +   \frac{\Delta
t}{\beta(m+n)} \Bigg]  \label{MFphi:thth:3} \\
&& \qquad\qquad +\cos(k\Delta t)\Big({\rm e}^{-\beta nk}-{\rm
e}^{-\beta mk}\Big)\Bigg[\frac{-\beta^2 n (n+m)+2(\Delta
t)^2}{[\beta(m+n)]^2+(2\Delta t)^2} +  \frac{\beta^2 n
(m-n)+2(\Delta t)^2}{[\beta(m-n)]^2+(2\Delta t)^2} - \frac{n}{m+n}
+  \frac{n}{m-n} \Bigg] \Bigg\}\nonumber\,.
\end{eqnarray}
This expression contains two singular terms when $m=n$. By
performing the integral (\ref{MFphi:thth:2}) in that case, they
are to be interpreted as:
\begin{equation}
\frac{{\rm e}^{-\beta nk}-{\rm e}^{-\beta mk}}{m-n}
\;\stackrel{m=n}{\longrightarrow}\; \beta k \, {\rm e}^{-\beta nk}
\,. \label{MFphi:thth:3b}
\end{equation}
This expression allows us to obtain the low temperature $\beta k
\gg 1$ limit of $M^{F}_{\phi,\mathrm{th-th}}(k,\Delta t)$. It then
suffices to consider three contributions in
equation~(\ref{MFphi:thth:3}) only. Firstly, there is the
contribution for $m=1=n$, for $n=1$ and $m\geq 2$, and finally for
$m=1$ and $n \geq 2$. The sum in the last two cases can be
evaluated in closed form, such that one obtains equation
(\ref{MFphi:thth:4}).

\subsection{High Temperature Contribution}

Let us now consider the high temperature limit. It is clear from
equation (\ref{MFphi:thth:3}) that when $\beta k \ll 1$ there is
unfortunately no small quantity to expand about as both $m$ and
$n$ can become arbitrarily large. Therefore, we go back to the
original expression~(\ref{selfMass:statistical3b}), proceed as
usual by making use of (\ref{angular integral:D-2}) and rewrite it
in terms of new $(u,v)$-coordinates (``lightcone coordinates''),
defined by:
\begin{subequations}
\label{new coords}
\begin{eqnarray}
u &=& k_1 - \omega \label{new coordsa} \\
v &=& k_1 + \omega \label{new coordsb}\,,
\end{eqnarray}
\end{subequations}
such that of course $k_1 = (v+u)/2$ and $ \omega = (v-u)/2$, in
terms of which the region of integration becomes:
\begin{subequations}
\label{new coords:region}
\begin{eqnarray}
-k &\leq & u \leq k\label{new coords:regiona} \\
k &\leq & v < \infty\label{new coords:regionb}\,.
\end{eqnarray}
\end{subequations}
Equation (\ref{selfMass:statistical3b}) thus transforms into:
\begin{equation}
M^{F}_{\phi,\mathrm{th-th}}(k,\Delta t) = \frac{h^2}{32\pi^2 k}
\int_{-k}^k \! \mathrm{d}u \frac{1}{2\sinh(\beta u/2)}
\int_k^{\infty} \! \mathrm{d}v \left[\cos(u\Delta t)+\cos(v\Delta
t) \right] \left\{ \! \frac{{\rm e}^{\frac{\beta u}{2}}} {{\rm
e}^{\beta(v+u)/2}-1} - \frac{{\rm e}^{-\frac{\beta u}{2}}} {{\rm
e}^{\beta(v-u)/2}-1} \right\}, \label{MFphi:thth:hiT:2}
\end{equation}
where we took account of the Jacobian $J=|\partial(k_1,
\omega)/\partial(u, v)|=1/2$. One can now perform the $v$-integral
involving the $\cos(u\Delta t)$-term. Secondly, since we are
interested in the limit $\beta k \ll 1$ and we moreover have $|u|
\leq k$, note that we also have $|\beta u| \ll 1$. The
$\cos(v\Delta t)$-term can thus be expanded around $| \beta u |
\ll 1$. An intermediate result reads:
\begin{eqnarray}
M^{F}_{\phi,\mathrm{th-th}}(k,\Delta t) &=& \frac{h^2}{32\pi^2k
\beta} \Bigg\{ \! - 2\int_{-k}^k \mathrm{d}u \cos(u\Delta t)
\Bigg[\frac{\log\left(1-\exp[-\beta(k+u)/2]\right)} {1-\exp(-\beta
u)} -\frac{\log\left(1-\exp[-\beta(k-u)/2]\right)}{\exp(\beta
u)-1} \Bigg] \nonumber\\
&& \qquad\qquad + \int_{-k}^k \! \mathrm{d}u \int_k^{\infty} \!
\mathrm{d}v \frac{\cos(v\Delta t)}{u}  \left\{ \! \frac{ 1+ \beta
u/2 } {{\rm e}^{\beta v/2}(1+\beta u/2)-1} - \frac{1-\beta u/2}
{{\rm e}^{\beta v/2}(1-\beta u/2)-1} \right\} \Bigg\}
\,.\label{MFphi:thth:hiT:3}
\end{eqnarray}
The reader can easily see that we deliberately not Taylor expand
the second line fully around $|\beta u| \ll 1$. The reason is that
the subsequent integration renders such a naive Taylor expansion
invalid. Let us first integrate the first line of equation
(\ref{MFphi:thth:hiT:3}). We now expand the $\cos(u\Delta
t)$-integral around $|\beta(k\pm u)|\ll 1$:
\begin{eqnarray}
&& \int_{-k}^k \mathrm{d}u \cos(u\Delta t)
\Bigg[\frac{\log\left(1-\exp[-\beta(k+u)/2]\right)} {1-\exp(-\beta
u)} -\frac{\log\left(1-\exp[-\beta(k-u)/2]\right)}{\exp(\beta
u)-1} \Bigg]\label{MFphi:thth:hiT:4}\\
&& \qquad\qquad =  \int_{-k}^k \mathrm{d}u \cos(u\Delta t)
\Bigg[\frac{1}{\beta u}\log\left(\frac{k+u}{k-u}\right) +
\frac{1}{2}\log\left(\frac{\beta^2}{4}(k^2-u^2)\right)-
\frac{1}{2}+ {\cal O}(\beta k,\beta u) \Bigg] \,. \nonumber
\end{eqnarray}
Let us first evaluate the simple $u$-integrals in equation
(\ref{MFphi:thth:hiT:4}):
\begin{equation}
\int_{-k}^k \! \mathrm{d}u \frac{\cos(u\Delta t)}{2} \!
\left[\log\! \left(\!\frac{\beta^2}{4}(k^2 \! - \! u^2)\!
\right)\!-\! 1\right] \! = \! \frac{\sin(k\Delta t)}{\Delta t
}\left(\! \mathrm{ci}(2|k\Delta t |) -\log\!\left(\frac{2|\Delta t
|}{k \beta^2}\right)\!-\!\gamma_{\mathrm{E}}\!-\! 1 \! \right)
\!-\! \frac{\cos(k\Delta t)}{| \Delta t | }\left(
\mathrm{si}(2|k\Delta t |)\!+\frac{\pi}{2}\right)
 \label{MFphi:thth:hiT:5}\!,
\end{equation}
where ${\rm ci}(z)$ and ${\rm si}(z)$ are the cosine and sine
integral functions, respectively, defined in equation
(\ref{ciandsi}). The more complicated integral
in~(\ref{MFphi:thth:hiT:4}) is:
\begin{equation}
\int_{-k}^{k} \! \mathrm{d}u \frac{\cos(u \Delta t)}{\beta u} \log
\! \left(\frac{k\! + \! u}{k \! -\! u} \! \right) =
\frac{1}{\beta} \sum_{n=0}^\infty \frac{2}{1+2n}\int_{-1}^1 \!
\mathrm{d}z \, z^{2n}\cos(k\Delta t z) = \sum_{n=0}^\infty
\frac{\beta^{-1}}{(\frac12+n)^2}\,
{}_1F_2\left(\frac12+n;\frac12,\frac32 \! +\! n;-\frac{(k\Delta t)
^2}{4}\right) .
 \label{MFphi:thth:hiT:6}
\end{equation}
By making use of its definition, we expand the hypergeometric
function ${}_1F_2$ as follows:
\begin{equation}
{}_1F_2\Big(\frac12+n;\frac12,\frac32+n;-\frac{(k\Delta t)
^2}{4}\Big) = \sqrt{\pi}\Big(\frac12+n\Big)
  \sum_{m=0}^\infty \frac{1}{m! (\frac12+n+m) \Gamma(\frac12+m)}\left(-\frac{(k\Delta t)^2}{4}\right)^m
\,.
 \label{MFphi:thth:hiT:7}
\end{equation}
Inserting this into~(\ref{MFphi:thth:hiT:6}) and performing the
$n$-sum we obtain:
\begin{equation}
\int_{-k}^{k} \mathrm{d}u \frac{\cos(u \Delta t)}{\beta u}
\log\left(\frac{k+u}{k-u}\right)  =  \frac{1}{\beta}\left[
\frac{\pi^2}{2} +\sqrt{\pi} \sum_{m=1}^\infty
\frac{\psi(\frac12+m)+2\ln(2)+\gamma_{\mathrm{E}}} {m\,m!\,
\Gamma(\frac12+m)} \left(-\frac{(k\Delta t)^2}{4}\right)^m
\right]\,,
 \label{MFphi:thth:hiT:8}
\end{equation}
where we performed the $n$-sum for $m=0$ separately. Note that:
\begin{equation}
\frac{\psi(m+1/2)}{\Gamma(m+1/2)} =
-\frac{\mathrm{d}}{\mathrm{d}\gamma}
\frac{1}{\Gamma(\gamma+m)}\Big|_{\gamma=1/2}
\label{MFphi:thth:hiT:9} \,.
\end{equation}
Finally, we can perform the $m$-sum appearing in equation
(\ref{MFphi:thth:hiT:8}) to yield:
\begin{equation}
\int_{-k}^{k} \mathrm{d}u \frac{\cos(u \Delta t)}{\beta u}
\log\left(\frac{k+u}{k-u}\right)  = \frac{\pi^2}{2\beta}
-\frac{4}{\beta} \left(\gamma_{\mathrm{E}} - \mathrm{ci}( |k
\Delta t|)+ \log(|k \Delta t|)\right) + \frac{(k\Delta
t)^2}{2\beta} \frac{\mathrm{d}}{\mathrm{d}\gamma} \,
{}_2F_3\left(1,1;2,2,1+\gamma;-\frac{(k\Delta
t)^2}{4}\right)\Bigg|_{\gamma=\frac12} . \label{MFphi:thth:hiT:10}
\end{equation}
It is useful to know the expansions of the hypergeometric function
in~(\ref{MFphi:thth:hiT:10}). The large time ($k\Delta t \gg 1$)
expansion of this function is:
\begin{subequations}
\label{MFphi:thth:hiT:11}
\begin{eqnarray}
{}_2F_3\left(1,1;2,2,1+\gamma;-\frac{(k\Delta t)^2}{4}\right) &=&
\frac{\Gamma(1+\gamma)}{\sqrt{\pi}} \frac{\cos\big[k\Delta t
-\frac{\pi}{2}\big(\gamma+\frac52\big)\big]} {(k \Delta t
/2)^{\gamma+\frac52}}\left(1+{\cal O}((k\Delta t)^{-1})\right)
\nonumber\\
&&+\, 4 \gamma \frac{\log((k\Delta t)
^2/4)-\psi(\gamma)+\gamma_{\mathrm{E}}}{(k\Delta t)^2}
\left(1+{\cal O}((k\Delta t)^{-2})\right)
\label{MFphi:thth:hiT:11a}\,,
\end{eqnarray}
whereas the small times ($k\Delta t \ll 1$) limit yields:
\begin{eqnarray}
{}_2F_3\left(1,1;2,2,1+\gamma;-\frac{(k\Delta t)^2}{4}\right) &=&
1-\frac{(k\Delta t)^{2}}{16(1+\gamma)}+\mathcal{O}((k\Delta
t)^{4})\label{MFphi:thth:hiT:11b}\,.
\end{eqnarray}
\end{subequations}
We still need to perform some more integrals in equation
(\ref{MFphi:thth:hiT:3}). The second line in equation
(\ref{MFphi:thth:hiT:3}) can be further simplified to:
\begin{equation}
\int_{-k}^k \! \mathrm{d}u \int_k^{\infty} \! \mathrm{d}v
\frac{\cos(v\Delta t) \mathrm{e}^{-\frac{\beta v}{2}}
}{1-\mathrm{e}^{-\frac{\beta v}{2}}} \left[ \beta - \frac{1+
\frac{\beta u}{2}}{u+\frac{2}{\beta}\left[1-\mathrm{e}^{-
\frac{\beta v}{2}}\right]} + \frac{1- \frac{\beta
u}{2}}{u-\frac{2}{\beta}\left[1-\mathrm{e}^{- \frac{\beta
v}{2}}\right]} \right] \,. \label{MFphi:thth:hiT:12}
\end{equation}
We can now perform the $u$-integral:
\begin{eqnarray}
&=& -2 \int_k^{\infty} \! \mathrm{d}v \frac{\cos(v\Delta t)
\mathrm{e}^{- \beta v}}{1-\mathrm{e}^{-\frac{\beta v}{2}}} \left\{
\log\left[ k+\frac{2}{\beta}\left(1-\mathrm{e}^{-\frac{\beta
v}{2}}\right)\right] - \log\left[ -
k+\frac{2}{\beta}\left(1-\mathrm{e}^{-\frac{\beta
v}{2}}\right)\right] \right\} \label{MFphi:thth:hiT:13}\\
&=&  -4  \sum_{m=1}^{\infty} \frac{\left( \frac{k\beta}{2}
\right)^{2m-1} }{2m-1} \int_{k}^{\infty} \! \mathrm{d} v
\frac{\cos(v \Delta t)  \mathrm{e}^{- \beta v}} { \left(
1-\mathrm{e}^{-\frac{\beta v}{2}} \right)^{2m}} = -4
\sum_{\begin{subarray}{c} m=1
\\ n=0 \end{subarray}}^{\infty} \frac{\left( \frac{k\beta}{2}
\right)^{2m-1}}{2m-1} \frac{\Gamma(2m+n)}{\Gamma(2m)\Gamma(n+1)}
\mathrm{Re} \int_{k}^{\infty}\mathrm{d} v \mathrm{e}^{-
\frac{\beta v}{2}(n+2)+ \imath v \Delta t} \nonumber \,,
\end{eqnarray}
where the reader can easily verify that the argument of both
logarithms in the first line is positive. In the second line we
have expanded the logarithm and made use of the binomial series.
Due to the cosine appearing in equation (\ref{MFphi:thth:hiT:13}),
we are only interested in the real part of the integral on the
second line. The $v$-integral can now trivially be performed. In
order to extract the high temperature limit correctly, it turns
out to be advantageous to perform the $m$-sum in equation
(\ref{MFphi:thth:hiT:12}) first:
\begin{equation}
= -4 k  \sum_{n=0}^{\infty}
\frac{\Gamma(n+2)}{\Gamma(n+1)(n+2-2\imath \Delta t/\beta)}
{}_{3}F_{2}\left(1,1+\frac{n}{2},\frac{n+3}{2};\frac{3}{2},\frac{3}{2};\left(\frac{k\beta}{2}\right)^{2}\right)
\mathrm{Re}\,\, \mathrm{e}^{- \frac{k \beta}{2}(n+2 - 2 \imath
\frac{\Delta t}{\beta})} \,. \label{MFphi:thth:hiT:14}
\end{equation}
The Hypergeometric function can be expanded in the high
temperature limit as:
\begin{equation}
{}_{3}F_{2}\left(1,1+\frac{n}{2},\frac{n+3}{2};\frac{3}{2},\frac{3}{2};\left(\frac{k\beta}{2}\right)^{2}\right)
= 1+ \frac{1}{72}(n+2)(n+3)(k \beta)^2
+\mathcal{O}\left((k\beta)^{4}\right)\,. \label{MFphi:thth:hiT:15}
\end{equation}
We have checked using direct numerical integration that the
analytic answers improves much if we keep also the second order
term in this expansion. Finally, we can perform the remaining sum
over $n$, yielding:
\begin{eqnarray}
&& \! -4 k  \sum_{n=0}^{\infty}
\frac{\Gamma(n+2)}{\Gamma(n+1)(n+2-2\imath \Delta t/\beta)}
\left(1+ \frac{1}{72}(n+2)(n+3)(k \beta)^2\right) \mathrm{e}^{-
\frac{k \beta}{2}(n+2 - 2 \imath \frac{\Delta t}{\beta})}
\label{MFphi:thth:hiT:16}\\
&& \quad = -2k \frac{\mathrm{e}^{-k\beta+\imath k \Delta
t}}{1-\imath \Delta t/\beta} \Bigg[ {}_{2}F_{1}\!\left(
2,2-\frac{2\imath \Delta t}{\beta};3- \frac{2\imath \Delta
t}{\beta}; \mathrm{e}^{-\frac{k\beta}{2}}\right) +
\frac{(k\beta)^2}{12} {}_{2}F_{1} \left(4, 2- \frac{2\imath \Delta
t}{\beta};3 - \frac{2\imath \Delta t}{\beta};
\mathrm{e}^{-\frac{k\beta}{2}}\right) \nonumber \Bigg]\nonumber ,
\end{eqnarray}
where of course we are interested in the real part of the
expression above. Having performed all the integrals needed to
calculate the high temperature limit of
$M^{F}_{\phi,\mathrm{th-th}}(k,\Delta t)$, we can collect the
results in equations (\ref{MFphi:thth:hiT:3}),
(\ref{MFphi:thth:hiT:5}), (\ref{MFphi:thth:hiT:10}),
(\ref{MFphi:thth:hiT:12}) and equation (\ref{MFphi:thth:hiT:16})
above, finding precisely equation
(\ref{MFphi:thth:hiT:FinalResult}).

\section{The Statistical Propagator in Fourier Space}
\label{AppendixB}

This appendix is devoted to calculating the statistical propagator
in Fourier space at finite temperature. The two Wightman
functions, needed to calculate the statistical propagator through
equation (\ref{propagatorsstatistical}), are given by:
\begin{subequations}
\label{FourierWightman}
\begin{eqnarray}
\imath \Delta^{-+}_{\phi}(k^{\mu}) &=& \frac{ - \imath
M^{-+}_{\phi}(k^{\mu}) \imath
\Delta^{\mathrm{a}}_{\phi}(k^{\mu})}{k_{\mu}k^{\mu}+m_{\phi}^{2}+\imath
M^{\mathrm{r}}_{\phi,\mathrm{ren}}(k^{\mu})}
 \label{FourierWightmana} \\
\imath \Delta^{+-}_{\phi}(k^{\mu}) &=& \frac{ - \imath
M^{+-}_{\phi}(k^{\mu}) \imath
\Delta^{\mathrm{a}}_{\phi}(k^{\mu})}{k_{\mu}k^{\mu}+m_{\phi}^{2}+\imath
M^{\mathrm{r}}_{\phi,\mathrm{ren}}(k^{\mu})}
\label{FourierWightmanb} \,,
\end{eqnarray}
\end{subequations}
where we have made use of the definition of the advanced
propagator:
\begin{equation}
\imath \Delta^{\mathrm{a}}_{\phi}(k^{\mu}) = \frac{-\imath}{
k_{\mu}k^{\mu}+m_{\phi}^{2}+\imath
M^{\mathrm{a}}_{\phi,\mathrm{ren}}(k^{\mu})}
 \label{FourierAdvanced} \,,
\end{equation}
and the definitions of the advanced and retarded self-masses:
\begin{subequations}
\label{FourierSelfMassRelation}
\begin{eqnarray}
\imath M^{\mathrm{r}}_{\phi,\mathrm{ren}}(k^{\mu}) &=& \imath
M^{++}_{\phi,\mathrm{ren}}(k^{\mu}) - \imath
M^{+-}_{\phi}(k^{\mu}) =  \imath
M^{-+}_{\phi}(k^{\mu}) - \imath M^{--}_{\phi,\mathrm{ren}}(k^{\mu})  \label{FourierSelfMassRelationa} \\
\imath M^{\mathrm{a}}_{\phi,\mathrm{ren}}(k^{\mu}) &=& \imath
M^{++}_{\phi,\mathrm{ren}}(k^{\mu}) - \imath
M^{-+}_{\phi}(k^{\mu}) =  \imath M^{+-}_{\phi}(k^{\mu}) - \imath
M^{--}_{\phi,\mathrm{ren}}(k^{\mu})
\label{FourierSelfMassRelationb} \,,
\end{eqnarray}
\end{subequations}
Our starting point is:
\begin{eqnarray}\label{SelfMassFourier1}
\imath M_{\phi}^{++}(k^{\mu}) &=& - \frac{\imath h^{2}}{2} \int
\mathrm{d}^{\scriptscriptstyle{D}}(x-x') \left(
\imath \Delta_{\chi}^{++}(x;x')\right)^{2} e^{-\imath k(x-x')} \\
&=& - \frac{\imath h^{2}}{2} \int
\frac{\mathrm{d}^{\scriptscriptstyle{D}}k'}{(2\pi)^{\scriptscriptstyle{D}}}
\imath \Delta_{\chi}^{++}(k^{\prime\mu})\imath
\Delta_{\chi}^{++}(k^{\mu}-k^{\prime\mu}) \nonumber\,.
\end{eqnarray}
The thermal propagators appearing in this equation are of course
given by (\ref{ThermalPropagator}), where $m_{\chi}\rightarrow 0$.
This calculation naturally splits again into three parts:
\begin{equation}\label{SelfMassFourier2}
\imath M_{\phi}^{++}(k^{\mu}) = \imath
M_{\phi,\mathrm{vac}}^{++}(k^{\mu}) + \imath
M_{\phi,\mathrm{vac-th}}^{++}(k^{\mu}) + \imath
M_{\phi,\mathrm{th-th}}^{++}(k^{\mu}) \,,
\end{equation}
where:
\begin{subequations}
\label{SelfMassFourier3}
\begin{eqnarray}
\imath M_{\phi,\mathrm{vac}}^{++}(k^{\mu}) &=&
\frac{h^{2}}{32\pi^{2}}\left[
\log\left(\frac{-k_{0}^{2}+k^{2}-\imath\epsilon}{4\mu^{2}}\right)+2\gamma_{\mathrm{E}}\right]
\label{SelfMassFourier3a} \\
\imath M_{\phi,\mathrm{vac-th}}^{++}(k^{\mu}) &=& - h^{2} \int
\frac{\mathrm{d}^{\scriptscriptstyle{D}}k'}{(2\pi)^{\scriptscriptstyle{D}}}
\frac{1}{k^{\prime\mu}k_{\mu} - \imath \epsilon} 2\pi
\delta\left((k^{\mu}-k^{\prime\mu})(k_{\mu}-k^{\prime}_{\mu})\right)
n_{\chi}^{\mathrm{eq}}(|k^{0}-k^{\prime 0}|)
\label{SelfMassFourier3b} \\
\imath M_{\phi,\mathrm{th-th}}^{++}(k^{\mu}) &=& - \frac{\imath
h^{2}}{2} \int
\frac{\mathrm{d}^{\scriptscriptstyle{D}}k'}{(2\pi)^{\scriptscriptstyle{D}}}
4\pi^2
\delta\left((k^{\mu}-k^{\prime\mu})(k_{\mu}-k^{\prime}_{\mu})\right)
\delta\left(k^{\prime\mu} k^{\prime}_{\mu}\right)
n_{\chi}^{\mathrm{eq}}(|k^{0}-k^{\prime 0}|)
n_{\chi}^{\mathrm{eq}}(|k^{\prime 0}|) \label{SelfMassFourier3c}
\,,
\end{eqnarray}
\end{subequations}
where the vacuum contribution (\ref{SelfMassFourier3a}) has
already been evaluated and renormalised in \cite{Koksma:2009wa}.
As all thermal contributions are finite, we can safely let
$D\rightarrow 4$ and make use of equation (\ref{angular
integral:D-2}):
\begin{subequations}
\label{SelfMassFourier4}
\begin{eqnarray}
\imath M_{\phi,\mathrm{vac-th}}^{++}(k^{\mu}) &=& - \frac{h^{2}}{8
\pi^2 k} \int_{0}^{\infty}\mathrm{d}k' \int_{|k-k'|}^{k+k'}
\mathrm{d}\omega \, k' n_{\chi}^{\mathrm{eq}}(\omega) \left(
\frac{1}{-(k^{0}+\omega)^2+k^{\prime 2}-\imath \epsilon} +
\frac{1}{-(k^{0}-\omega)^2+k^{\prime 2}-\imath \epsilon} \right)
\label{SelfMassFourier4a} \\
\imath M_{\phi,\mathrm{th-th}}^{++}(k^{\mu}) &=& - \frac{\imath
h^{2}}{16 \pi k} \int_{0}^{\infty}\mathrm{d}k'
\int_{|k-k'|}^{k+k'} \mathrm{d}\omega
n_{\chi}^{\mathrm{eq}}(k')\sum_{\pm} n_{\chi}^{\mathrm{eq}}(|k^0
\pm  k'|) \left[ \delta(k^0 \pm k' +\omega) + \delta(k^0 \pm k'
-\omega) \right] \label{SelfMassFourier4b} \,.
\end{eqnarray}
\end{subequations}
Here, $k=\|\vec{k}\|$ as before. Transforming to
$(u,v)$-coordinates already used in equation (\ref{new coords})
now yields:
\begin{subequations}
\label{SelfMassFourier5}
\begin{eqnarray}
\imath M_{\phi,\mathrm{vac-th}}^{++}(k^{\mu}) &=& -
\frac{h^{2}}{32 \pi^2 k} \int_{-k}^{k}\mathrm{d}u
\int_{k}^{\infty} \mathrm{d}v
\frac{u+v}{\mathrm{e}^{\frac{\beta}{2}(u+v)}-1} \left(
\frac{1}{(v+k^{0})(u-k^0)-\imath \epsilon} +
\frac{1}{(v-k^{0})(u+k^0)-\imath \epsilon} \right)
\label{SelfMassFourier5a} \\
\imath M_{\phi,\mathrm{th-th}}^{++}(k^{\mu}) &=& - \frac{\imath
h^{2}}{32 \pi k} \int_{-k}^{k}\mathrm{d}u \int_{k}^{\infty}
\mathrm{d}v n_{\chi}^{\mathrm{eq}}\left(\frac{1}{2}(u+v) \right
)\sum_{\pm} n_{\chi}^{\mathrm{eq}}\left(\left|k^0 \pm
\frac{1}{2}(u+v)\right|\right) \left[ \delta(k^0 \pm v) +
\delta(k^0 \pm u) \right] \label{SelfMassFourier5b} \,.
\end{eqnarray}
\end{subequations}
Let us firstly calculate $\imath
M_{\phi,\mathrm{th-th}}^{++}(k^{\mu})$. The Dirac delta-functions
trivially reduce equation (\ref{SelfMassFourier5b}) further and
moreover, we can make use of:
\begin{subequations}
\label{SelfMassFourier6}
\begin{eqnarray}
\int_{-k}^{k}\mathrm{d}u
\frac{1}{\mathrm{e}^{\frac{\beta}{2}(u-k^0)} -1}
\frac{1}{\mathrm{e}^{-\frac{\beta}{2}(u+k^0)} -1} &=&
\frac{2}{\beta} \frac{1}{\mathrm{e}^{-\beta k^0}-1} \left[2
\log\left(
\frac{1-\mathrm{e}^{-\frac{\beta}{2}(k-k^0)}}{1-\mathrm{e}^{\frac{\beta}{2}(k+k^0)}}\right)
+ k\beta\right]
\label{SelfMassFourier6a} \\
\int_{k}^{\infty} \mathrm{d}v
\frac{1}{\mathrm{e}^{\frac{\beta}{2}(v-k^0)}-1}
\frac{1}{\mathrm{e}^{\frac{\beta}{2}(v+k^0)}-1} &=& \frac{2}{\beta}
\left[ \frac{1}{1-\mathrm{e}^{\beta k^0}} \log
\left(1-\mathrm{e}^{-\frac{\beta}{2}(k-k^0)}\right) +
\frac{1}{1-\mathrm{e}^{-\beta k^0}} \log
\left(1-\mathrm{e}^{-\frac{\beta}{2}(k+k^0)}\right)\right]
\label{SelfMassFourier6b} \,.
\end{eqnarray}
\end{subequations}
The final result for $\imath
M_{\phi,\mathrm{th-th}}^{++}(k^{\mu})$ thus reads:
\begin{eqnarray}
&& \! \imath M_{\phi,\mathrm{th-th}}^{++}(k^{\mu}) =  -
\frac{\imath h^{2}}{16 \pi k \beta}\Bigg [ \sum_{\pm}
\frac{\theta(\mp k^{0}-k)}{\mathrm{e}^{\mp \beta k^0}-1} \left\{ 2
\log \left( \frac{1-\mathrm{e}^{-\frac{\beta}{2}(k\mp
k^0)}}{1-\mathrm{e}^{\frac{\beta}{2}(k\pm k^0)}}\right) +
k\beta\right\}
\label{SelfMassFourier7}\\
&& \qquad\qquad + \sum_{\pm} \left[ \theta( \mp k^0 +k) - \theta
(\mp k^0-k)\right] \left\{ \frac{1}{1-\mathrm{e}^{\pm \beta k^0}}
\log \left(1-\mathrm{e}^{-\frac{\beta}{2}(k\mp k^0)}\right) +
\frac{1}{1-\mathrm{e}^{\mp \beta k^0}} \log
\left(1-\mathrm{e}^{-\frac{\beta}{2}(k\pm k^0)}\right)\right\}
\Bigg]\nonumber  \,.
\end{eqnarray}
Since this contribution to $\imath M_{\phi}^{++}(k^{\mu})$ does
not depend on the pole prescription, it completely fixes similar
contributions to the other self-masses, e.g. $\imath
M_{\phi,\mathrm{th-th}}^{+-}(k^{\mu})= \imath
M_{\phi,\mathrm{th-th}}^{++}(k^{\mu})$. It turns out that equation
(\ref{SelfMassFourier5a}) is not most advantageous to derive
$\imath M_{\phi,\mathrm{vac-th}}^{++}(k^{\mu})$.

Let us therefore firstly evaluate $\imath M^{\pm \mp}(k^{\mu})$.
Let us thus start just as in equation (\ref{SelfMassFourier1}) and
set:
\begin{equation}\label{SelfMassFourier8}
\imath M_{\phi}^{\pm \mp}(k^{\mu}) = \imath
M_{\phi,\mathrm{vac}}^{\pm \mp}(k^{\mu}) + \imath
M_{\phi,\mathrm{vac-th}}^{\pm \mp}(k^{\mu}) + \imath
M_{\phi,\mathrm{th-th}}^{\pm \mp}(k^{\mu}) \,.
\end{equation}
The vacuum-vacuum contribution has been evaluated in
\cite{Koksma:2009wa} and is given by:
\begin{equation}\label{SelfMassFourier9}
\imath M_{\phi,\mathrm{vac}}^{\pm \mp}(k^{\mu})= - \frac{\imath
h^2}{16 \pi} \theta(\mp k^0 - k) \,.
\end{equation}
The thermal-thermal contributions are given above in equation
(\ref{SelfMassFourier7}), so we only need to determine the
vacuum-thermal contributions. Hence, we perform an analogous
calculation as for $\imath M^{++}$ and transform to the familiar
lightcone coordinates $u$ and $v$ to find the following
intermediate result:
\begin{equation}\label{SelfMassFourier10}
\imath M_{\phi,\mathrm{vac-th}}^{\pm \mp}(k^{\mu})= - \frac{\imath
h^2}{16 \pi k} \int_{-k}^{k}\mathrm{d}u
\int_{k}^{\infty}\mathrm{d}v \, n_{\chi}^{\mathrm{eq}}(|k^{0}\pm
(u+v)/2|) \left[ \delta (k^0 \pm u) + \delta (k^0 \pm v)
\right]\,.
\end{equation}
The delta functions allow us to perform one of the two integrals
trivially. The remaining integral can also be obtained
straightforwardly:
\begin{equation}\label{SelfMassFourier11}
\imath M_{\phi,\mathrm{vac-th}}^{\pm \mp}(k^{\mu})= \frac{\imath
h^2}{8 \pi k \beta} \left[ \left\{\theta(k \pm k^0) - \theta (-k
\pm k^0)\right\} \log\left(1-\mathrm{e}^{-\frac{\beta}{2}(k \pm
k^0)} \right) - \theta(\mp k^0 -k) \log
\left(\frac{1-\mathrm{e}^{-\frac{\beta}{2} (\mp k^0 +k)}}{
1-\mathrm{e}^{-\frac{\beta}{2} (\mp k^0 -k)}}\right) \right]\,.
\end{equation}
By subtracting and adding the above self-masses, we can obtain the
vacuum-thermal contributions to the causal and statistical
self-masses in Fourier space from equations
(\ref{selfMass:causal1}) and (\ref{selfMass:statistical1}),
respectively. The vacuum-thermal contribution to the causal
self-mass reads:
\begin{eqnarray}\label{SelfMassFourier12}
M_{\phi,\mathrm{vac-th}}^{c}(k^{\mu})  &=&  \imath
M_{\phi,\mathrm{vac-th}}^{+-}(k^{\mu}) - \imath
M_{\phi,\mathrm{vac-th}}^{-+}(k^{\mu})\\ &=& \frac{\imath
h^2}{8\pi k \beta} \mathrm{sgn}(k^0)\left[
\log\left(1-\mathrm{e}^{-\frac{\beta}{2} (k+|k^0|)} \right) -
\log\left(1-\mathrm{e}^{-\frac{\beta}{2} |k-|k^0||} \right)
\right] \nonumber \,,
\end{eqnarray}
where we have made use of the theta functions to bring this result
in a particularly compact form. Likewise, the vacuum-thermal contribution
to the statistical self-mass now reads:%
\begin{eqnarray}\label{SelfMassFourier13}
M_{\phi,\mathrm{vac-th}}^{F}(k^{\mu})  &=&  \frac{1}{2} \left[
M_{\phi,\mathrm{vac-th}}^{+-}(k^{\mu}) +
M_{\phi,\mathrm{vac-th}}^{-+}(k^{\mu}) \right]\\ &=&
\frac{h^2}{16\pi k \beta} \left[ \mathrm{sgn}(k-|k^0|)
\log\left(1-\mathrm{e}^{-\frac{\beta}{2} (k+|k^0|)} \right) +
\log\left(1-\mathrm{e}^{-\frac{\beta}{2} |k-|k^0||} \right)
\right] \nonumber \,.
\end{eqnarray}
As a check of the results above, we performed the inverse Fourier
transforms of the causal and statistical self-masses in equations
(\ref{selfMass:causal3}) and (\ref{selfMass:statistical4}),
respectively, and found agreement with the results presented
above.

The most convenient way of solving the vacuum-thermal contribution
to $\imath M^{++}_{\phi}(k^\mu)$ is by making use of equation
(\ref{reduction:selfMass:MF+Mc2}) and (\ref{selfMass:causal3}).
Let us set the imaginary part of $\imath M^{++}_{\phi}(k^\mu)$
equal to:
\begin{eqnarray}\label{DecomposeiM++}
M^{++}_{\mathrm{sgn}}(k,\Delta t) &\equiv&
\frac{1}{2}\mathrm{sgn}(\Delta t ) M_{\phi}^{c}(k,\Delta t) =
\frac{h^{2}}{32\pi^{2}} \frac{\sin(k\Delta t)}{k (\Delta
t)^{2}} \mathrm{sgn}(\Delta t) \left[\frac{2\pi \Delta t}
{\beta}\coth\left(\frac{2\pi\Delta t}{\beta}\right)-1\right]\nonumber\\
&  \stackrel{\beta k \ll 1}{\longrightarrow} & \frac{h^{2}}{16 \pi
\beta}\frac{\sin(k\Delta t)}{k \Delta t}\,,
\end{eqnarray}
where we have taken the high temperature limit on the second line.
We thus have:
\begin{eqnarray}\label{DecomposeiM++2}
M^{++}_{\mathrm{sgn}}(k^\mu) &=& \int_{-\infty}^{\infty}
\mathrm{d}(\Delta t) M^{++}_{\mathrm{sgn}}(k,\Delta t)
\mathrm{e}^{\imath k^0 \Delta t} \nonumber\\ & \stackrel{\beta k
\ll 1}{\longrightarrow} &  \frac{h^2}{8 \pi k \beta}
\int_{0}^{\infty}
 \mathrm{d}(\Delta t) \frac{\sin(k \Delta t) \cos(k^0 \Delta t)}{\Delta t}
 \nonumber
 \\
&=& \frac{h^2}{32\beta k}\Big[{\rm sgn}(k^0+k)+{\rm sgn}(-k^0+k)\Big]\,,
\end{eqnarray}
where have performed the remaining integral on the second line
straightforwardly. Using equation
(\ref{reduction:selfMass:MF+Mc2}) and (\ref{SelfMassFourier13}) we
find:
\begin{subequations}
\label{M++vacthFourier}
\begin{eqnarray}
M^{++}_{\phi, \rm vac-th}(k^\mu) &\stackrel{\beta k \ll
1}{\longrightarrow} & M^F_{\phi, \rm vac-th}(k^\mu) + \imath M^{++}_{\rm sgn}(k^\mu)
\label{M++vacthFouriera}\\
M^{--}_{\phi, \rm vac-th}(k^\mu) & \stackrel{\beta k \ll
1}{\longrightarrow} & M^F_{\phi, \rm vac-th}(k^\mu) - \imath
M^{++}_{\rm sgn}(k^\mu) \label{M++vacthFourierb}\,,
\end{eqnarray}
\end{subequations}
such that:
\begin{subequations}
\label{M++vacthFourierHighT}
\begin{eqnarray}
\imath M^{++}_{\phi, \rm vac-th}(k^\mu)  & \stackrel{\beta k \ll
1}{\longrightarrow} &  \frac{\imath h^2}{16\pi k \beta} \Bigg[ \mathrm{sgn}(k- |k^0|)
\log\left(1-\mathrm{e}^{-\frac{\beta}{2} (k+|k^0|)} \right) +
 \log\left(1-\mathrm{e}^{-\frac{\beta}{2} |k-|k^0||} \right)\! \nonumber \\
&& \qquad\qquad  + \frac{\imath \pi}{2} \left[ {\rm sgn}(k^0+k)+
{\rm sgn}(-k^0+k)\right] \Bigg] \label{M++vacthFourierHighTa}\\
\imath M^{--}_{\phi, \rm vac-th}(k^\mu)  & \stackrel{\beta k \ll
1}{\longrightarrow} &  \frac{\imath h^2}{16\pi k \beta}  \Bigg[ \mathrm{sgn}(k-|k^0|)
\log \left(1-\mathrm{e}^{-\frac{\beta}{2} (k+|k^0|)} \right) + \log\left(1 -
\mathrm{e}^{-\frac{\beta}{2} |k-|k^0||} \right) \nonumber \\
&& \qquad\qquad  - \frac{\imath \pi}{2}  \left[{\rm sgn}(k^0+k)
+ {\rm sgn}(-k^0 + k)\right]\Bigg]\,.\label{M--vacthFourierHighT}
\end{eqnarray}
\end{subequations}
One can check equation (\ref{DecomposeiM++2}) by means of an
alternative approach. The starting point is the first line of
equation (\ref{selfMass:causal3}) and one can furthermore realise
that differentiating $M^{++}_{\mathrm{sgn}}(k^\mu)$ with respect
to $k^0$ brings down a factor of $\imath \Delta t$ which
conveniently cancels the factor of $\Delta t $ that is present in
the denominator. One can then integrate the resulting expressions
(introducing $\epsilon$ regulators and UV cutoffs where necessary)
confirming expression (\ref{DecomposeiM++2}).

In the low temperature limit, equation (\ref{DecomposeiM++}) reduces to:
\begin{equation}\label{DecomposeiM++3}
M^{++}_{\mathrm{sgn}}(k,\Delta t) \stackrel{\beta k \gg
1}{\longrightarrow} \frac{h^{2}}{24 k \beta^2} \sin(k\Delta t)
\mathrm{sgn}(\Delta t)\,.
\end{equation}
We can introduce an $\epsilon$ regulator:
\begin{eqnarray}\label{DecomposeiM++4}
M^{++}_{\mathrm{sgn}}(k^\mu) & \stackrel{\beta k \gg 1}{\longrightarrow} &
\frac{h^2}{12 k \beta^2}\int_{0}^{\infty} \mathrm{d}(\Delta t) \sin(k\Delta t)
 \cos(k^0 \Delta t)  \mathrm{e}^{- \epsilon \Delta t} \nonumber\\
&= &  \frac{h^2}{12 \beta^2} \frac{1}{k^2-k_0^2}\,.
\end{eqnarray}
Analogously, we can derive the following expressions for the
vacuum-thermal contributions to $\imath M^{\pm\pm}_{\phi}(k^\mu)$
in the low temperature limit:
\begin{subequations}
\label{M++vacthFourierLowT}
\begin{eqnarray}
\imath M^{++}_{\phi, \rm vac-th}(k^\mu)  & \stackrel{\beta k \gg
1}{\longrightarrow} &  \frac{\imath h^2}{16\pi k \beta} \Bigg[
\mathrm{sgn}(k- |k^0|) \log\left(1-\mathrm{e}^{-\frac{\beta}{2}
(k+|k^0|)} \right) + \log\left(1-\mathrm{e}^{-\frac{\beta}{2}
|k-|k^0||} \right)\!
+ \frac{\imath 4\pi k}{3\beta} \frac{1}{k^2-k_0^2} \Bigg] \label{M++vacthFourierLowTa}\\
\imath M^{--}_{\phi, \rm vac-th}(k^\mu)  & \stackrel{\beta k \gg
1}{\longrightarrow} &  \frac{\imath h^2}{16\pi k \beta}  \Bigg[
\mathrm{sgn}(k-|k^0|) \log \left(1-\mathrm{e}^{-\frac{\beta}{2}
(k+|k^0|)} \right) +
 \log\left(1 - \mathrm{e}^{-\frac{\beta}{2} |k-|k^0||} \right)
 - \frac{\imath 4\pi k}{3\beta} \frac{1}{k^2-k_0^2}\Bigg]\,.\label{M--vacthFourierLowT}
\end{eqnarray}
\end{subequations}
We have now all self-masses at our disposal necessary to calculate
$F(k^\mu)$. To numerically evaluate the integrals in equation
(\ref{Fconstantmass2}) we do not rely on the high and low
temperature expressions in equations (\ref{M++vacthFourierLowT})
or (\ref{M++vacthFourierHighT}) but we rather use exact numerical
methods, i.e.: the first line of equation (\ref{DecomposeiM++}).


\begin{thebibliography}{99}
\bibitem{Koksma:2009wa}
  J.~F.~Koksma, T.~Prokopec, M.~G.~Schmidt,
  Decoherence in an Interacting Quantum Field Theory: The Vacuum Case,
  Phys.\ Rev.\  {\bf D81 } (2010)  065030.
  [arXiv:0910.5733 [hep-th]].
\bibitem{Koksma:2010zi}
  J.~F.~Koksma, T.~Prokopec, M.~G.~Schmidt,
  Entropy and Correlators in Quantum Field Theory,
  Annals Phys.\  {\bf 325 } (2010)  1277-1303.
  [arXiv:1002.0749 [hep-th]].
\bibitem{Koksma:2010dt}
  J.~F.~Koksma, T.~Prokopec, M.~G.~Schmidt,
  Decoherence in Quantum Mechanics,
  [arXiv:1012.3701 [quant-ph]].
\bibitem{Koksma:2011fx}
  J.~F.~Koksma, T.~Prokopec, M.~G.~Schmidt,
  Decoherence and Dynamical Entropy Generation in Quantum Field Theory,
  [arXiv:1101.5323 [quant-ph]].
\bibitem{Giraud:2009tn}
  A.~Giraud and J.~Serreau,
  Decoherence and Thermalization of a Pure Quantum State in Quantum Field
  Theory,
  arXiv:0910.2570 [hep-ph].
\bibitem{Prokopec:1992ia}
  T.~Prokopec,
  Entropy of the Squeezed Vacuum,
  Class.\ Quant.\ Grav.\  {\bf 10} (1993) 2295.
\bibitem{Brandenberger:1992jh}
  R.~H.~Brandenberger, T.~Prokopec and V.~F.~Mukhanov,
  The Entropy of the Gravitational Field,
  Phys.\ Rev.\  D {\bf 48} (1993) 2443
  [arXiv:gr-qc/9208009].
\bibitem{Kiefer:2006je}
  C.~Kiefer, I.~Lohmar, D.~Polarski and A.~Starobinsky,
  Pointer States for Primordial Fluctuations in Inflationary Cosmology,
  Class.\ Quant.\ Grav.\  {\bf 24 } (2007)  1699-1718.
  [astro-ph/0610700].
\bibitem{Campo:2008ju}
  D.~Campo and R.~Parentani,
  Decoherence and Entropy of Primordial Fluctuations. I: Formalism and
  Interpretation,
  Phys.\ Rev.\  D {\bf 78} (2008) 065044
  [arXiv:0805.0548 [hep-th]].
\bibitem{Campo:2008ij}
  D.~Campo and R.~Parentani,
  Decoherence and Entropy of Primordial Fluctuations II. The Entropy
  Budget,
  Phys.\ Rev.\  D {\bf 78} (2008) 065045
  [arXiv:0805.0424 [hep-th]].
\bibitem{Zeh:1970}
  H.~D.~Zeh,
  On the Interpretation of Measurement in Quantum Theory,
  Found.\ Phys. {\bf 1} (1970) 69.
\bibitem{Zurek:1981xq}
  W.~H.~Zurek,
  Pointer Basis of Quantum Apparatus: Into What Mixture Does the Wave Packet
  Collapse?,
  Phys.\ Rev.\  D {\bf 24} (1981) 1516.
\bibitem{Joos:1984uk}
  E.~Joos and H.~D.~Zeh,
  The Emergence of Classical Properties through Interaction with the
  Environment,
  Z.\ Phys.\  B {\bf 59} (1985) 223.
\bibitem{Joos:book}
  E.~Joos, H.~D.~Zeh, C.~Kiefer, D.~Giulini, J.~Kupsch and
  I.~O.~Stamatescu,
  Decoherence and the Appearance of a Classical World in Quantum Theory,
  Springer (2003).
\bibitem{Zurek:2003zz}
  W.~H.~Zurek,
  Decoherence, Einselection, and the Quantum Origins of the Classical,
  Rev.\ Mod.\ Phys.\  {\bf 75} (2003) 715.
\bibitem{Paz:2000le}
  J.~P.~Paz and W.~H.~Zurek,
  Environment-induced Decoherence and the Transition from Quantum to
  Classical,
  Lectures given at the 72nd Les Houches Summer School on `Coherent Matter
  Waves' (1999) quant-ph/0010011.
\bibitem{Burgess:2006jn}
  C.~P.~Burgess, R.~Holman and D.~Hoover,
  On the Decoherence of Primordial Fluctuations during Inflation,
  Phys.\ Rev.\  D {\bf 77} (2008) 063534
  [arXiv:astro-ph/0601646].
\bibitem{Hu:1993vs}
  B.~L.~Hu, J.~P.~Paz, Y.~Zhang,
  Quantum Brownian Motion in a General Environment. 2: Nonlinear Coupling and Perturbative Approach,
  Phys.\ Rev.\  {\bf D47 } (1993)  1576-1594.
\bibitem{Hu:1993qa}
  B.~L.~Hu, A.~Matacz,
  Quantum Brownian Motion in a Bath of Parametric Oscillators: A Model for System - Field Interactions,
  Phys.\ Rev.\  {\bf D49 } (1994)  6612-6635.
  [gr-qc/9312035].
\bibitem{Calzetta:book}
  E.~A.~Calzetta and B.~L.~Hu,
  Nonequilibrium Quantum Field Theory,
  Cambridge University Press (2008).
\bibitem{Farrar:1993hn}
  G.~R.~Farrar and M.~E.~Shaposhnikov,
  Baryon Asymmetry of the Universe in the Standard Electroweak Theory,
  Phys.\ Rev.\  D {\bf 50} (1994) 774
  [arXiv:hep-ph/9305275].
\bibitem{Farrar:1993sp}
  G.~R.~Farrar and M.~E.~Shaposhnikov,
  Baryon Asymmetry of the Universe in the Minimal Standard Model,
  Phys.\ Rev.\ Lett.\  {\bf 70} (1993) 2833
  [Erratum-ibid.\  {\bf 71} (1993) 210]
  [arXiv:hep-ph/9305274].
\bibitem{Gavela:1993ts}
  M.~B.~Gavela, P.~Hernandez, J.~Orloff and O.~Pene,
  Standard Model CP-violation and Baryon asymmetry,
  Mod.\ Phys.\ Lett.\  A {\bf 9} (1994) 795
  [arXiv:hep-ph/9312215].
\bibitem{Huet:1994jb}
  P.~Huet and E.~Sather,
  Electroweak Baryogenesis and Standard Model CP Violation,
  Phys.\ Rev.\  D {\bf 51} (1995) 379
  [arXiv:hep-ph/9404302].
\bibitem{Gavela:1994ds}
  M.~B.~Gavela, M.~Lozano, J.~Orloff and O.~Pene,
  Standard Model CP Violation and Baryon Asymmetry. Part 1: Zero
  Temperature,
  Nucl.\ Phys.\  B {\bf 430} (1994) 345
  [arXiv:hep-ph/9406288].
\bibitem{Gavela:1994dt}
  M.~B.~Gavela, P.~Hernandez, J.~Orloff, O.~Pene and C.~Quimbay,
  Standard Model CP Violation and Baryon Asymmetry. Part 2: Finite
  Temperature,
  Nucl.\ Phys.\  B {\bf 430} (1994) 382
  [arXiv:hep-ph/9406289].
\bibitem{Balazs:2004bu}
  C.~Balazs, M.~S.~Carena, C.~E.~M.~Wagner,
  Dark Matter, Light Stops and Electroweak Baryogenesis,
  Phys.\ Rev.\  {\bf D70 } (2004)  015007.
  [hep-ph/0403224].
\bibitem{Konstandin:2005cd}
  T.~Konstandin, T.~Prokopec, M.~G.~Schmidt {\it et al.},
  MSSM Electroweak Baryogenesis and Flavor Mixing in Transport Equations,
  Nucl.\ Phys.\  {\bf B738 } (2006)  1-22.
  [hep-ph/0505103].
\bibitem{Huber:2006wf}
  S.~J.~Huber, T.~Konstandin, T.~Prokopec {\it et al.},
  Electroweak Phase Transition and Baryogenesis in the nMSSM,
  Nucl.\ Phys.\  {\bf B757 } (2006)  172-196.
  [hep-ph/0606298].
\bibitem{Fromme:2006cm}
  L.~Fromme, S.~J.~Huber, M.~Seniuch,
  Baryogenesis in the Two-Higgs Doublet Model,
  JHEP {\bf 0611 } (2006)  038.
  [hep-ph/0605242].
\bibitem{Chung:2009qs}
  D.~J.~H.~Chung, B.~Garbrecht, M.~.J.~Ramsey-Musolf {\it et al.},
  Supergauge Interactions and Electroweak Baryogenesis,''
  JHEP {\bf 0912 } (2009)  067.
  [arXiv:0908.2187 [hep-ph]].
\bibitem{Herranen:2008hu}
  M.~Herranen, K.~Kainulainen and P.~M.~Rahkila,
  Quantum Kinetic Theory for Fermions in Temporally Varying Backrounds,
  JHEP {\bf 0809} (2008) 032
  [arXiv:0807.1435 [hep-ph]].
\bibitem{Herranen:2008hi}
  M.~Herranen, K.~Kainulainen and P.~M.~Rahkila,
  Towards a Kinetic Theory for Fermions with Quantum Coherence,
  Nucl.\ Phys.\  B {\bf 810} (2009) 389
  [arXiv:0807.1415 [hep-ph]].
\bibitem{Herranen:2008di}
  M.~Herranen, K.~Kainulainen and P.~M.~Rahkila,
  Kinetic Theory for Scalar Fields with Nonlocal Quantum Coherence,
  JHEP {\bf 0905} (2009) 119
  [arXiv:0812.4029 [hep-ph]].
\bibitem{Garbrecht:2003mn}
  B.~Garbrecht, T.~Prokopec and M.~G.~Schmidt,
  Coherent baryogenesis,
  Phys.\ Rev.\ Lett.\  {\bf 92} (2004) 061303
  [arXiv:hep-ph/0304088].
\bibitem{Garbrecht:2004gv}
  B.~Garbrecht, T.~Prokopec and M.~G.~Schmidt,
  Coherent Baryogenesis and Nonthermal Leptogenesis: a Comparison,
  arXiv:hep-ph/0410132.
\bibitem{Garbrecht:2005rr}
  B.~Garbrecht, T.~Prokopec and M.~G.~Schmidt,
  SO(10) - GUT coherent baryogenesis,
  Nucl.\ Phys.\  B {\bf 736} (2006) 133
  [arXiv:hep-ph/0509190].
\bibitem{Berges:2004yj}
  J.~Berges,
  Introduction to Nonequilibrium Quantum Field Theory,
  AIP Conf.\ Proc.\  {\bf 739} (2005) 3
  [arXiv:hep-ph/0409233].
\bibitem{Aarts:2002dj}
  G.~Aarts, D.~Ahrensmeier, R.~Baier, J.~Berges and J.~Serreau,
  Far-from-equilibrium Dynamics with Broken Symmetries from the 2PI-1/N
  Expansion,
  Phys.\ Rev.\  D {\bf 66} (2002) 045008
  [arXiv:hep-ph/0201308].
\bibitem{Juchem:2004cs}
  S.~Juchem, W.~Cassing and C.~Greiner,
  Nonequilibrium Quantum-field Dynamics and Off-shell Transport for
  phi**4-theory in 2+1 Dimensions,
  Nucl.\ Phys.\  A {\bf 743} (2004) 92
  [arXiv:nucl-th/0401046].
\bibitem{Arrizabalaga:2005tf}
  A.~Arrizabalaga, J.~Smit and A.~Tranberg,
  Equilibration in phi**4 Theory in 3+1 Dimensions,
  Phys.\ Rev.\  D {\bf 72} (2005) 025014
  [arXiv:hep-ph/0503287].
\bibitem{Berges:2002wr}
  J.~Berges, S.~Borsanyi and J.~Serreau,
  Thermalization of Fermionic Quantum Fields,
  Nucl.\ Phys.\  B {\bf 660} (2003) 51
  [arXiv:hep-ph/0212404].
\bibitem{Anisimov:2008dz}
  A.~Anisimov, W.~Buchmuller, M.~Drewes and S.~Mendizabal,
  Nonequilibrium Dynamics of Scalar Fields in a Thermal Bath,
  Annals Phys.\  {\bf 324} (2009) 1234
  [arXiv:0812.1934 [hep-th]].
\bibitem{Jackson:2010cw}
  M.~G.~Jackson, K.~Schalm,
  Model Independent Signatures of New Physics in the Inflationary Power Spectrum,
  [arXiv:1007.0185 [hep-th]].
\bibitem{vanHees:2002bv}
  H.~van Hees and J.~Knoll,
  Renormalization in Self-consistent Approximation Schemes at Finite
  Temperature. III: Global Symmetries,
  Phys.\ Rev.\  D {\bf 66} (2002) 025028
  [arXiv:hep-ph/0203008].
\bibitem{Borsanyi:2009zza}
  S.~Borsanyi and U.~Reinosa,
  Renormalisation of out-of-equilibrium Quantum Fields,
  Nucl.\ Phys.\  A {\bf 820} (2009) 147C.
\bibitem{Blaizot:2003br}
  J.~P.~Blaizot, E.~Iancu and U.~Reinosa,
  Renormalizability of Phi-derivable Approximations in Scalar phi**4
  Theory,
  Phys.\ Lett.\  B {\bf 568} (2003) 160
  [arXiv:hep-ph/0301201].
\bibitem{Calzetta:2003dk}
  E.~A.~Calzetta and B.~L.~Hu,
  Correlation Entropy of an Interacting Quantum Field and H-theorem for the
  O(N) model,
  Phys.\ Rev.\  D {\bf 68} (2003) 065027
  [arXiv:hep-ph/0305326].
\bibitem{Nishiyama:2010wc}
  A.~Nishiyama and A.~Ohnishi,
  Entropy Current for the Relativistic Kadanoff-Baym Equation and H-theorem
  in $O(N)$ Theory with NLO Self-energy of $1/N$ Expansion,
  arXiv:1006.1124 [nucl-th].
\bibitem{Garny:2009ni}
  M.~Garny and M.~M.~Muller,
  Kadanoff-Baym Equations with Non-Gaussian Initial Conditions: The
  Equilibrium Limit,
  arXiv:0904.3600 [hep-ph].
\bibitem{Brandenberger:1990bx}
  R.~H.~Brandenberger, R.~Laflamme and M.~Mijic,
  Classical Perturbations from Decoherence of Quantum Fluctuations in the
  Inflationary Universe,
  Mod.\ Phys.\ Lett.\  A {\bf 5} (1990) 2311.
\bibitem{Polarski:1995jg}
  D.~Polarski and A.~A.~Starobinsky,
  Semiclassicality and Decoherence of Cosmological Perturbations,
  Class.\ Quant.\ Grav.\  {\bf 13} (1996) 377
  [arXiv:gr-qc/9504030].
\bibitem{Calzetta:1995ys}
  E.~Calzetta and B.~L.~Hu,
  Quantum Fluctuations, Decoherence of the Mean Field, and Structure
  Formation in the Early Universe,
  Phys.\ Rev.\  D {\bf 52} (1995) 6770
  [arXiv:gr-qc/9505046].
\bibitem{Lesgourgues:1996jc}
  J.~Lesgourgues, D.~Polarski and A.~A.~Starobinsky,
  Quantum-to-classical Transition of Cosmological Perturbations for
  Non-vacuum Initial States,
  Nucl.\ Phys.\  B {\bf 497} (1997) 479
  [arXiv:gr-qc/9611019].
\bibitem{Kiefer:1998qe}
  C.~Kiefer, D.~Polarski and A.~A.~Starobinsky,
  Quantum-to-classical Transition for Fluctuations in the Early Universe,
  Int.\ J.\ Mod.\ Phys.\  D {\bf 7} (1998) 455
  [arXiv:gr-qc/9802003].
\bibitem{Campo:2005sy}
  D.~Campo and R.~Parentani,
  Inflationary Spectra and Partially Decohered Distributions,
  Phys.\ Rev.\  D {\bf 72} (2005) 045015
  [arXiv:astro-ph/0505379].
\bibitem{Lombardo:2005iz}
  F.~C.~Lombardo and D.~Lopez Nacir,
  Decoherence during Inflation: The Generation of Classical
  Inhomogeneities,
  Phys.\ Rev.\  D {\bf 72} (2005) 063506
  [arXiv:gr-qc/0506051].
\bibitem{Martineau:2006ki}
  P.~Martineau,
  On the Decoherence of Primordial Fluctuations during Inflation,
  Class.\ Quant.\ Grav.\  {\bf 24} (2007) 5817
  [arXiv:astro-ph/0601134].
\bibitem{Lyth:2006qz}
  D.~H.~Lyth and D.~Seery,
  Classicality of the Primordial Perturbations,
  Phys.\ Lett.\  B {\bf 662} (2008) 309
  [arXiv:astro-ph/0607647].
\bibitem{Prokopec:2006fc}
  T.~Prokopec and G.~I.~Rigopoulos,
  Decoherence from Isocurvature Perturbations in Inflation,
  JCAP {\bf 0711} (2007) 029
  [arXiv:astro-ph/0612067].
\bibitem{Sharman:2007gi}
  J.~W.~Sharman and G.~D.~Moore,
  Decoherence due to the Horizon after Inflation,
  JCAP {\bf 0711} (2007) 020
  [arXiv:0708.3353 [gr-qc]].
\bibitem{Kiefer:2007zza}
  C.~Kiefer, I.~Lohmar, D.~Polarski and A.~A.~Starobinsky,
  Origin of Classical Structure in the Universe,
  J.\ Phys.\ Conf.\ Ser.\  {\bf 67} (2007) 012023.
\bibitem{Sudarsky:2009za}
  D.~Sudarsky,
  Shortcomings in the Understanding of Why Cosmological Perturbations Look
  Classical,
  arXiv:0906.0315 [gr-qc].
\bibitem{Schwinger:1960qe}
  J.~S.~Schwinger,
  Brownian Motion of a Quantum Oscillator,
  J.\ Math.\ Phys.\  {\bf 2} (1961) 407.
\bibitem{Keldysh:1964ud}
  L.~V.~Keldysh,
  Diagram Technique for Nonequilibrium Processes,
  Zh.\ Eksp.\ Teor.\ Fiz.\  {\bf 47} (1964) 1515
  [Sov.\ Phys.\ JETP {\bf 20} (1965) 1018].
\bibitem{Koksma:2007uq}
  J.~F.~Koksma, T.~Prokopec and G.~I.~Rigopoulos,
  The Scalar Field Kernel in Cosmological Spaces,
  Class.\ Quant.\ Grav.\  {\bf 25} (2008) 125009
  [arXiv:0712.3685 [gr-qc]].
\bibitem{LeBellac:1996}
  M.~Le~Bellac,
  Thermal Field Theory,
  Cambridge monographs on Mathematical Physics,
  Cambridge University Press (1996).
\bibitem{Hu:1991di}
  B.~L.~Hu, J.~P.~Paz, Y.~-h.~Zhang,
  Quantum Brownian Motion in a General Environment: 1. Exact Master Equation with Nonlocal Dissipation and Colored Noise,
  Phys.\ Rev.\  {\bf D45 } (1992)  2843-2861.
\bibitem{Blume-Kohout:2003}
  R.~Blume-Kohout, W.~H.~Zurek
  Decoherence from a Chaotic Environment: an Upside-down "Oscillator" as a Model,
  Phys.\ Rev.\  A {\bf 68} (2003) 032104
  [quant-ph/0212153].
\bibitem{Baacke:1998di}
  J.~Baacke, K.~Heitmann and C.~Patzold,
  Nonequilibrium Dynamics of Fermions in a Spatially Homogeneous Scalar
  Background Field,
  Phys.\ Rev.\  D {\bf 58} (1998) 125013
  [arXiv:hep-ph/9806205].
\bibitem{Baacke:1999nq}
  J.~Baacke and C.~Patzold,
  Renormalization of the Nonequilibrium Dynamics of Fermions in a Flat  FRW
  Universe,
  Phys.\ Rev.\  D {\bf 62} (2000) 084008
  [arXiv:hep-ph/9912505].
\bibitem{Zurek:1991vd}
  W.~H.~Zurek,
  Decoherence and the Transition from Quantum to Classical,
  Phys.\ Today {\bf 44N10} (1991) 36.
\bibitem{Boyanovsky:2004dj}
  D.~Boyanovsky, K.~Davey and C.~M.~Ho,
  Particle Abundance in a Thermal Plasma: Quantum Kinetics vs. Boltzmann
  Equation,
  Phys.\ Rev.\  D {\bf 71} (2005) 023523
  [arXiv:hep-ph/0411042].
\bibitem{Herranen:2010mh}
  M.~Herranen, K.~Kainulainen, P.~M.~Rahkila,
  Coherent Quantum Boltzmann Equations from cQPA,
  JHEP {\bf 1012 } (2010)  072.
  [arXiv:1006.1929 [hep-ph]].
\bibitem{Birrell:1982ix}
  N.~D.~Birrell and P.~C.~W.~Davies,
  Quantum Fields in Curved Space,
  Cambridge monographs on Mathematical Physics,
  Cambridge University Press (1982).
\bibitem{Koksma:2010zy}
  J.~F.~Koksma, W.~Westra,
  A Causal Alternative to Feynman's Propagator,
  [arXiv:1012.3473 [hep-th]].
\bibitem{Westra:2010zx}
  W.~Westra,
  Localization of Particles in Quantum Field Theory,
  [arXiv:1012.3472 [hep-th]].
\bibitem{QuantumComputing}
  M.~A.~Nielsen, I.~L.~Chuang,
  Quantum Computation and Quantum Information,
  Cambridge University Press (2000);
\bibitem{Knill}
  E.~Knill, R.~Laflamme, and G.~J.~Milburn,
  A Scheme for Efficient Quantum Computation with Linear Optics,
  Nature {\bf 409} (2001) 46.
\end{thebibliography}
\end{document}